\documentclass[preprint,3p]{elsarticle}

\usepackage{amssymb}

\usepackage{graphicx}	
\usepackage{amsmath}	
\usepackage{comment}
\usepackage{float}
\usepackage{graphicx}
\usepackage{caption}
\usepackage{subcaption}
\usepackage{breqn}
\usepackage{amsmath}
\usepackage{flushend}
\usepackage[acronym]{glossaries}

\newacronym{isco}{ISCO}{Innermost Stable Circular Orbit}

\newacronym{bh}{BH}{Black Hole}

\newacronym{cop}{COP}{Compact Object Primary}

\newacronym{pnp}{PNP}{Pseudo-Newtonian Potential}

\newacronym{pw}{PW}{Paczynsky-Wiita pseudo-potential in \citet{paczynsky1980thick}}

\newacronym{abn}{ABN}{Artemova-Bj\"ornsson-Novikov pseudo-potential in \citet{artemova1996modified}}

\newacronym{gm}{GM}{Ghosh-Mukhopadhyay pseudo-potential in \citet{ghosh2007}}

\newacronym{mle}{MLE}{Maximum Lyapunov Exponent}

\newacronym{gmf}{GMf}{Ghosh-Mukhopadhyay pseudo-potential after fitting with $\left( V_{\text{GMf}} \right)$}

\newacronym{gms}{GMS}{Ghosh-Mukhopadhyay pseudo-potential for Schwarzschild-like geometry with $a=0$}

\journal{Chaos, Solitons \& Fractals}

\begin{document}

\begin{frontmatter}



\title{\textbf{Chaotic dynamics of off-equatorial orbits around pseudo-Newtonian compact objects with dipolar halos}}


\author[inst1]{Saikat Das\footnote{E-mail: saikatdas@iisc.ac.in (Corresponding author)}}

\affiliation[inst1]{organization={Joint Astronomy Programme, Department of Physics, Indian Institute of Science},
            city={Bangalore},
            postcode={560012}, 
            country={India}}

\author[inst2]{and Suparna Roychowdhury \footnote{E-mail: suparna@sxccal.edu}}

\affiliation[inst2]{organization={Department of Physics, St. Xavier’s College},
            addressline={30 Park Street}, 
            city={Kolkata},
            postcode={700016}, 
            country={India}}

\begin{abstract}
In this paper, we implement a generalised pseudo-Newtonian potential to study the off-equatorial orbits inclined at a certain angle with the equatorial plane around Schwarzschild and Kerr-like compact object primaries surrounded by a dipolar halo of matter. The chaotic dynamics of the orbits are detailed for both non-relativistic and special-relativistic test particles. The dependence of the degree of chaos on the Kerr parameter $a$ and the inclination angle $i$ is established individually using widely used indicators, such as the Poincar\'e Maps and the Maximum Lyapunov Exponents. Although the orbits' chaoticity has a positive correlation with $i$, the growth in the chaotic behaviour is not systematic. There is a threshold value of the inclination angle $i_{\text{c}}$, after which the degree of chaos sharply increases. On the other hand, the chaoticity of the inclined orbits anti-correlates with $a$ throughout its entire range. However, the negative correlation is systematic at lower values of the inclination angle. At higher values of $i$, the degree of chaos increases rapidly below a threshold value of the Kerr parameter, $a_{\text{c}}$. Above this threshold value, the correlation becomes weak. Furthermore, we establish a qualitative correlation between the threshold values and the overall chaoticity of the system. The studies performed with different orbital parameters and several initial conditions reveal the intricate nature of the system.
\end{abstract}

\begin{keyword}
Chaos \sep Black hole \sep Pseudo-Newtonian potential \sep Core-shell model \sep Off-equatorial orbit \sep Dragging of inertial frames

\textit{Received 6 November 2023; Accepted 17 December 2023}


\end{keyword}

\end{frontmatter}


\flushbottom


\section*{Acronyms}
\begin{tabular}{ll}
\textbf{ABN} & Artemova-Bj\"ornsson-Novikov pseudo-potential in \citet{artemova1996modified} \\

\textbf{BH} & Black Hole \\

\textbf{COP} & Compact Object Primary \\

\textbf{GM} & Ghosh-Mukhopadhyay pseudo-potential in \citet{ghosh2007} \\

\textbf{GMf} & Ghosh-Mukhopadhyay pseudo-potential after fitting with $\left( V_{\text{GMf}} \right)$ \\

\textbf{GMS} & Ghosh-Mukhopadhyay pseudo-potential for Schwarzschild-like geometry with $a=0$ \\

\textbf{ISCO} & Innermost Stable Circular Orbit \\

\textbf{MLE} & Maximum Lyapunov Exponent \\

\textbf{PNP} & Pseudo-Newtonian Potential \\

\textbf{PW} & Paczynsky-Wiita pseudo-potential in \citet{paczynsky1980thick} \\

\end{tabular}

\section{Introduction}
In the last two decades, the study of the dynamics of orbits around a single \acrshort{bh} \citep{nag2017influence, polcar2019free, dubeibe2021effect} or a \acrshort{bh} binary \citep{dubeibe2017pseudo, de2021beyond, alrebdi2022equilibrium} has become very popular. With the instrumental development of the detection and measuring techniques, it is already established that most of the galaxies consist of supermassive \acrshort{bh}s at their galactic centres \citep{kormendy1995inward, beckmann2012active}. In almost all cases, the \acrshort{bh}s are usually surrounded by a hollow spherical halo of matter and large accretion disks around them \citep{panagia1996nature, meyer1997formation}. Any particle travelling within the disk region or the corona will be influenced by both the central \acrshort{bh} and the halo around it, and both of them need to be considered while looking for the locus of the particle. This is where the Core-Shell Model comes into the picture while modelling systems like these \citep{vieira1999relativistic}. Not only the stand-alone \acrshort{bh}s or neutron stars with accretion disks around them but even the galaxies can be modelled using the core-shell scheme because of the observational evidence of their structures which consist of huge rings and shells around the supermassive \acrshort{bh}s located at their individual galactic centres \citep{sackett1990dark, arnaboldi1993studies, reshetnikov1997global, malin1983catalog, quinn1984formation, dupraz1987dynamical}.

In the present work, we are focusing on the two-body system, where one of them is a nonrotating or a rotating \acrshort{cop} surrounded by a hollow halo of matter, and the other one is a test particle with a unit mass. While studying the dynamics of the orbit of the test particle around the \acrshort{cop}, the phenomenon of chaos naturally comes into the picture. Several works have already been done on the chaotic behaviour of the orbits under the influence of the fully relativistic gravitational field of a \acrshort{bh}-halo system \citep{vieira1996chaos, letelier1997chaos, vieira1999relativistic, gueron2002geodesic, vogt2003exact, semerak2010free, semerak2012free, sukova2013free, janiuk2011different, witzany2015free}. Incorporating the charge into the system also brings out interesting results where the charged test particle is influenced by the gravitational field as well as the magnetic force of the central \acrshort{bh} along with its magnetosphere \citep{kopavcek2014inducing, kopavcek2015regular, takahashi2009chaotic, vittorio2021detection, vittorio2021timescales}. The spin of the \acrshort{cop} plays an essential role in the orbital dynamics around it. As most of the \acrshort{bh}s are rotating in nature \citep{miller2009stellar, ziolkowski2010population, daly2011estimates, reynolds2012probing, dotti2012orientation, tchekhovskoy2012prograde, garofalo2013retrograde, healy2014remnant, sesana2014linking}, we are more interested in studying the chaotic dynamics of the orbits around a Kerr-like \acrshort{bh}, or a \acrshort{cop} in general. 

While many research groups encounter the problem of accretion dynamics with a fully relativistic approach, it is very intensive and computationally demanding to simulate these systems exactly. Therefore, we will follow a beyond-Newtonian approach or a pseudo-Keplerian formalism, where a Newtonian-like potential is designed in such a way that it mimics the actual potential by sustaining the essential aspects of the spacetime around the \acrshort{cop} within a feasible limit of error. Indeed, the mimicking potential, known as the \acrshort{pnp}, will not precisely reproduce the fully relativistic scenario because of the nonlinearity in spacetime close to the event horizon. Nevertheless, it reproduces spacetime far from the event horizon with a high degree of accuracy as the relativistic nonlinearity weakens in this region. Furthermore, it simplifies the calculation and the computation significantly. Such a \acrshort{pnp} was first developed by \citet{paczynsky1980thick} (\acrshort{pw} \acrshort{pnp}), which is applicable for the orbits on or near the equatorial plane of a Schwarzschild-like \acrshort{cop}. After the successful implementation of this \acrshort{pnp}, many pseudo-potentials were introduced for both Schwarzschild and Kerr-type \acrshort{cop}s, each with different sets of advantages and drawbacks \citep{nowak1991diskoseismology, chakrabarti1992newtonian, Semerak1999pseudo, mukhopadhyay2002description, mukhopadhyay2003pseudo, ghosh2004rotating, chakrabarti2006studies}. Till now, one of the most popular and widely implemented \acrshort{pnp}s for a Kerr-like \acrshort{cop} is the one developed in \citet{artemova1996modified} (\acrshort{abn} \acrshort{pnp}). However, all of the mentioned \acrshort{pnp}s apply to the orbits near the equatorial plane. Even if they are prescribed for thick accretion disks, they do not include the inclination of the orbit in their models. For this reason, hardly any study has been done on the chaotic behaviour of inclined orbits in a pseudo-Newtonian framework. Nonetheless, a few studies have been performed on the dynamics of the off-equatorial orbits in a fully relativistic formalism \citep{kovavr2008off, kovavr2010off, barausse2007circular}. However, the chaotic nature of the equatorial orbits and their correlation with the rotation parameter has been widely studied using the previously available \acrshort{pnp}s \citep{gueron2001chaos, gueron2001chaotic, chen2003chaotic, letelier2011chaotic, wang2012dynamics, nag2017influence}. 

The potential developed in \citet{ghosh2007} (\acrshort{gm} \acrshort{pnp}) allows us to look for the chaotic dynamics of the off-equatorial orbits inclined at a certain angle $i$ with the equatorial plane. The generalised \acrshort{pnp} is useful to study the accretion dynamics around rotating and nonrotating \acrshort{bh}s in the off-equatorial planes \citep{ghosh2014newtonian}. The vector potential, derived from the generalised pseudo-Keplerian gravitational force prescribed in \citep{ghosh2007}, is also suitable for the hydrodynamical accretion studies of thick accretion disks \citep{bhattacharya2010disk}. Besides its applicability over a wide range of inclination angles ($0^\circ \leq i \leq 30^\circ$), the potential is valid for the entire range of the rotation parameter of the \acrshort{cop} ($-1 \leq a < 1$). As the \acrshort{pnp} is directly developed from the spacetime metric, it reproduces the values of the radius of marginally bound orbit $R_{\text{mb}}$, and the efficiency of unit mass at marginally stable orbit $E_{\text{ms}}$ with minimal error, and the radius of the \acrshort{isco}, $R_{\text{ISCO}}$, with no error at all. This helps us implement this \acrshort{pnp} to look at how the chaoticity of the off-equatorial orbits depends on the entire range of the Kerr parameter and the angle of inclination. The \acrshort{pnp} sustains the frame-dragging effect \citep{steklain2009stability, wang2012dynamics}, an important characteristic of the Kerr geometry, which is a direct consequence of the Lense-Thirring Precession of the orbits around a Kerr-like \acrshort{cop} \citep{lense1918influence}. 

As mentioned earlier, we have used the core-shell model in the present work because it closely resembles the realistic picture of an actual astrophysical \acrshort{bh} \citep{vieira1999relativistic}. We have used a dipolar perturbative term signifying the hollow halo of matter around the \acrshort{cop}. As opposed to the quadrupolar term used in this regard \citep{wang2012dynamics}, the dipolar term will be predominant when the halo is somewhat asymmetrically placed about the equatorial plane of the \acrshort{cop}, even if it is axially symmetric about the rotations axis of the system. The asymmetric mass distribution of the halo about the equatorial plane is more practical \citep{binney2008}. The dipolar term, corresponding to this asymmetric halo, has been used several times in literature \citep{gueron2001chaos, nag2017influence}.

In the current piece of work, we have implemented the \acrshort{gm} \acrshort{pnp} to study the off-equatorial orbits around a Schwarzschild and a Kerr-like \acrshort{cop}. The paper is organised as follows. In section \ref{ch:description_of_PNP}, we briefly describe the \acrshort{pnp}s and a numerical scheme to study the off-equatorial orbits. We have also presented the required equations of motion for both non-relativistic and special relativistic test particles. In section \ref{ch:chaos_poincare}, we study the stability of both circular and generic orbits. In the case of the latter, we use the previously developed equations of motion to study the chaotic dynamics of the orbits by generating the Poincar\'e Maps of sections of their phase-space trajectories. Through the analysis, we establish a correlation of the degree of chaos with the inclination angle $i$ of the orbit and the Kerr parameter $a$ of the \acrshort{cop} in a qualitative manner. In section \ref{ch:lyapunov}, we quantify the chaoticity using \acrshort{mle}, one of the widely used chaotic indicators implemented in this context. We implement them to corroborate the previously obtained qualitative results and to analyse the chaotic correlations in more depth. Finally, in section \ref{ch:conclusion}, we conclude our findings and discuss the future directions of our research.

\section{Formulation of the pseudo-Newtonian potentials and the equations of motion} \label{ch:description_of_PNP}
We begin our study by providing a mathematical description of the \acrshort{pnp}s, along with the necessary equations of motion to study the system's orbital dynamics. For the special case of equatorial orbits, we compare the effective potential consisting of \acrshort{gm} \acrshort{pnp} with that of \acrshort{abn} \acrshort{pnp}, another well-established \acrshort{pnp} widely used in this regard.

\subsection{Pseudo-Newtonian potential} \label{sch:PNP}
\begin{figure*}
        \centering
	\includegraphics[width=0.5\columnwidth]{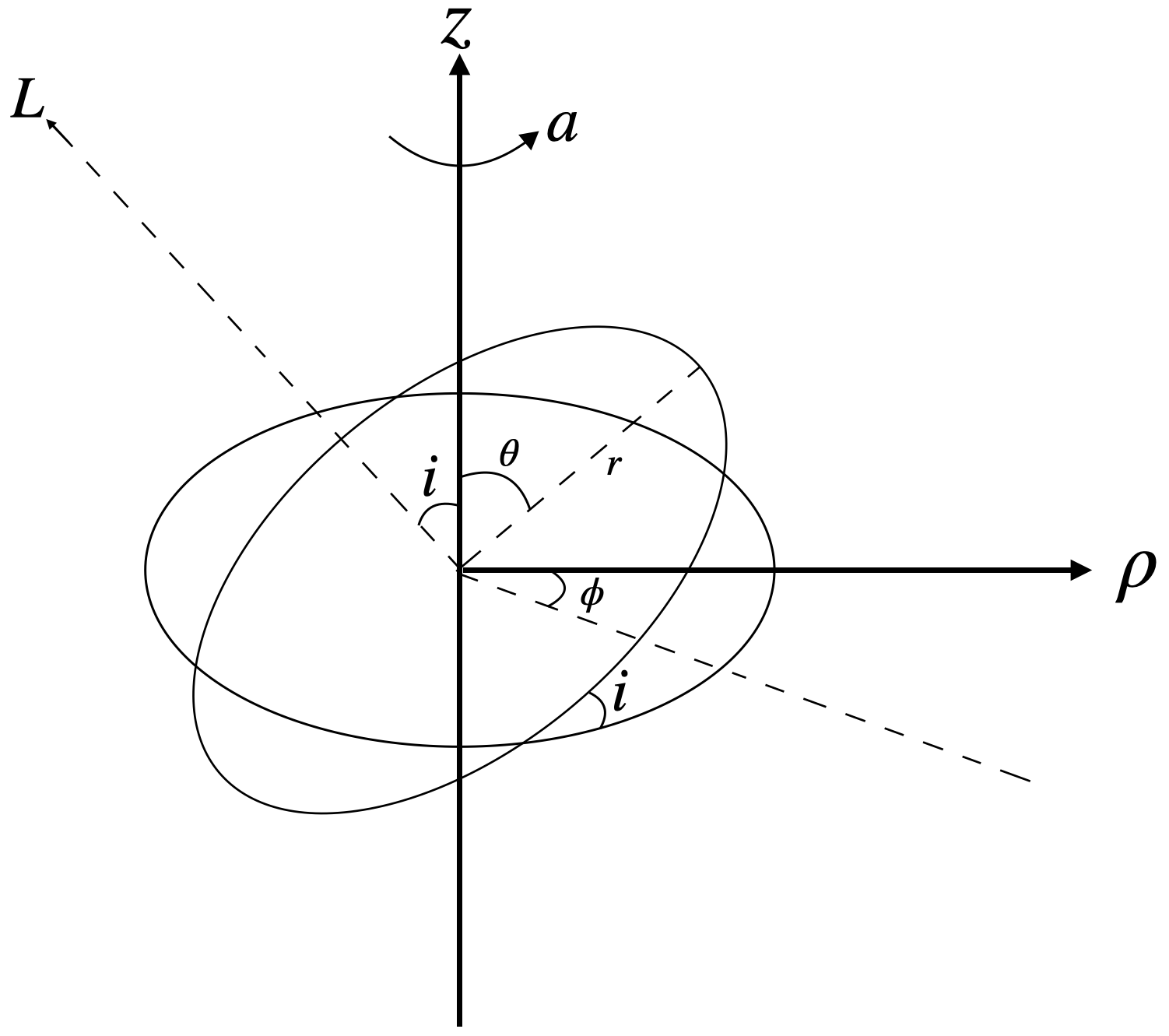}
    \caption{Different dynamical parameters of an off-equatorial orbit around a \acrshort{cop}. The inclined plane consists of a circular orbit along which the test particle rotates. The orbit is inclined at an angle $i$ with the symmetry axis of the \acrshort{cop} (the $z$-axis) and its equatorial plane. The rotation parameter of the \acrshort{cop} is $a$. $L$ denotes the total angular momentum of the test particle, and its direction is along the rotating axis of the inclined orbit.}
    \label{fig:off-equatorial plane}
\end{figure*}

If a particle of unit mass is accreting around a rotating \acrshort{cop} of mass $M_{\text{BH}}$ and angular momentum $J_\phi$ on a plane inclined at an angle $i$ with the equatorial plane of the \acrshort{cop} (Figure \ref{fig:off-equatorial plane}), the generalised gravitational force on the particle, as prescribed in \citet{ghosh2007}, can be mimicked by the expression  
\begin{equation} \label{eq:GMPNF}
    F_{\text{GM} r}=\frac{2 \mathcal{A}^2\sec^2i}{\left[a\sqrt{2}r^{3/2}\left\{\Delta+2r\left(r-1\right)\right\}+r\Delta\sqrt{ \left\{\mathcal{A}+r^4-a^2(\Delta+r^2-3r)\right\} \cos2i\sec^2i}\right]^2}
\end{equation}
where $\mathcal{A}=a^4+r^4+2a^2r(r-2)$ and $\Delta=r^2+a^2-2r$. The Kerr parameter $a=J_\phi / M_{\text{BH}}$ denotes the rotation parameter of the \acrshort{cop}, such that $a\in[-1,1)$. Here, $r$ is the radial distance of the rotating particle from the origin in the spherical polar coordinate $(r,\theta,\phi)$. The gravitational force on the particles, given in equation (\ref{eq:GMPNF}), can be applied successfully by taking the value of $i$ in the range $i\in[0^\circ,30^\circ]$ while keeping the error within a reasonable limit, and maintaining a resemblance with the actual physical scenario \citep{ghosh2007}. The corresponding pseudo-Kerr potential $V_{\text{GM}}$, for particular values of $a$ and $i$, can be evaluated from the generalised force using the relation
\begin{equation} \label{eq:Pot_Def}
    V_{\text{GM}}(r,a,i)=-\int_r^\infty F_{\text{GM} r}(r',a,i)dr'
\end{equation}

For Schwarzschild-like \acrshort{cop}s, the value of $a$ becomes $0$. Thus, by putting $a=0$ in the equation (\ref{eq:Pot_Def}), we get the \acrshort{pnp} for the inclined orbits around a Schwarzschild-like \acrshort{cop}, which is given by
\begin{equation} \label{eq:GMPW}
    V_{\text{GMS}}(r)=-\frac{\sec(2i)}{r-2}
\end{equation}
It should be mentioned that throughout the work, the velocities have been scaled by $c$, the speed of light in vacuum, and the distances are scaled by $r_{\text{g}}=GM_{\text{BH}}/c^2$, where $G$ is the gravitational constant. We have scaled all the parameters such that $G=M_{\text{BH}}=c=1$.

The potential, given in equation (\ref{eq:GMPW}), is strikingly similar to the \acrshort{pnp} presented in \citet{paczynsky1980thick}. By putting $i=0$, $V_{\text{GMS}}$ gets equal to the \acrshort{pw} potential. The fact that $V_{\text{GM}}$ comes down to the \acrshort{pw} \acrshort{pnp} for $a=i=0$ signifies the generic nature of the \acrshort{gm} \acrshort{pnp}. Similarly, if we put $i=0$ but $a\neq 0$, we get the \acrshort{pnp} for the orbits on the equatorial plane in the Kerr geometry. 

The analytically closed form of the generalised \acrshort{pnp} $V_{\text{GM}}$, however, is complicated to derive by a straightforward integration, as per the expression given in equation (\ref{eq:Pot_Def}), because of the complex mathematical form of the gravitational force, given in equation (\ref{eq:GMPNF}). Therefore, we take a numerical approach where we integrate the generalised force $F_{\text{GM} r}$ numerically for several values of $r$ and fit the data with a fitting function of the form 
\begin{equation} \label{eq:GMPNP}
    V_{\text{GMf}}(r)=-\frac{\exp\left(\frac{\gamma_1}{(r-\gamma_2)^{\gamma_3}}+\gamma_4\right)}{(r-\gamma_2)^{\gamma_5}}
\end{equation}
where $\gamma_1,\gamma_2,\gamma_3,\gamma_4$, and $\gamma_5$ are fitting parameters. It is evident that the potential $V_{\text{GMf}}$ is spherically symmetric, similar to the potential $V_{\text{GM}}$. The fitting function asymptotically approaches zero when the test particle moves towards infinity $\left( V_{\text{GMf}} \rightarrow 0 \text{, when } r \rightarrow \infty \right)$. Also, it tends to infinity when the test particle approaches the event horizon at the radial distance $r=\gamma_2 r_{\text{g}}$ $\left( V_{\text{GMf}} \rightarrow - \infty \text{, when } r \rightarrow \gamma_2 \right)$. Moreover, the function $V_{\text{GMf}}$ fits with the \acrshort{pnp} $V_{\text{GM}}$ efficiently with a minimal relative error, keeping the radii of the \acrshort{isco} $R_{\text{ISCO}}$, and the marginally bound orbit $R_{\text{mb}}$ to be almost same as that in the Kerr geometry. These facts satisfy the conditions of a \acrshort{pnp}, and it leads us to implement $V_{\text{GMf}}$, given in equation (\ref{eq:GMPNP}), as a suitable fitting function. For particular values of $a$ and $i$, the fitting parameters are evaluated so that $V_{\text{GMf}}$ fits with $V_{\text{GM}}$, and after that we can use the fitting function as an independent \acrshort{pnp}. 

As there is azimuthal symmetry in the orbital motion of the particle revolving around a \acrshort{cop}, it is convenient to write equation (\ref{eq:GMPNP}) in the cylindrical polar coordinates $(\rho,\phi,z)$ as 
\begin{equation} \label{eq:monopole}
    V_{\text{GMf}}(\rho,z)=-\frac{\exp\left(\frac{\gamma_1}{\left(\sqrt{\rho^2+z^2}-\gamma_2\right)^{\gamma_3}}+\gamma_4\right)}{\left(\sqrt{\rho^2+z^2}-\gamma_2\right)^{\gamma_5}}
\end{equation}
The absence of any $\phi$ term in the potential signifies the azimuthal symmetry. Section \ref{ch:rotating_cop} presents more details about the fitting function.

As the \acrshort{gm} \acrshort{pnp} is generic in nature, it is worthwhile to ask how this potential will behave for the orbits on the equatorial plane around a rotating \acrshort{cop} ($i=0$ but $a \neq 0$) and how this potential differs from any other \acrshort{pnp} which is being applied in this scenario. We consider the well-known \acrshort{abn} \acrshort{pnp} to draw a comparison with \acrshort{gmf} \acrshort{pnp} for $i=0$. 

The \acrshort{abn} \acrshort{pnp}, after integrating the free-fall acceleration prescribed in \citet{artemova1996modified}, comes out to be
\begin{equation} \label{eq:APNP}
    V_{\text{ABN}}(r)=-\frac{1}{r_1 (\beta-1)} \left[ \frac{r^{\beta-1}}{(r-r_1)^{\beta -1}} -1 \right]
\end{equation}
Here, $r_1$ represents the radius of the event horizon, given as follows.
\begin{equation} \label{eq:ABN_event_horizon}
    r_1 = 1+\sqrt{1-a^2}
\end{equation}
The value of $\beta$ is given by
\begin{equation} \label{eq:ABN_beta}
    \beta=\frac{r_{\text{in}}}{r_1}-1
\end{equation}
where $r_{\text{in}}$ is the radius of \acrshort{isco} such that
\begin{equation} \label{eq:ABN_r_in}
    r_{\text{in}}=3+Z_2 \mp \sqrt{(3-Z_1)(3+Z_1+2Z_2)}
\end{equation}
The values of the parameters $Z_1$ and $Z_2$ are defined as
\begin{equation}
    Z_1=1+(1-a^2)^{\frac{1}{3}} \left[ (1+a)^{\frac{1}{3}}+(1-a)^{\frac{1}{3}} \right] \text{ and }
    Z_2=\sqrt{3a^2+Z_1^2}
\end{equation}
The negative and positive signs in equation (\ref{eq:ABN_r_in}) signify the co-rotating ($a>0$) and counter-rotating ($a<0$) \acrshort{cop}s, respectively.

Along with the monopole term represented by any of the \acrshort{pnp}s, which is conservative, we use a dipolar, non-central, perturbative term, which is introduced to simulate a halo far from the central object. The distribution of the halo is such that it is axially symmetric about the rotating axis of the \acrshort{cop}. If the halo is also symmetric about the equatorial plane, the first non-central, contributing term in the potential will be quadrupolar in nature \citep{vieira1999relativistic, wang2012dynamics}. However, we consider the halo to be asymmetrically placed about the equatorial plane, somewhat deformed in the polar regions \citep{binney2008}. In this case, the non-central, leading term will be dipolar, which can be represented with a potential 
\begin{equation}
    \psi_{\text{dip}} = \mathcal{D} r P_1(\cos \theta) = \mathcal{D} z
\end{equation}
Here, $\mathcal{D}$ is the dipole coefficient, $P_n(x)$ is the Legendre polynomial of degree $n$, $\cos \theta = z/r$, and $r^2 = \rho^2 + z^2$. The value of $\mathcal{D}$ will depend on the asymmetry in the mass distribution of the halo, and it will be small for any practical system \citep{vieira1999relativistic, nag2017influence}. Therefore, the net potential that the particle experiences while revolving around a \acrshort{cop} surrounded by an asymmetrically distributed halo can be represented by a linear superposition of the potentials, \textit{i.e.},
\begin{equation} \label{eq:phi_g}
    \Phi_{\text{g}}(\rho,z)=V_{\text{PNP}}(\rho,z)+\psi_{\text{dip}}
\end{equation}
The pseudo-potential term $V_{\text{PNP}}$ in equation (\ref{eq:phi_g}) will be replaced by either of the monopole terms, which is $V_{\text{GM}}$, $V_{\text{GMS}}$, $V_{\text{GMf}}$, or $V_{\text{ABN}}$, depending on the nature of the \acrshort{cop} under consideration. The linear nature of the Newtonian framework allows the superposition of the potentials, in this case \citep{wang2012dynamics}. 

The orbit of the test particle, with specific angular momentum $L$, is inclined with respect to the equatorial plane with an inclination angle $i$, and it is precessing about the spinning axis of the \acrshort{cop}. This phenomenon is known as the Lense-Thirring Precession \citep{lense1918influence}. The $z$-component of the angular momentum $L_z$ is conserved due to the azimuthal symmetry of the orbit. The locus of the particle will not always make an angle $i$ with the equatorial plane. Still, the plane of orbital precession will always be inclined at $i$ with the spinning axis of the \acrshort{cop} (Figure \ref{fig:off-equatorial plane}). Therefore, $L_z=L \cos i$, which implies that $L$ is also conserved \citep{ghosh2007}. 

The motion of the orbiting particle in the $\phi$ direction will induce a centrifugal force on it along the radial direction. This can be taken into account by adding a term $\left( \dfrac{L^2}{2\rho^2} \right)$ with the potential $\Phi_{\text{g}}(\rho,z)$ in equation (\ref{eq:phi_g}). Hence, the following expression can describe the effective potential in Kerr geometry.  
\begin{equation} \label{eq:overall_potential_general}
    \left.V_{\text{eff}}(\rho,z)\right|_{a \neq 0}=-\frac{\exp\left(\frac{\gamma_1}{\left(\sqrt{\rho^2+z^2}-\gamma_2\right)^{\gamma_3}}+\gamma_4\right)}{\left(\sqrt{\rho^2+z^2}-\gamma_2\right)^{\gamma_5}}+\mathcal{D} z+\frac{L^2}{2\rho^2}
\end{equation}
The net potential for an off-equatorial orbit in Schwarzschild geometry $(a=0)$ is more straightforward. We can use the potential given in equation (\ref{eq:GMPW}) and write down the net effective potential as follows.
\begin{equation} \label{eq:overall_potential_Sch}
    \left.V_{\text{eff}}(\rho,z)\right|_{a=0}=-\frac{\sec(2i)}{\sqrt{\rho^2+z^2}-2}+\mathcal{D} z+\frac{L^2}{2\rho^2}
\end{equation}
The dynamics of the test particle will be governed by these effective potentials, the choice of which depends on the type of \acrshort{cop} it is rotating around. 

\subsection{Equations of motion} \label{sch:EOM}
We move on to formulating the equations of motion for our system. The equations of motion of the test particle can be derived from the non-dimensional Hamiltonian \citep{wang2011gravitational, wang2012dynamics}, given by
\begin{equation}
    H = \frac{1}{2} \left( p_\rho^2 + p_z^2 \right) + V_{\text{eff}} (\rho,z)
\end{equation}
where $V_{\text{eff}}$ can be replaced by either the potential given in equation (\ref{eq:overall_potential_general}) for a Kerr-like \acrshort{cop} or the one given in equation (\ref{eq:overall_potential_Sch}) for a Schwarzschild-like \acrshort{cop}. Thus, the equations of motion can be formulated as follows.
\begin{subequations} \label{eq:eom_nrel}
	\begin{alignat} {4}
		&\label{eq:rho_dot_nrel}\dot{\rho}=p_\rho\\
		&\label{eq:prho_dot_nrel}\dot{p_\rho}=-\frac{\partial V_{\text{eff}}}{\partial \rho}=-\frac{\partial \Phi_{\text{g}}}{\partial \rho}+\frac{L^2}{\rho^3}\\ 
		&\label{eq:z_dot_nrel}\dot{z}=p_z\\
		&\label{eq:pz_dot_nrel}\dot{p_z}=-\frac{\partial V_{\text{eff}}}{\partial z}=-\frac{\partial \Phi_{\text{g}}}{\partial z}
	\end{alignat}
\end{subequations}
Here, there is no special relativistic correction as the particle's speed is assumed to be much less than $c$. The energy of the test particle is given by $E=\sqrt{1+2E_{\text{mech}}}$, where 
\begin{equation} \label{eq:energy_mechanical}
    E_{\text{mech}} = \frac{1}{2} \left( p_\rho^2 + p_z^2 \right) + \Phi_{\text{g}} (\rho,z) + \frac{L^2}{2\rho^2}
\end{equation}
Therefore, the conservation equation can be written as 
\begin{equation} \label{eq:energy_cons_newt}
    E^2 - p_\rho^2 - p_z^2 - 2\Phi_{\text{g}} - \frac{L^2}{\rho^2} = 1
\end{equation}
From equation (\ref{eq:energy_cons_newt}), it can be concluded that the motion of the test particle will be restricted by $E^2 - 2\Phi_{\text{g}} - \dfrac{L^2}{\rho^2} - 1 \geq 0$, which is a direct consequence of the conservation of energy and angular momentum. 

However, if the speed of the particle is comparable to the speed of light, special relativistic corrections to the equations in (\ref{eq:rho_dot_nrel}-\ref{eq:pz_dot_nrel}) have to be performed \citep{nag2017influence,gueron2001chaos}. We consider the non-dimensional Lagrangian of the relativistic particle $\mathcal{L}=-\dfrac{1}{\gamma}-\Phi_{\text{g}}$, where the term $\gamma$ is given by
\begin{equation*}
    \gamma = \frac{1}{\sqrt{1-\left( \dot{\rho}^2 + \rho^2 \dot{\phi}^2 + \dot{z}^2 \right)}}
\end{equation*}
The energy of the particle will be $E=\gamma+\Phi_{\text{g}}$, and the angular momentum of the particle will be $L=\dfrac{\partial \mathcal{L}}{\partial \dot{\phi}}=\gamma \rho^2 \dot{\phi}$. The Lagrange’s equations for $\rho$ and $z$ can be found as
\begin{subequations} \label{eq:lagrange_eqn}
	\begin{alignat} {2}
		&\label{eq:lagrange_eqn_rho}\left( \Phi_{\text{g}} - E \right)\Ddot{\rho} = \frac{\partial \Phi_{\text{g}}}{\partial \rho} \left( 1- \dot{\rho}^2 \right) - \frac{\partial \Phi_{\text{g}}}{\partial z} \dot{\rho} \dot{z} - \frac{L^2}{(E-\Phi_{\text{g}}) \rho^3}\\
		&\label{eq:lagrange_eqn_z}\left( \Phi_{\text{g}} - E \right)\Ddot{z} = \frac{\partial \Phi_{\text{g}}}{\partial z} \left( 1- \dot{z}^2 \right) - \frac{\partial \Phi_{\text{g}}}{\partial \rho} \dot{\rho} \dot{z}
	\end{alignat}
\end{subequations}
As we have considered the test particle of unit mass, we can write $\dot{\rho}=p_\rho$ and $\dot{z}=p_z$. Thus, the relativistic equations of motion turn out to be
\begin{subequations} \label{eq:eom_rel}
	\begin{alignat} {4}
		&\label{eq:rho_dot_rel}\dot{\rho}=p_\rho\\
		&\label{eq:prho_dot_rel}\dot{p_\rho}=\frac{1}{\Phi_{\text{g}}-E}\left[ \frac{\partial \Phi_{\text{g}}}{\partial \rho} \left(1-p_\rho^2 \right)-\frac{\partial \Phi_{\text{g}}}{\partial z}p_z p_\rho - \frac{L^2}{(E-\Phi_{\text{g}}) \rho^3} \right]\\ 
		&\label{eq:z_dot_rel}\dot{z}=p_z\\
		&\label{eq:pz_dot_rel}\dot{p_z}=\frac{1}{\Phi_{\text{g}}-E}\left[ \frac{\partial \Phi_{\text{g}}}{\partial z} \left(1-p_z^2 \right)-\frac{\partial \Phi_{\text{g}}}{\partial \rho}p_z p_\rho \right]
	\end{alignat}
\end{subequations}
From the expressions of energy $E$ and angular momentum $L$, along with the relation $u^\mu u_\mu = 1$, the conservation equation can be evaluated as
\begin{equation} \label{eq:energy_cons}
    p_\rho^2+p_z^2+\frac{L^2}{(E-\Phi_{\text{g}})^2 \rho^2}+\frac{1}{(E-\Phi_{\text{g}})^2}=1
\end{equation}
Similar to the non-relativistic case, the motion of the relativistic test particle will be restricted within the region governed by $1-\dfrac{L^2}{(E-\Phi_{\text{g}})^2 \rho^2}-\dfrac{1}{(E-\Phi_{\text{g}})^2} \geq 1$.

It is worth mentioning that similar to the velocities and distances, all the terms signifying the physical quantities have been scaled to make them non-dimensional. The time $t$ has been scaled as $ct/r_{\text{g}}$, the momentum $p$ has been scaled as $p/(\mu c)$, and the angular momentum has been scaled as $L/(\mu c r_{\text{g}})$, where $\mu$ is the mass of the test particle taken to be unity in our analysis. 

The energy conservation equations, given in equation (\ref{eq:energy_cons_newt}) for non-relativistic case and equation (\ref{eq:energy_cons}) for relativistic case, converts the 4-dimensional phase space consisting of $(\rho,p_\rho,z,p_z)$ into an effective 3-dimensional hypersurface consisting of $(\rho,p_\rho,z)$. The value of $p_z$ is bound to satisfy equation (\ref{eq:energy_cons}) for a particular set of the values of $(\rho,p_\rho,z)$. This provides comprehensibility in analysing the system and also gives the advantage of minimising errors in numerical calculations.

\subsection{Comparison between \acrshort{gmf} \acrshort{pnp} and \acrshort{abn} \acrshort{pnp} for equatorial orbits} \label{sch:comp_GMf_ABN}
Before implementing the \acrshort{gmf} \acrshort{pnp} in studying the off-equatorial orbits, it is essential to consider the \acrshort{pnp} for the equatorial orbits by putting the inclination angle $i=0^\circ$. In this context, we compare the two pseudo-potentials, namely the \acrshort{gmf} \acrshort{pnp} and the \acrshort{abn} \acrshort{pnp}, and investigate which is more appropriate for sustaining the essential aspects of general relativity. Both potentials reproduce the radii of the \acrshort{isco} to be the same as that in the fully relativistic case. Both of them reproduce the values of the radii of marginally bound orbits ($R_{\text{mb}}$) and the energy of unit mass at the marginally stable orbits ($E_{\text{ms}}$) with similar accuracy. The margin of dissimilarity being significantly less, both the \acrshort{pnp}s can be implemented efficiently to study the accretion dynamics of equatorial disks. However, the simpler mathematical form of \acrshort{abn} \acrshort{pnp} makes it more comprehensive and convenient to implement in complicated mathematical problems, for example, the Hill problem \cite{steklain2009stability}. For off-equatorial orbits, the \acrshort{gm} \acrshort{pnp} is possibly one of the mandatory choices despite having a complex mathematical form.  

\begin{figure*} 
    
     \centering
     \begin{subfigure}[b]{0.495\textwidth}
         \centering
         \includegraphics[width=\textwidth]{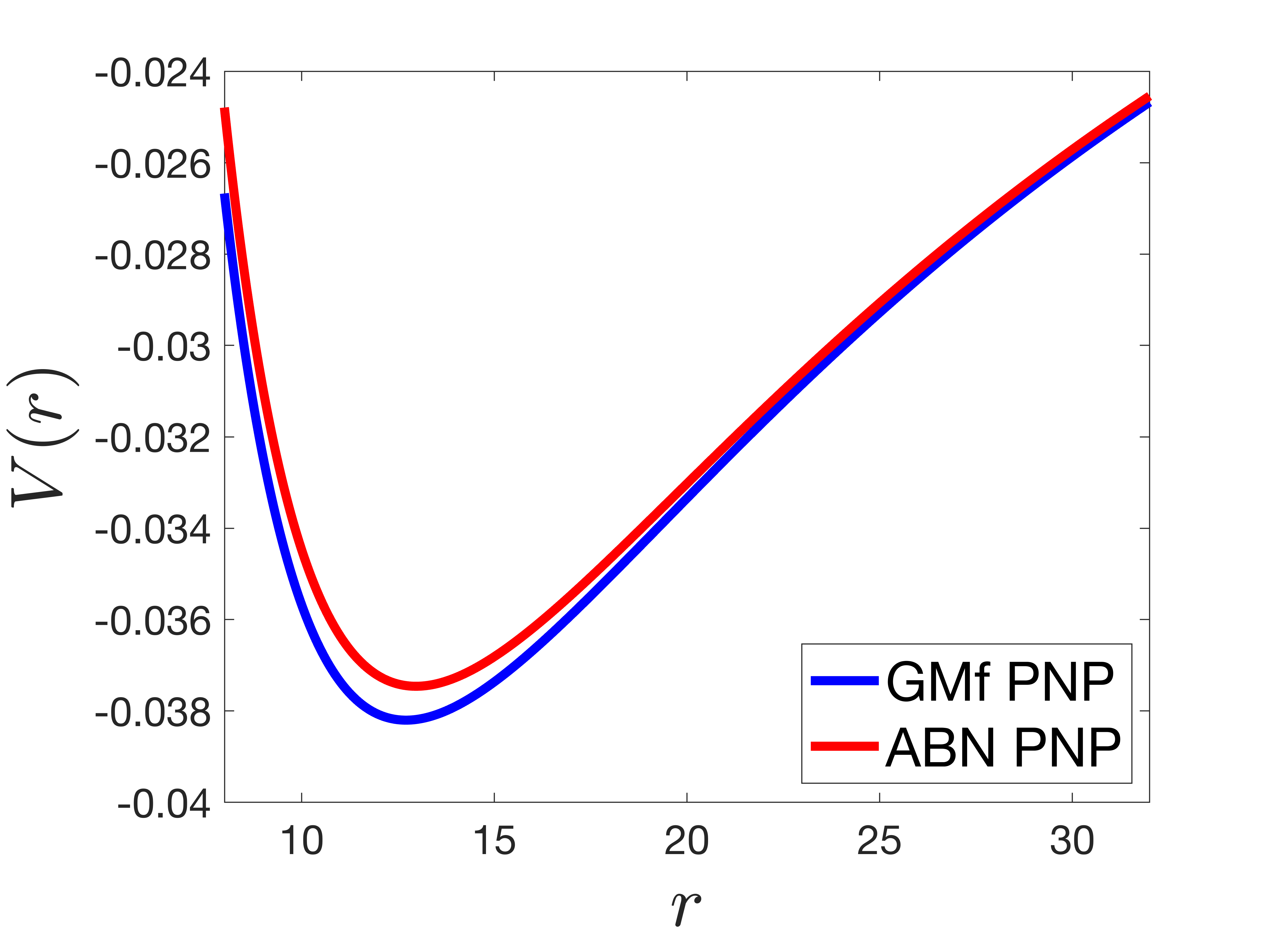}
         \caption{$a$=0.1}
         \label{fig:comparison_of_potentials_a_1}
     \end{subfigure}
     \hfill
     \begin{subfigure}[b]{0.495\textwidth}
         \centering
         \includegraphics[width=\textwidth]{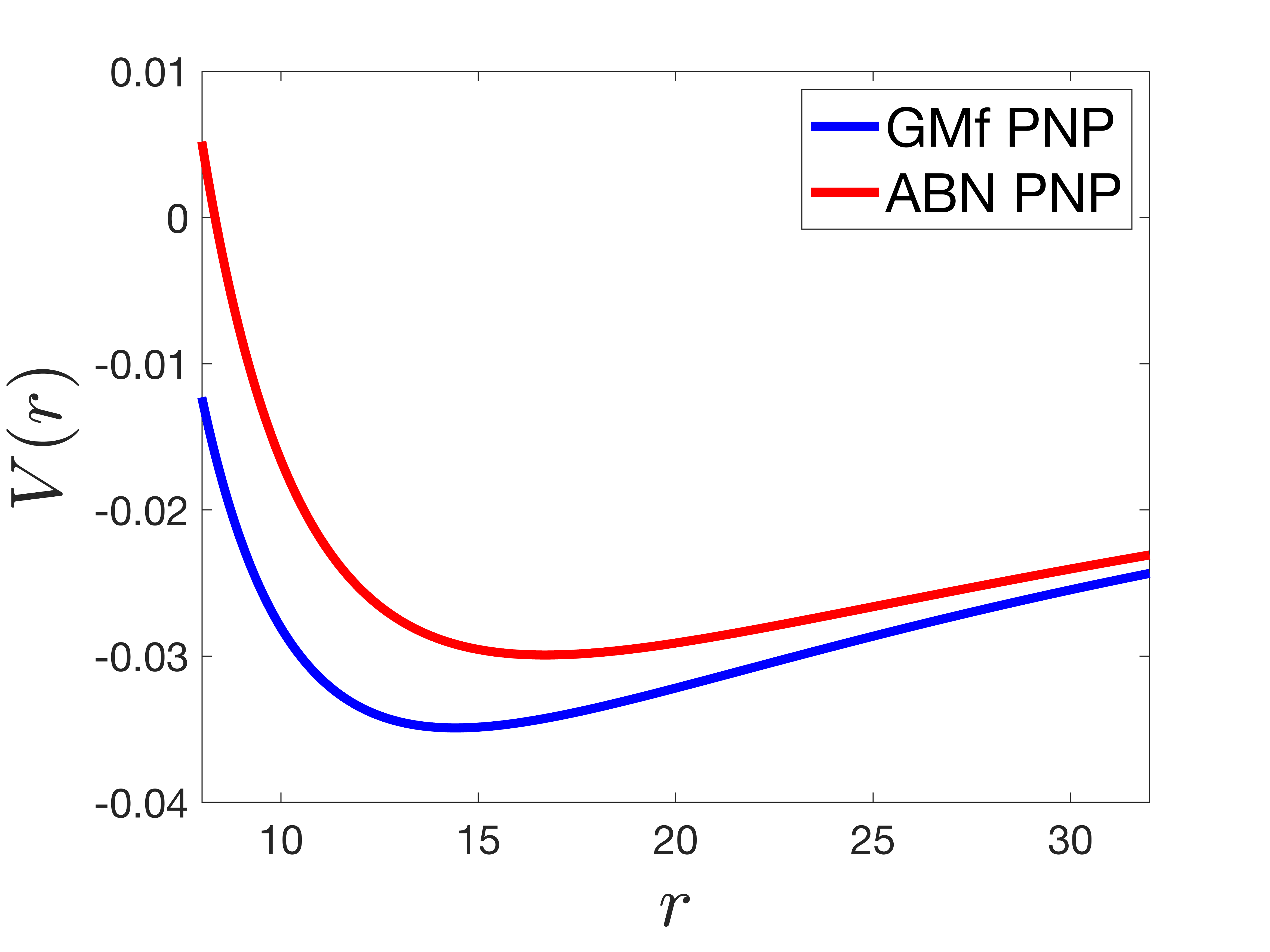}
         \caption{$a$=0.9}
         \label{fig:comparison_of_potentials_a_2}
     \end{subfigure}
     \hfill
     \begin{subfigure}[b]{0.495\textwidth}
         \centering
         \includegraphics[width=\textwidth]{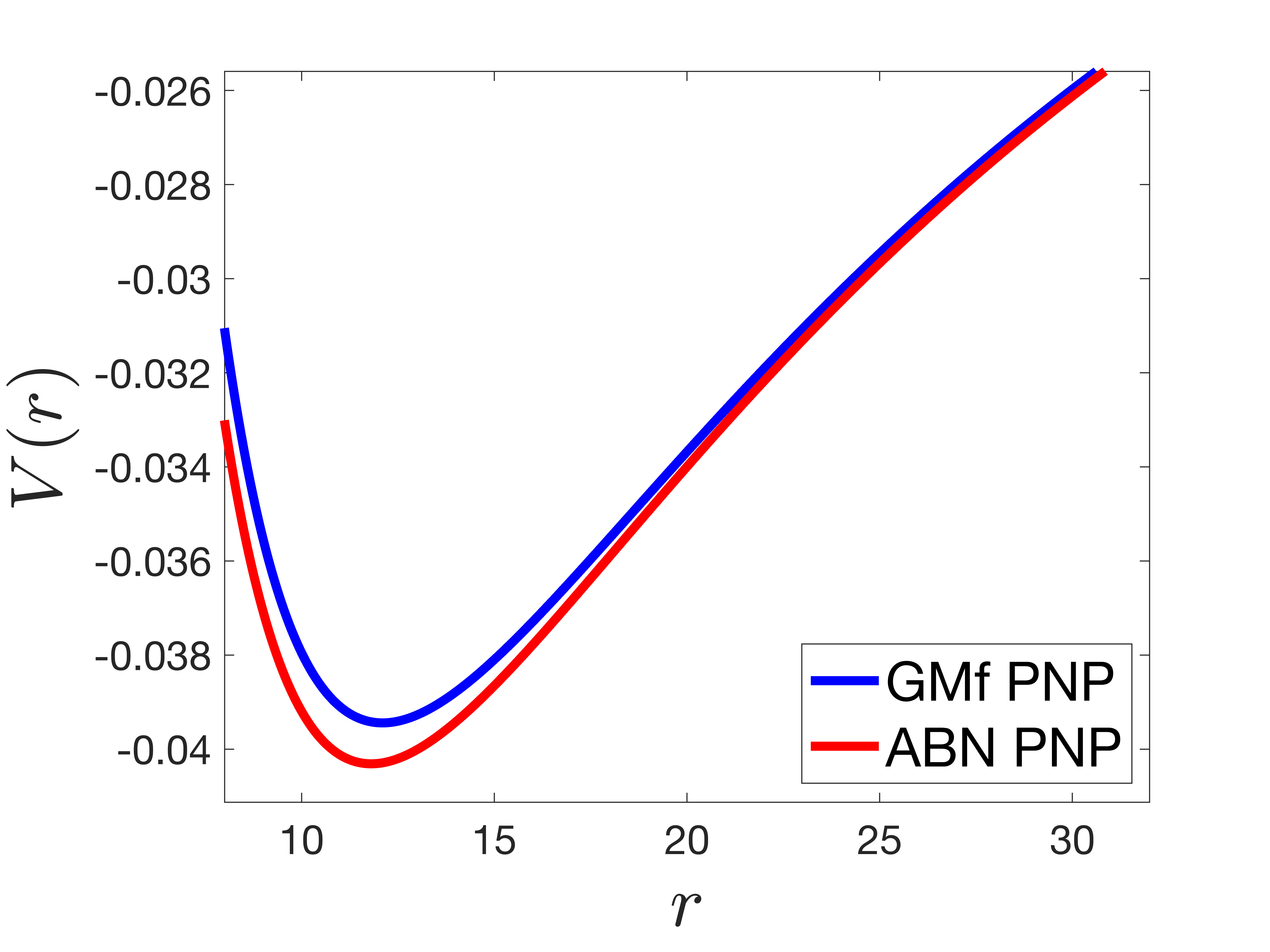}
         \caption{$a$=-0.1}
         \label{fig:comparison_of_potentials_a_3}
     \end{subfigure}
     \hfill
     \begin{subfigure}[b]{0.495\textwidth}
         \centering
         \includegraphics[width=\textwidth]{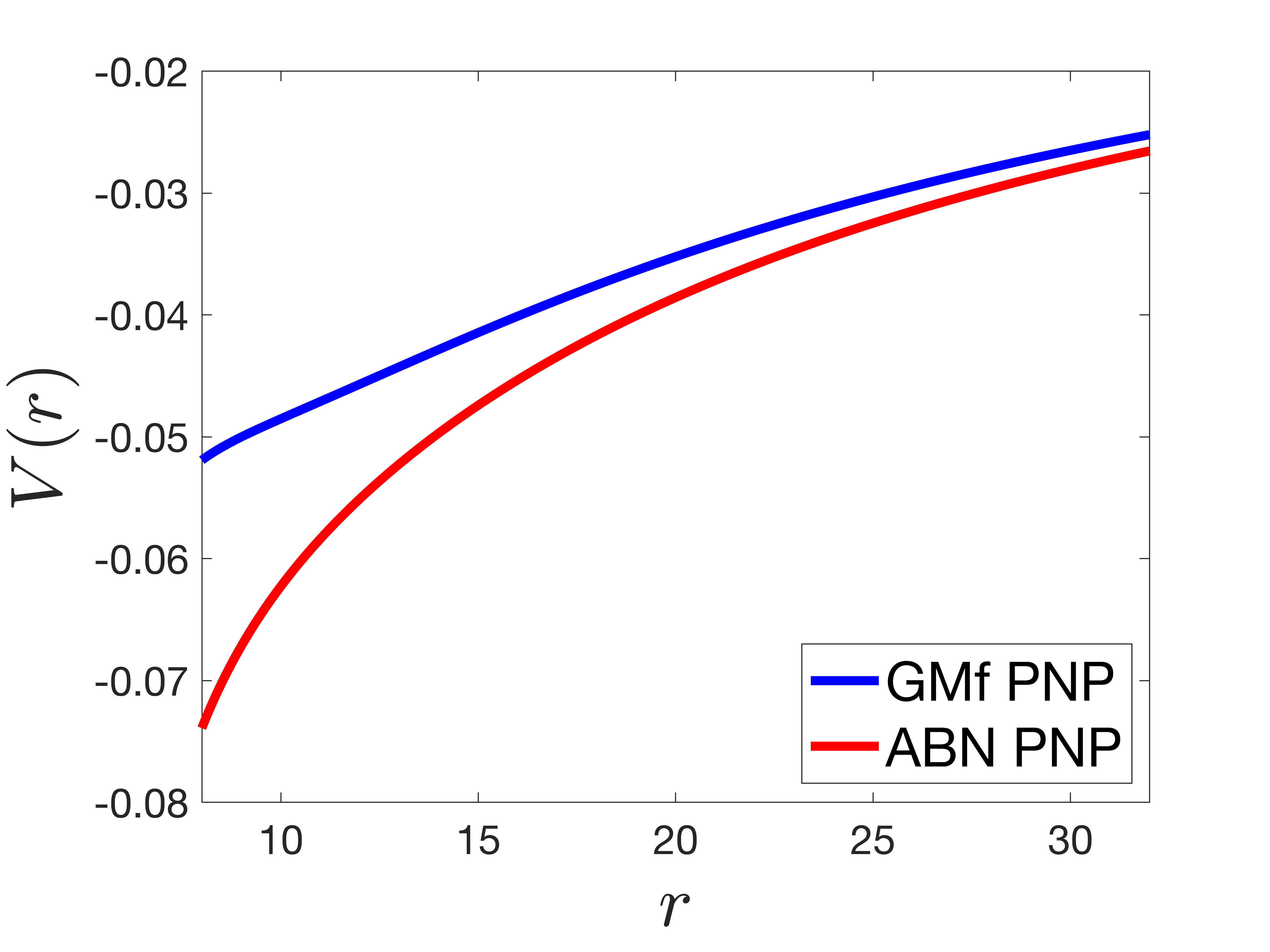}
         \caption{$a$=-0.9}
         \label{fig:comparison_of_potentials_a_4}
     \end{subfigure}
        \caption{Comparison between the \acrshort{gmf} \acrshort{pnp} $\left(V(r)=V_{\text{GMf}}(r)+L^2/2r^2\right)$ and \acrshort{abn} \acrshort{pnp} $\left(V(r)=V_{\text{ABN}}(r)+L^2/2r^2\right)$, for $z=0$, $L=4.2$, and different values of the Kerr parameter $a$. }
        \label{fig:comparison_of_potentials_a}
\end{figure*}

If we find out the differences between the two \acrshort{pnp}s at higher radial distances far from the event horizon, we can see a trend in their behaviour, which has a correlation with the rotation parameter $a$. Let us consider the effective potential $V(r)$ consisting of the \acrshort{gmf} \acrshort{pnp} $V_{\text{GMf}}(r)$, or the \acrshort{abn} \acrshort{pnp} $V_{\text{ABN}}(r)$, along with the centrifugal contribution $\left( \dfrac{L^2}{2r^2} \right)$. At $a=0$, which represents a Schwarzschild-like \acrshort{cop}, both the \acrshort{pnp}s come down to the \acrshort{pw} \acrshort{pnp}. That is why both \acrshort{abn} \acrshort{pnp} and \acrshort{gmf} \acrshort{pnp} are equal to each other for $a=0$. For $a>0$, which represents a co-rotating Kerr-like \acrshort{cop}, the effective potential consisting of the \acrshort{abn} \acrshort{pnp} is more than that of the \acrshort{gmf} \acrshort{pnp} (Figure \ref{fig:comparison_of_potentials_a_1}-\ref{fig:comparison_of_potentials_a_2}). For $a<0$, which represents a counter-rotating Kerr-like \acrshort{cop}, the situation is opposite (Figure \ref{fig:comparison_of_potentials_a_3}-\ref{fig:comparison_of_potentials_a_4}). In this case, the effective potential consisting of the \acrshort{abn} \acrshort{pnp} is less compared to that of the \acrshort{gmf} \acrshort{pnp}. This impacts the chaotic behaviour of the equatorial orbits in a certain way, which has been studied in section \ref{ch:chaos_poincare}.

\section{Stability of orbits and visualization of chaos} \label{ch:chaos_poincare}

In this section, we analyse the dynamical behaviours of the orbits around Schwarzschild or Kerr-like \acrshort{cop}. We begin with the special case of stable circular orbits. Thereafter, we move on to the behaviours of generic orbits, where the coexistence of order and chaos occurs naturally. We look for the transitions from order to chaos and establish how the chaoticity correlates with the orbital parameters qualitatively using the Poincar\'e sections of the orbits.

\subsection{Dynamical analysis of circular orbits}
A circular orbit essentially satisfies the relations $\dot{r}=0$ and $\dot{p_r}=0$. As the gravitational force corresponding to the \acrshort{pnp}s is central in nature, we can expect stable circular orbits to exist around the \acrshort{cop}s. However, the dipolar perturbation term in equation (\ref{eq:phi_g}) restricts the stable circular orbits to exist on any plane other than the equatorial plane where the dipolar term vanishes as $z=0$. In the off-equatorial planes, the dipolar term disrupts the spherical symmetry of the system. However, we are also interested in considering the off-equatorial planes. Therefore, we suppress the dipolar term for the time being by putting $\mathcal{D}=0$ in equation (\ref{eq:phi_g}).  

For an orbit around a \acrshort{cop} with specific values of $a$ and $i$ to be circular, the relation $\dfrac{\partial V_{\text{eff}}}{\partial r}=0$ is required to be satisfied. For it to be stable, the extremum has to be a minimum. Therefore, the condition for an orbit to be a stable circular orbit turns out to be
\begin{equation} \label{eq:stability_cond}
    \frac{\partial V_{\text{eff}}}{\partial r}=0 \text{, and } \frac{\partial^2 V_{\text{eff}}}{\partial r^2} \geq 0
\end{equation}
Instead of the inequality, if we equate the second relation in equation (\ref{eq:stability_cond}) with 0 and solve the equations simultaneously, the solution essentially gives the radius of the \acrshort{isco} for given values of $a$ and $i$ \cite{shapiro1983black}.

\begin{figure*} 
    
     \centering
     \begin{subfigure}[b]{0.495\textwidth}
         \centering
         \includegraphics[width=\textwidth]{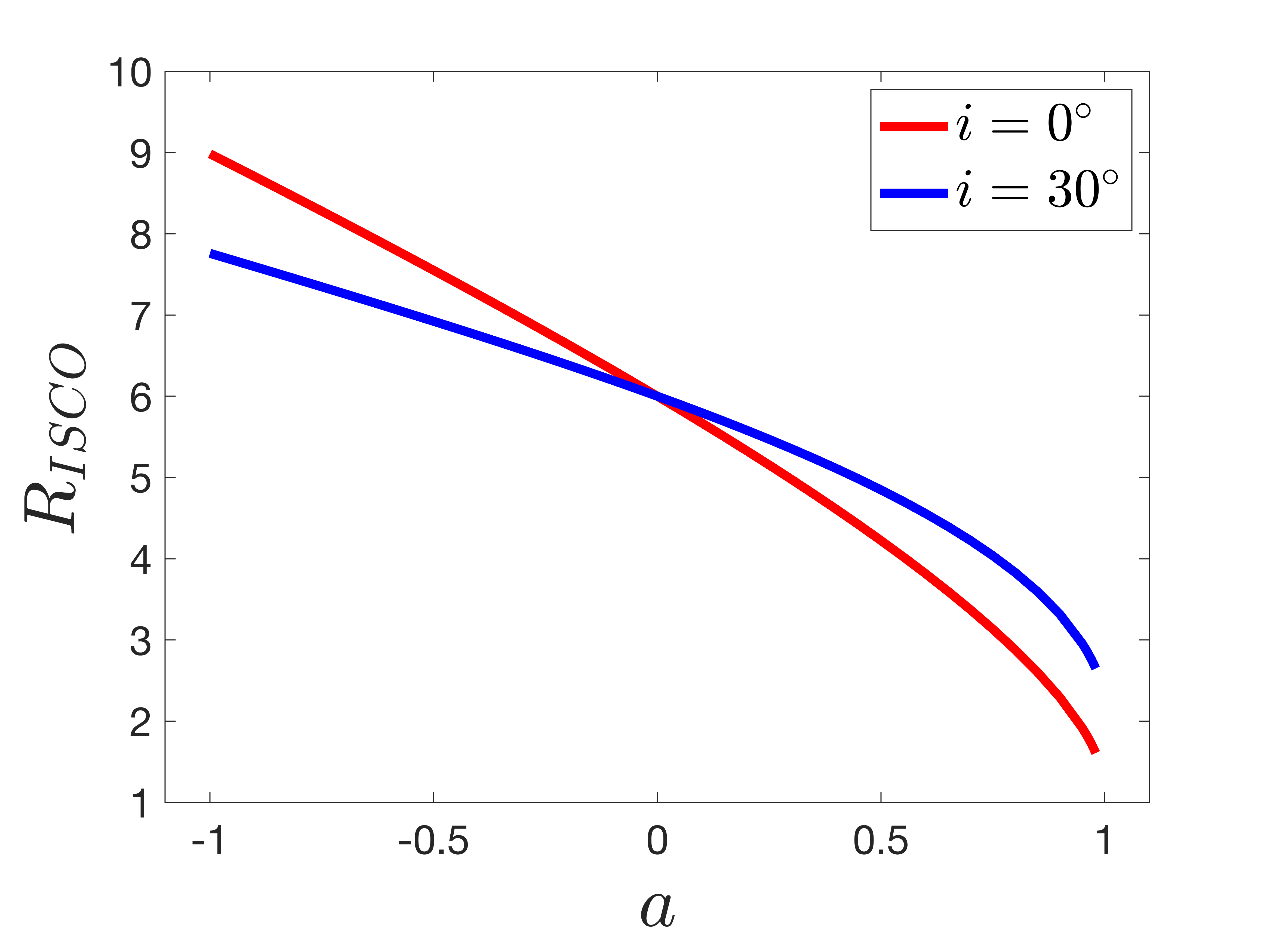}
         \caption{$R_{\text{ISCO}}$ vs $a$, $i=0^\circ$, $30^\circ$}
         \label{fig:ISCO_with_a_i_1}
     \end{subfigure}
     \hfill
     \begin{subfigure}[b]{0.495\textwidth}
         \centering
         \includegraphics[width=\textwidth]{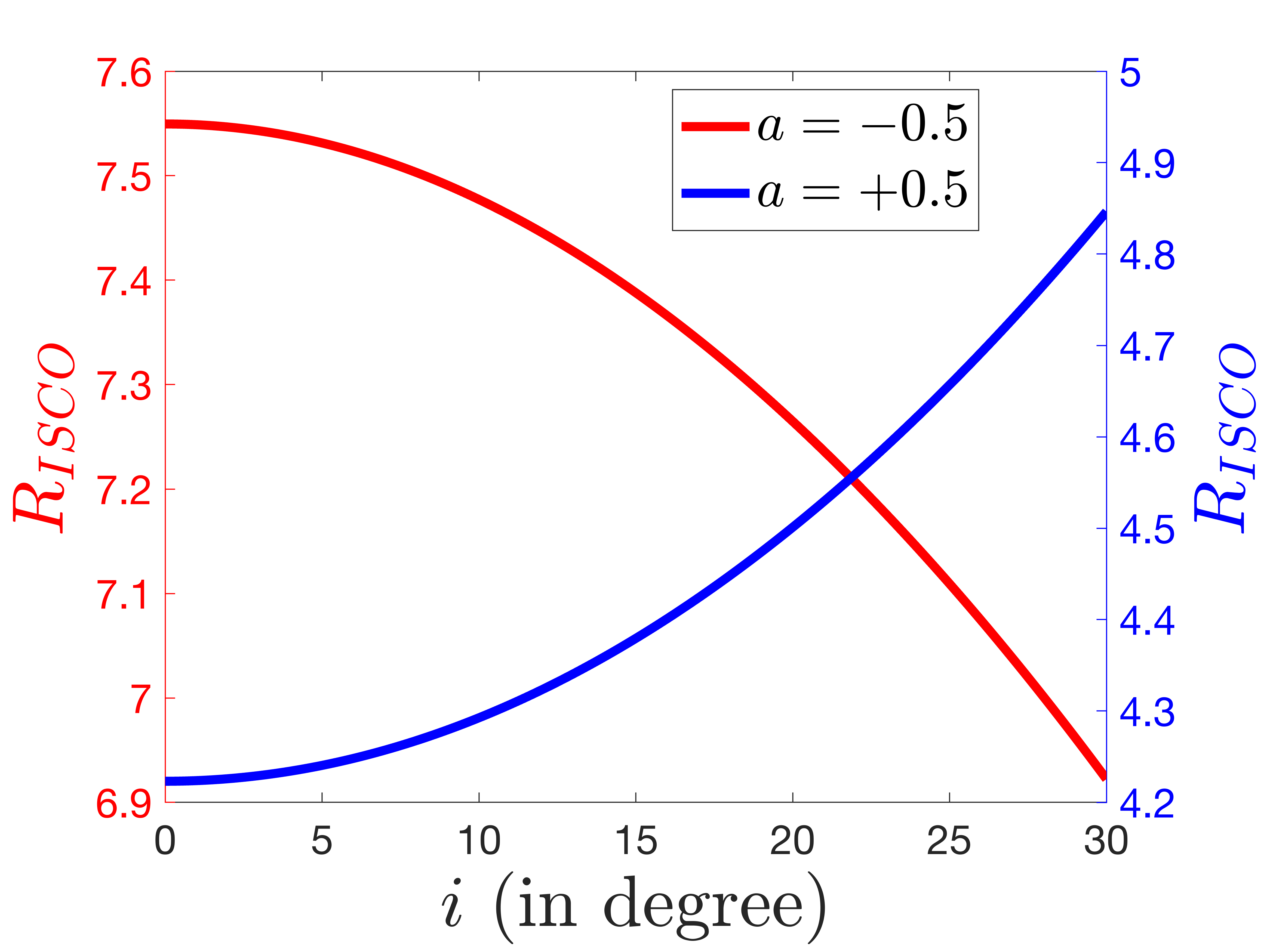}
         \caption{$R_{\text{ISCO}}$ vs $i$, $a=\pm 0.5$}
         \label{fig:ISCO_with_a_i_2}
     \end{subfigure}
        \caption{Correlations of the radii of \acrshort{isco}s $\left( R_{\text{ISCO}} \right)$ with the Kerr parameter $a$ and the inclination angle $i$. The orbits are governed by the \acrshort{gmf} \acrshort{pnp}.}
        \label{fig:ISCO_with_a_i}
\end{figure*}

In Figure \ref{fig:ISCO_with_a_i}, we study how the radii of \acrshort{isco}s vary with the rotation parameter of the \acrshort{cop} $a$ and the inclination of the orbit $i$. At first, we observe that the change in $R_{\text{ISCO}}$ with respect to the Kerr parameter $a$ is significant for any angle of inclination, be it $i=0^\circ$, or $i=30^\circ$ (Figure \ref{fig:ISCO_with_a_i_1}). For the equatorial orbits with $i=0^\circ$, the radii change from $9 r_{\text{g}}$ to $\approx 1 r_{\text{g}}$ in the span of $a \in [-1,1)$. The change in $R_{\text{ISCO}}$ with respect to $a$ for \acrshort{abn} \acrshort{pnp} is precisely the same as that for \acrshort{gmf} \acrshort{pnp}. The value of $R_{\text{ISCO}}$ is almost linearly decreasing for most of the range of $a$, except for the orbits around the maximally co-rotating \acrshort{cop}s (when $a \rightarrow 1$), in which case the change in $R_{\text{ISCO}}$ is more drastic. For $i=30^\circ$, the value of $R_{\text{ISCO}}$ varies in a similar way with $a$, although it changes from $7.76 r_{\text{g}}$ to $\approx 2.5 r_{\text{g}}$ in the span of $a \in [-1,1)$. For $a=0$, there is no change in the radius of \acrshort{isco} as the inclination angle varies, and it remains unchanged at $R_{\text{ISCO}}=6 r_{\text{g}}$. The result is expected for the Schwarzschild-like \acrshort{cop}, as there is no rotation axis to identify the equatorial plane. Hence, all the planes are equivalent, and there is no special plane to impose larger or smaller stability on the circular orbits. 

In Figure \ref{fig:ISCO_with_a_i_2}, we can see the dependence of $R_{\text{ISCO}}$ on the inclination angle $i$ for both co-rotating ($a>0$) and counter-rotating ($a<0$) \acrshort{cop}s. In both cases, the change in the value of $R_{\text{ISCO}}$ in the range $i \in [0^\circ,30^\circ]$ is small. For the orbits around the counter-rotating \acrshort{cop}, the value of $R_{\text{ISCO}}$ decreases monotonically with $i$. On the contrary, it increases monotonically with the inclination angle for the orbits around the co-rotating \acrshort{cop}s. The radii change more rapidly at the higher values of $i$ than at its lower values. 

With the above analysis, we get a picture of the stable circular orbits in our system, including their dependence on the Kerr parameter $a$ and inclination angle $i$. We observed that the radii of \acrshort{isco}s are significantly more influenced by $a$ than by $i$. For orbits around counter-rotating \acrshort{cop}s, $R_{\text{ISCO}}$ decreases with $i$. On the other hand, it increases with $i$ for the orbits around the co-rotating \acrshort{cop}s. For the Schwarzschild-like \acrshort{cop} ($a=0$), $R_{\text{ISCO}}$ does not change with $i$. However, generic orbits are much more complicated and nuanced, requiring more detailed analysis and special tools to study their stability. 

\subsection{Dynamical analysis of generic orbits}
We move on to the analysis of the generic orbits. We have implemented the method of Poincar\'e Map of the sections of phase space to visualise the change in chaos and order in the system qualitatively. In this method, instead of looking at the entire three-dimensional hypersurface $(\rho,p_\rho,z)$, we look at a two-dimensional cross-section of the phase space, which consists of the points of intersection of the trajectories on a fixed plane (such as $z=0$) while moving in a particular direction along the trajectory \citep{berry1978topics}. The periodic orbits give rise to one or more isolated points on the sectional map, whereas the chaotic orbits give rise to a sea of scattered points \citep{berry1978topics, goldstein2001classical}. There can be other stable orbits, quasi-periodic regular orbits, which form systematic patterns in the map corresponding to the concentric Kolmogorov-Arnold-Moser (KAM) tori. 

In this paper, we have plotted the Poincar\'e sections on the $z=0$ plane, while the orbits are moving out of the plane along their trajectories in the phase space. For a particular set of orbital parameters, the initial conditions, \textit{i.e.}, the coordinates $\rho (t=0)$, $z (t=0)$ and the momentum $p_{\rho} (t=0)$, are varied, and the value of $p_z (t=0) > 0$ is evaluated from the conservation equation (\ref{eq:energy_cons_newt}) for non-relativistic test particles and conservation equation (\ref{eq:energy_cons}) for relativistic test particles. We have studied the dependence of chaos on a particular parameter through a comparative study of the Poincar\'e maps by varying that parameter while keeping the rest fixed.

\subsubsection{nonrotating compact object primaries}

\begin{figure*} 
    
     \centering
     \begin{subfigure}[b]{0.495\textwidth}
         \centering
         \includegraphics[width=\textwidth]{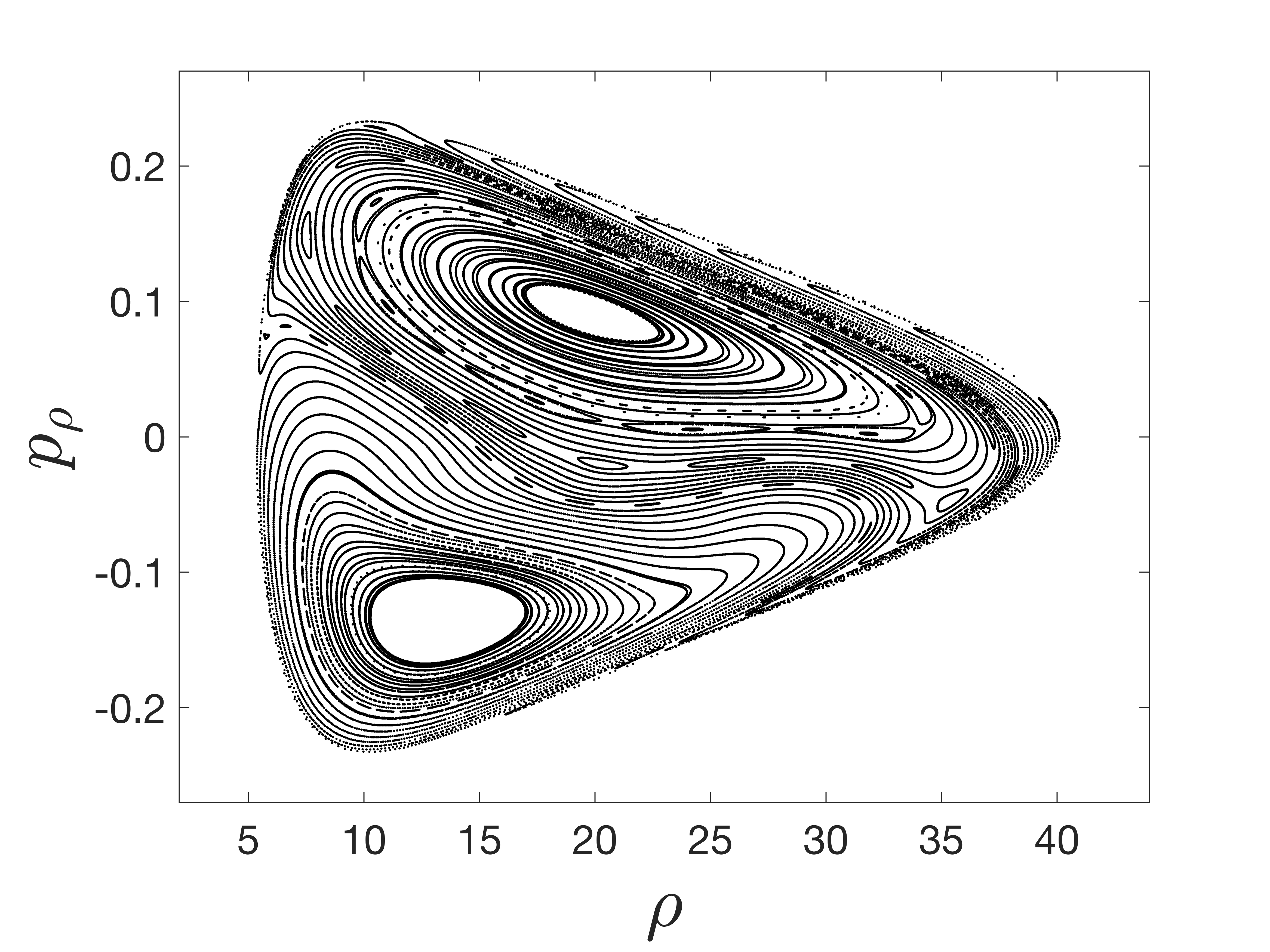}
         \caption{$i=13^\circ$, Non-relativistic}
         \label{fig:PM_with_i13_NonRot}
     \end{subfigure}
     \hfill
     \begin{subfigure}[b]{0.495\textwidth}
         \centering
         \includegraphics[width=\textwidth]{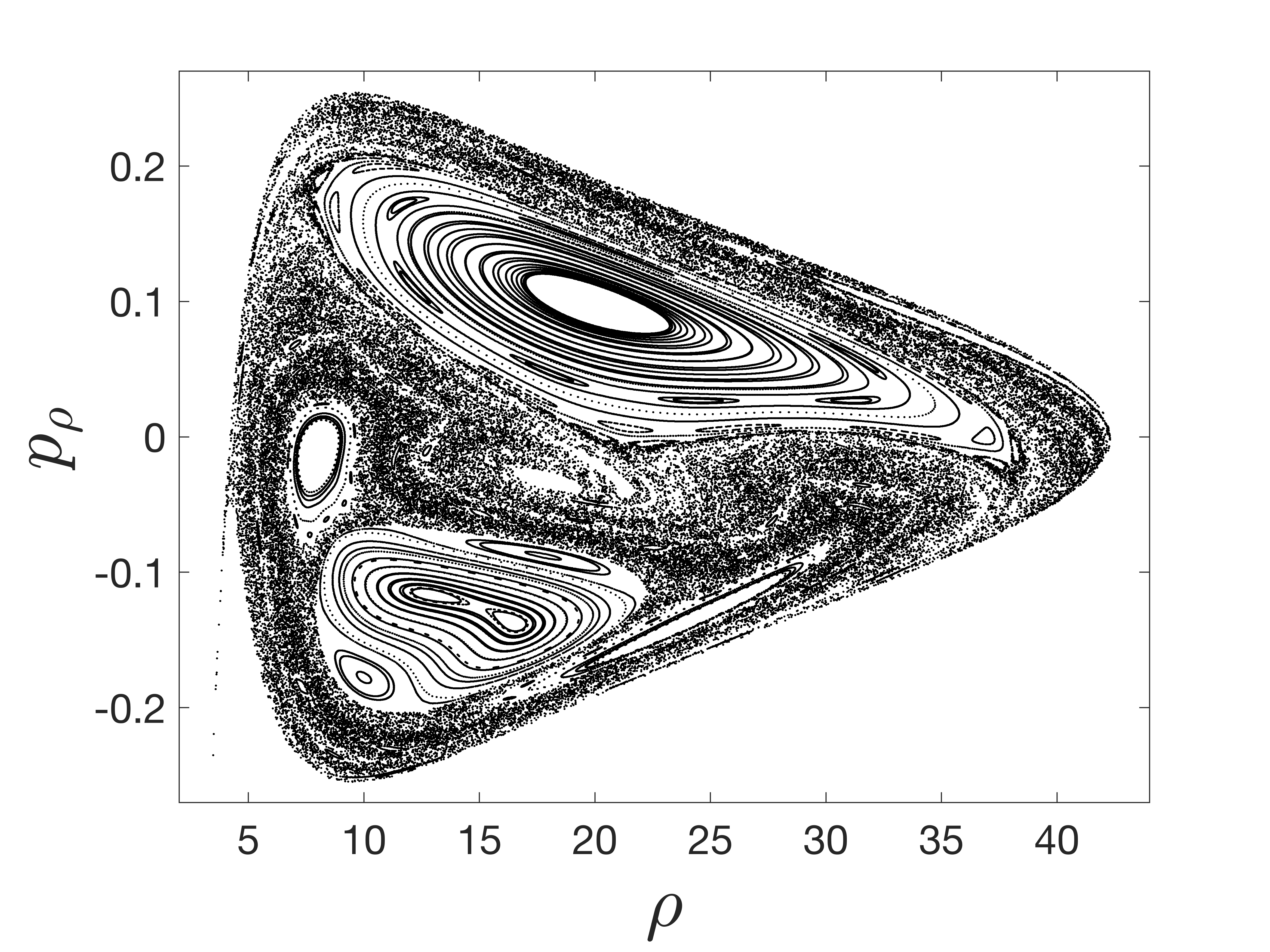}
         \caption{$i=15^\circ$, Non-relativistic}
         \label{fig:PM_with_i15_NonRot}
     \end{subfigure}
     \hfill
     \begin{subfigure}[b]{0.495\textwidth}
         \centering
         \includegraphics[width=\textwidth]{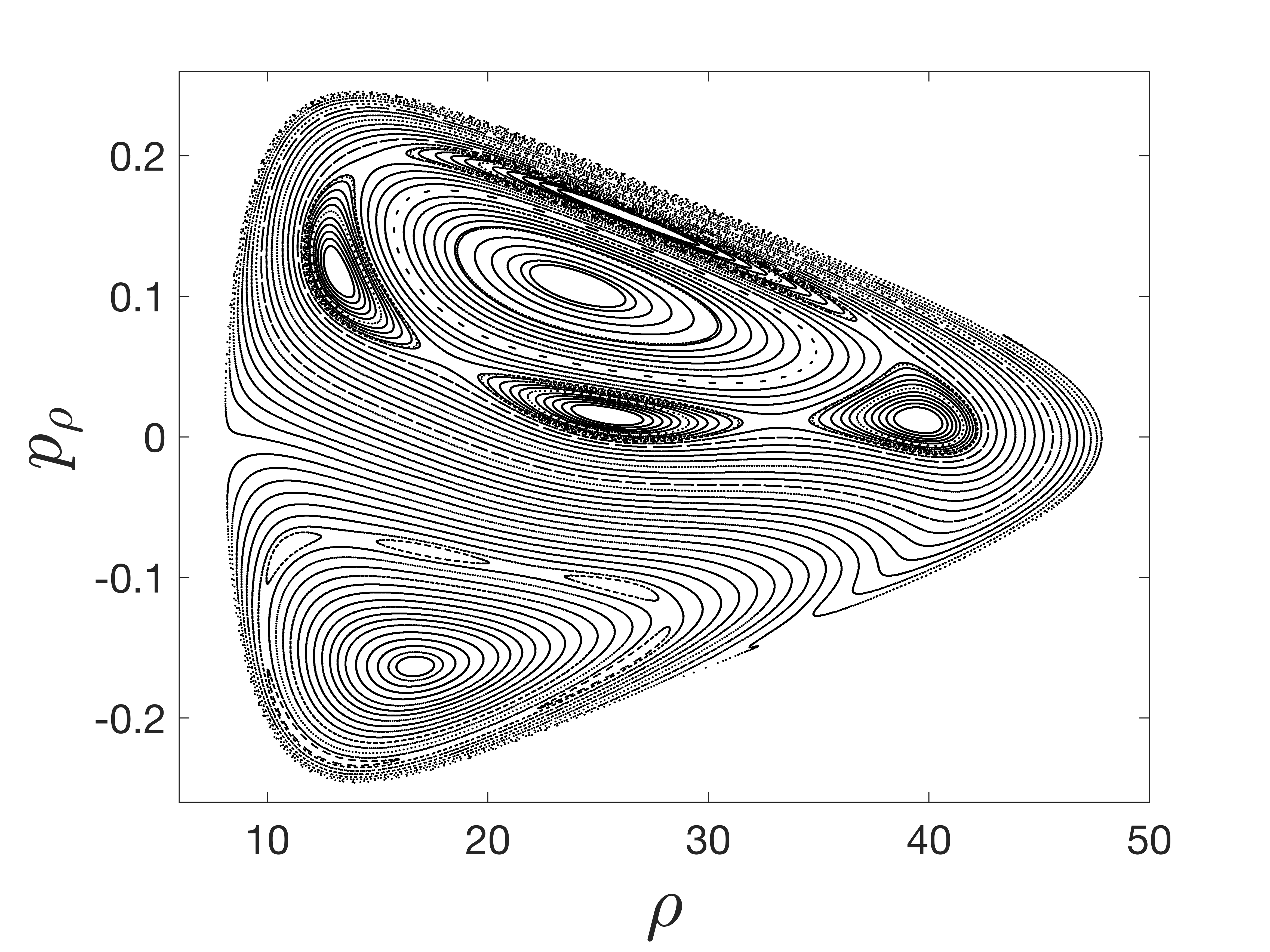}
         \caption{$i=28^\circ$, Non-relativistic}
         \label{fig:PM_with_i28_NonRot}
     \end{subfigure}
     \hfill
     \begin{subfigure}[b]{0.495\textwidth}
         \centering
         \includegraphics[width=\textwidth]{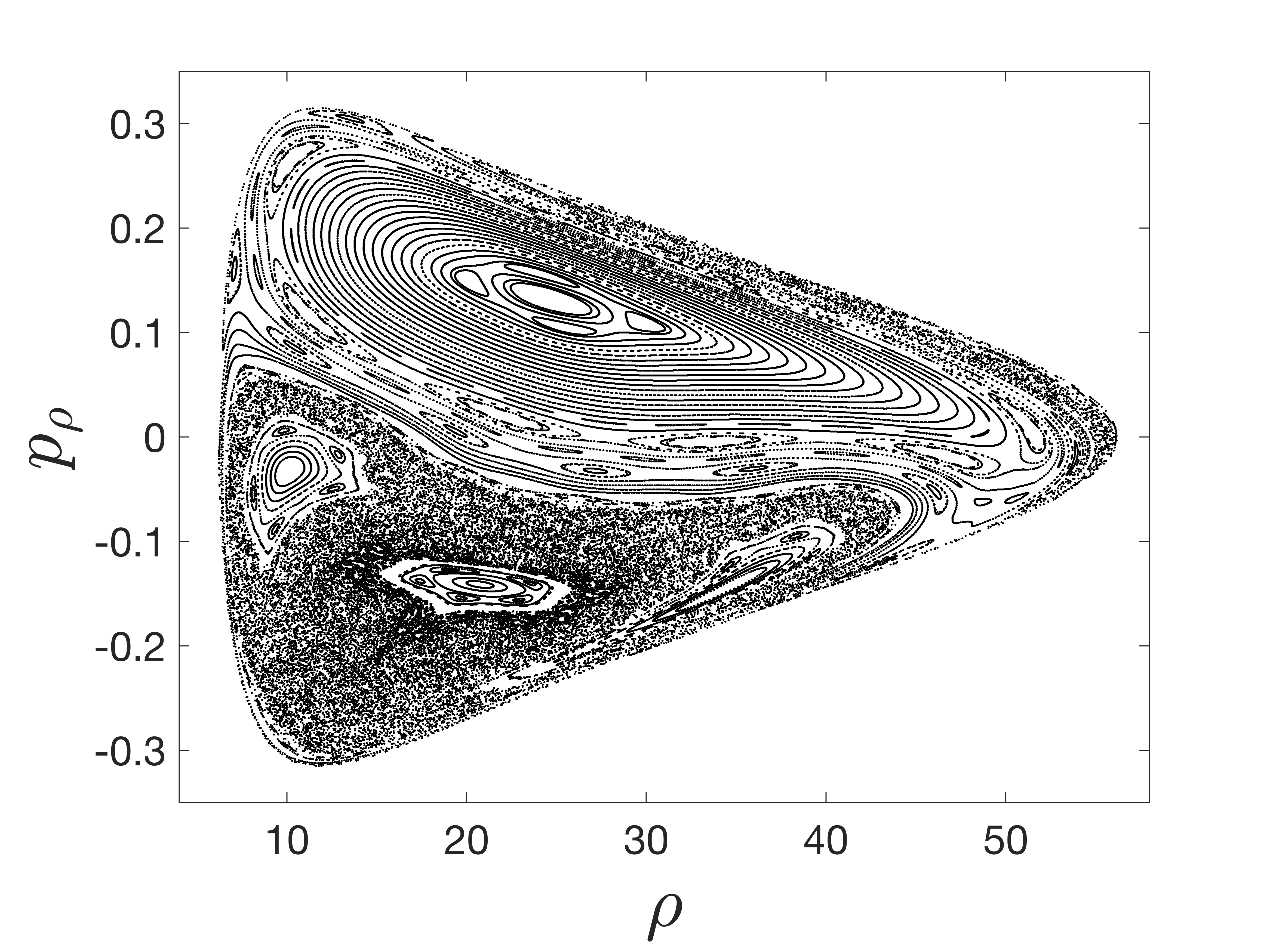}
         \caption{$i=30^\circ$, Non-relativistic}
         \label{fig:PM_with_i30_NonRot}
     \end{subfigure}
     \hfill
     \begin{subfigure}[b]{0.495\textwidth}
         \centering
         \includegraphics[width=\textwidth]{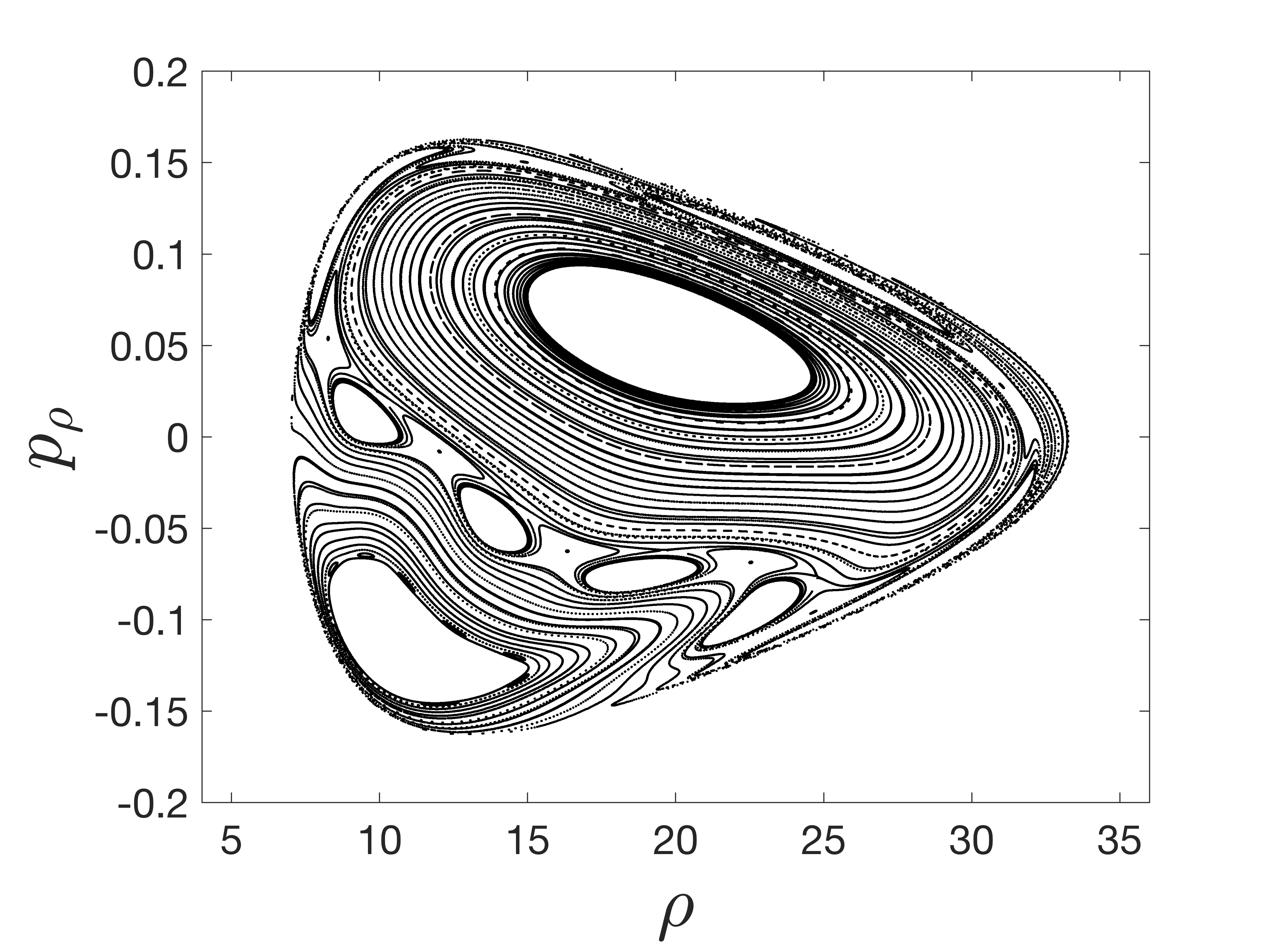}
         \caption{$i=13^\circ$, Relativistic}
         \label{fig:PM_with_i13_rel_NonRot}
     \end{subfigure}
     \hfill
     \begin{subfigure}[b]{0.495\textwidth}
         \centering
         \includegraphics[width=\textwidth]{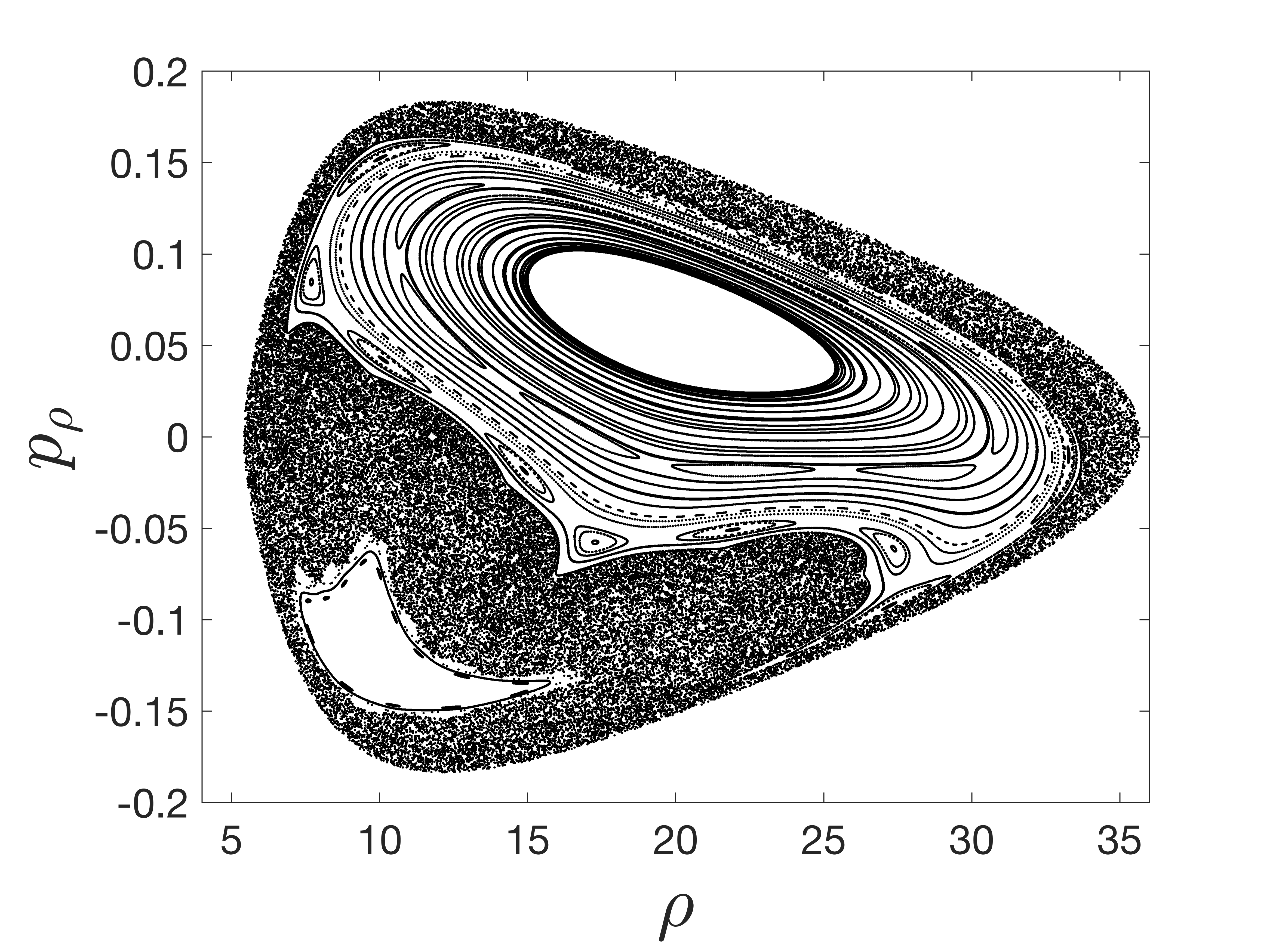}
         \caption{$i=15^\circ$, Relativistic}
         \label{fig:PM_with_i15_rel_NonRot}
     \end{subfigure}
        \caption{Poincar\'e maps on the cross-sectional plane $z=0$ for the orbits governed by \acrshort{gms} \acrshort{pnp} around nonrotating \acrshort{cop}s ($a=0$). The inclination angle $i$ is varied. The parameters are (a)-(b) $E=0.976$, $L=4.20$, $\mathcal{D}=2\times 10^{-4}$, (c)-(d) $E=0.968$, $L=5.85$, $\mathcal{D}=2\times 10^{-4}$, and (e)-(f) $E=0.974$, $L=4.60$, $\mathcal{D}=2\times 10^{-4}$.}
        \label{fig:PM_with_i_NonRot}
\end{figure*}

As the gravitational force, given in equation (\ref{eq:GMPNF}), is generic in nature, we can put $a=0$ in the expression and study the off-equatorial orbits around a static and stationary \acrshort{cop}. We have implemented the effective potential consisting of \acrshort{gms} \acrshort{pnp}, as given in equation (\ref{eq:overall_potential_Sch}), and worked out the equations of motion as per the equations given in (\ref{eq:rho_dot_nrel}-\ref{eq:pz_dot_nrel}) which does not consider any relativistic correction. We have varied the inclination angle $i$ and generated the Poincar\'e Maps for each of the angles. We have shown the maps for $i=13^\circ,\text{ and } 15^\circ$ by keeping the total energy $E$, total angular momentum $L$, and dipole coefficient $\mathcal{D}$ unchanged (Figures \ref{fig:PM_with_i13_NonRot}-\ref{fig:PM_with_i15_NonRot}). Evidently, the Poincar\'e Map of the phase space for $i=15^\circ$ consists of a more chaotic region than that for $i=13^\circ$, in which case most of the orbits are regular in nature. We have also shown the cross-sectional maps for $i=28^\circ,\text{ and } 30^\circ$ by increasing the angular momentum and decreasing the total energy in order to suppress the chaos for convenience in visualisation (Figures \ref{fig:PM_with_i28_NonRot}-\ref{fig:PM_with_i30_NonRot}). We can observe more chaotic regions on the map for $i=30^\circ$ compared to that for $i=28^\circ$. Therefore, this provides qualitative evidence that the degree of chaos has a positive correlation with the angle of inclination $i$ of the orbit.

To incorporate the relativistic corrections into the equations, we have derived the equations of motion using equations given in (\ref{eq:rho_dot_rel}-\ref{eq:pz_dot_rel}). We solved these equations too, to get the Poincar\'e Maps corresponding to the phase spaces of the orbits of relativistic test-particle and studied the trend of the degree of chaos with the inclination angle $i$ (Figures \ref{fig:PM_with_i13_rel_NonRot}-\ref{fig:PM_with_i15_rel_NonRot}). Similar to the non-relativistic case, we have kept the parameters $E$, $L$, and $\mathcal{D}$ to be constant. We can observe the exact similar behaviour of the orbits, as we have seen in the non-relativistic case. The phase space becomes more chaotic when the inclination angle $i$ increases from $13^\circ$ to $15^\circ$. The only difference is that we have used a smaller value of $E$ and a higher value of $L$ to suppress the nonlinearity of the relativistic equations of motion (\ref{eq:rho_dot_rel}-\ref{eq:pz_dot_rel}) in comparison to the corresponding non-relativistic case (Figures \ref{fig:PM_with_i13_NonRot}-\ref{fig:PM_with_i15_NonRot}). This is a consequence of the fact that the special-relativistic phase-space trajectories are more chaotic than the corresponding non-relativistic counterparts, which is well-established in the literature \citep{gueron2001chaos,nag2017influence}. 

\subsubsection{Rotating compact object primaries} \label{ch:rotating_cop}
After considering the Schwarzschild geometry, we move on to the Kerr geometry, where the Kerr parameter $a$ is non-zero. In this case, for particular values of $a$ and $i$, we first evaluate the \acrshort{pnp} $V_{\text{GM}}$ from equation (\ref{eq:Pot_Def}) for several values of $r$, using the generalised force $F_{\text{GM}r}$ given in equation (\ref{eq:GMPNF}). After that, we take the natural log of the potential values to fit the data with the logarithmic form of the fitting function in equation (\ref{eq:GMPNP}), which is given by
\begin{equation} \label{eq:log_GMPNP}
    \ln\left(-V_{\text{GMf}}(r)\right)=\left( \frac{\gamma_1}{(r-\gamma_2)^{\gamma_3}}+\gamma_4 \right) - \gamma_5\ln\left(r-\gamma_2\right)
\end{equation}
This allows us to fit the curve more conveniently with minimal error and evaluate the fitting parameters. After we estimate the values of the fitting parameters, we can use them in equation (\ref{eq:overall_potential_general}) to get the overall potential consisting of the \acrshort{gmf} \acrshort{pnp} term for particular values of $a$ and $i$, along with the dipolar perturbation term and the centrifugal contribution as well. Thereafter, we implement this potential in the equations (\ref{eq:rho_dot_nrel})-(\ref{eq:pz_dot_nrel}) and equations (\ref{eq:rho_dot_rel})-(\ref{eq:pz_dot_rel}) to get the equations of motion for non-relativistic and relativistic cases, respectively. 

\begin{figure*} 
    
     \centering
     \begin{subfigure}[b]{0.49\textwidth}
         \centering
         \includegraphics[width=\textwidth]{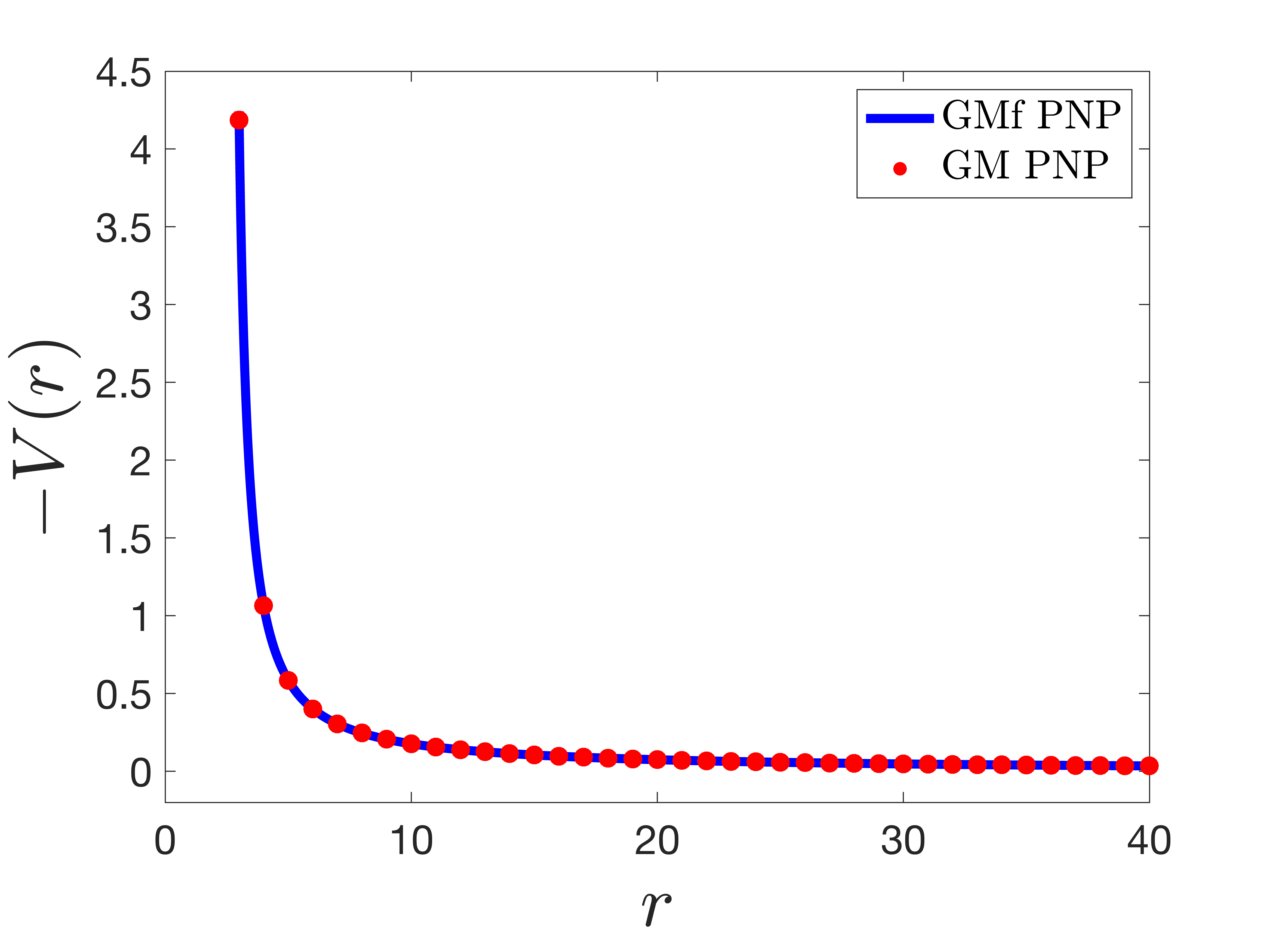}
         \caption{$V_{\text{GM}}$ and $V_{\text{GMf}}$ for $a=-0.8$, $i=20^\circ$}
         \label{fig:GMPNP_fitting_1}
     \end{subfigure}
     \hfill
     \begin{subfigure}[b]{0.49\textwidth}
         \centering
         \includegraphics[width=\textwidth]{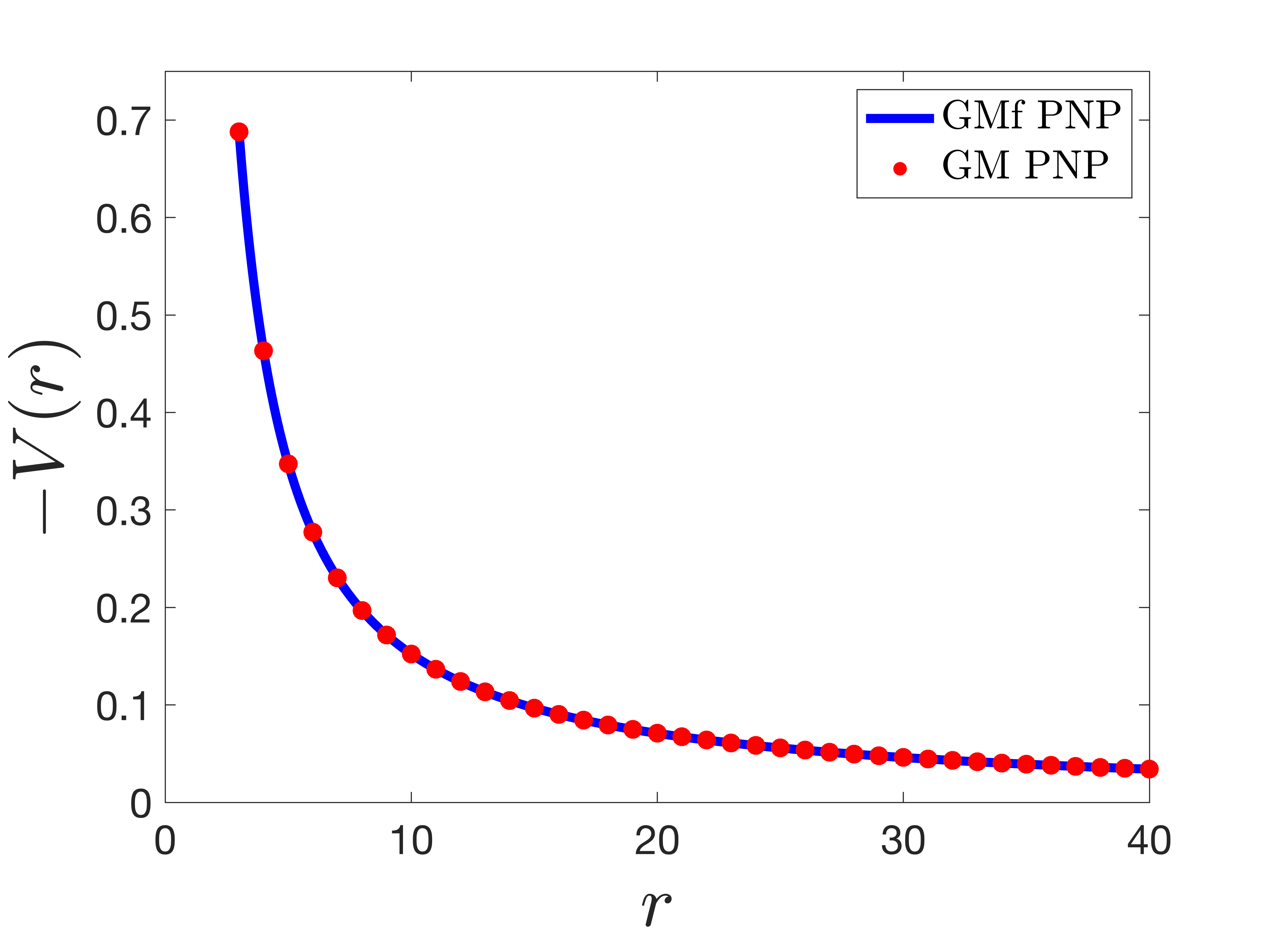}
         \caption{$V_{\text{GM}}$ and $V_{\text{GMf}}$ for $a=0.8$, $i=20^\circ$}
         \label{fig:GMPNP_fitting_2}
     \end{subfigure}
     \begin{subfigure}[b]{0.49\textwidth}
         \centering
         \includegraphics[width=\textwidth]{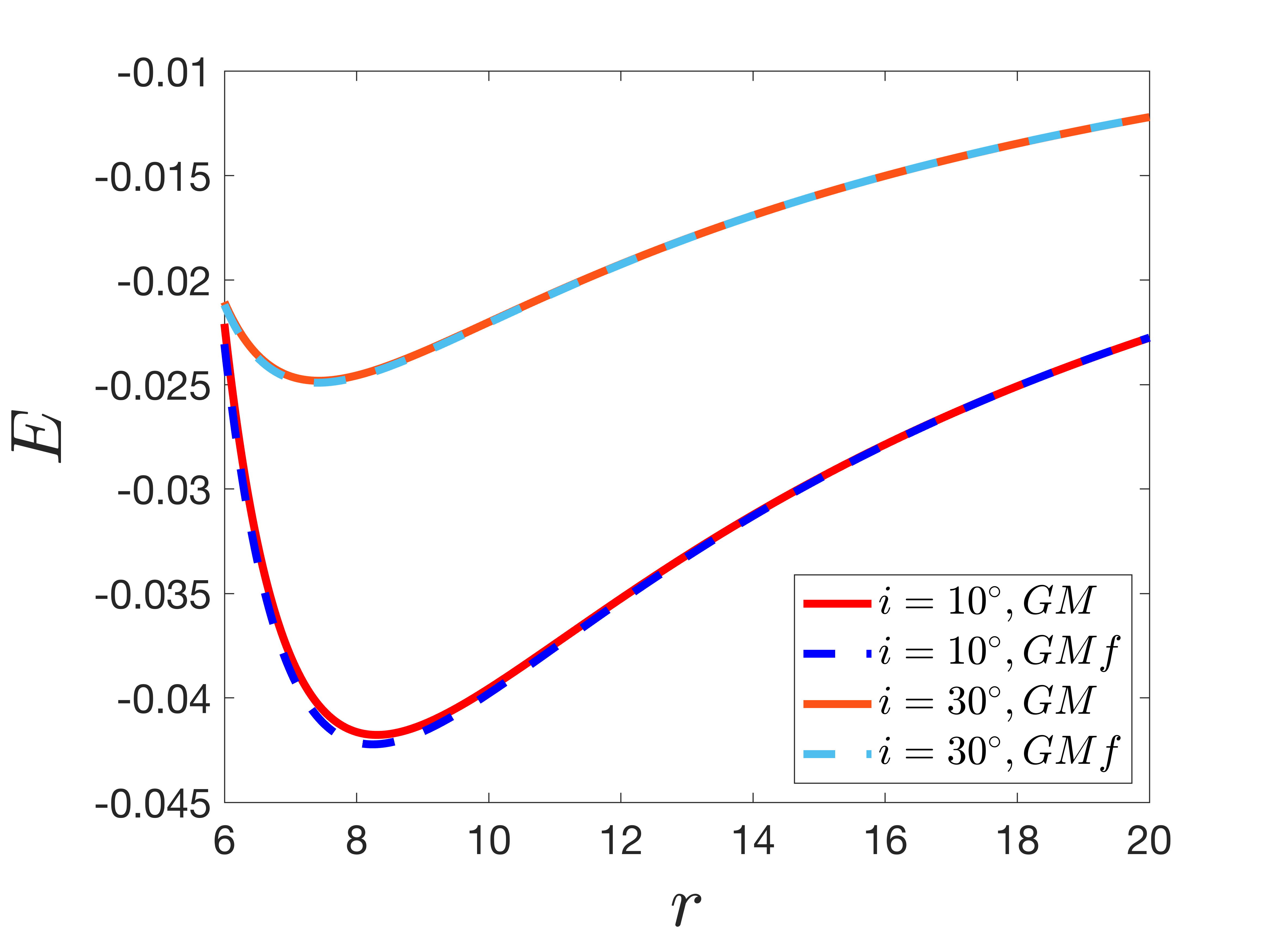}
         \caption{Specific energy curves for $a=-0.8$, $i=10^\circ,30^\circ$}
         \label{fig:GMPNP_fitting_3}
     \end{subfigure}
     \hfill
     \begin{subfigure}[b]{0.49\textwidth}
         \centering
         \includegraphics[width=\textwidth]{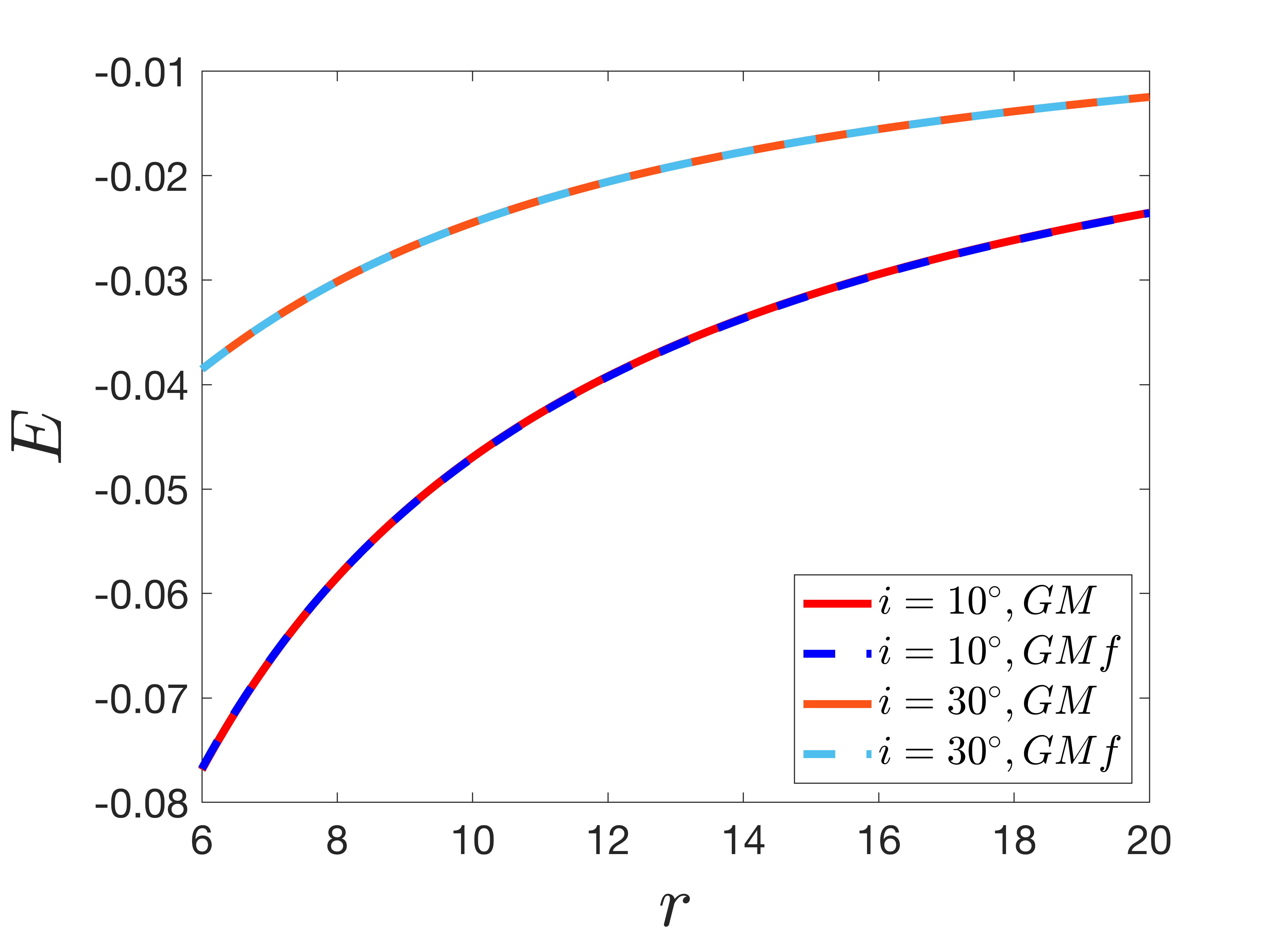}
         \caption{Specific energy curves for $a=0.8$, $i=10^\circ,30^\circ$}
         \label{fig:GMPNP_fitting_4}
     \end{subfigure}
     \hfill
     \begin{subfigure}[b]{0.49\textwidth}
         \centering
         \includegraphics[width=\textwidth]{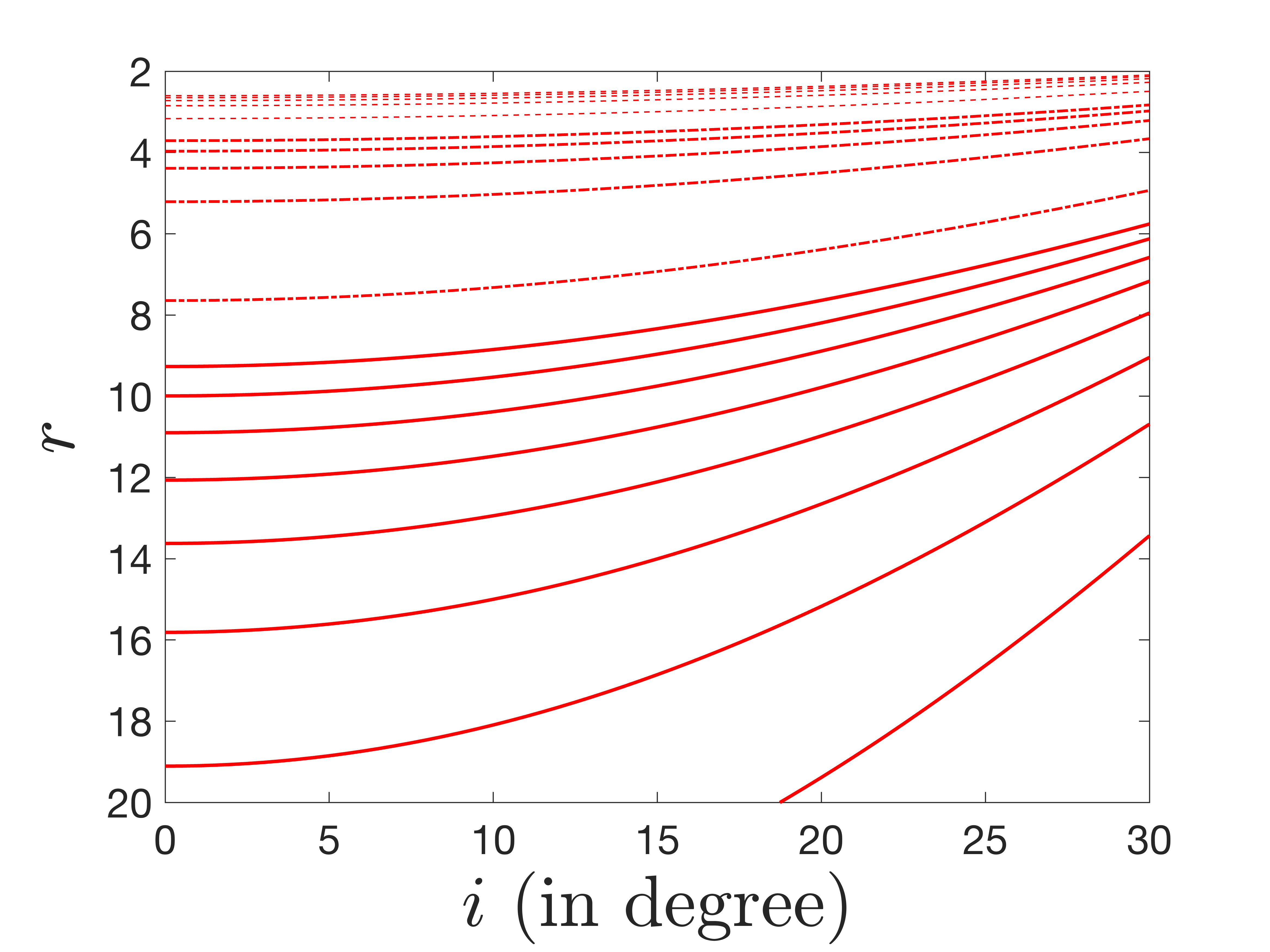}
         \caption{Contours of constant potentials $V_{\text{GM}}$ for $a=-0.8$}
         \label{fig:GMPNP_fitting_5}
     \end{subfigure}
     \hfill
     \begin{subfigure}[b]{0.49\textwidth}
         \centering
         \includegraphics[width=\textwidth]{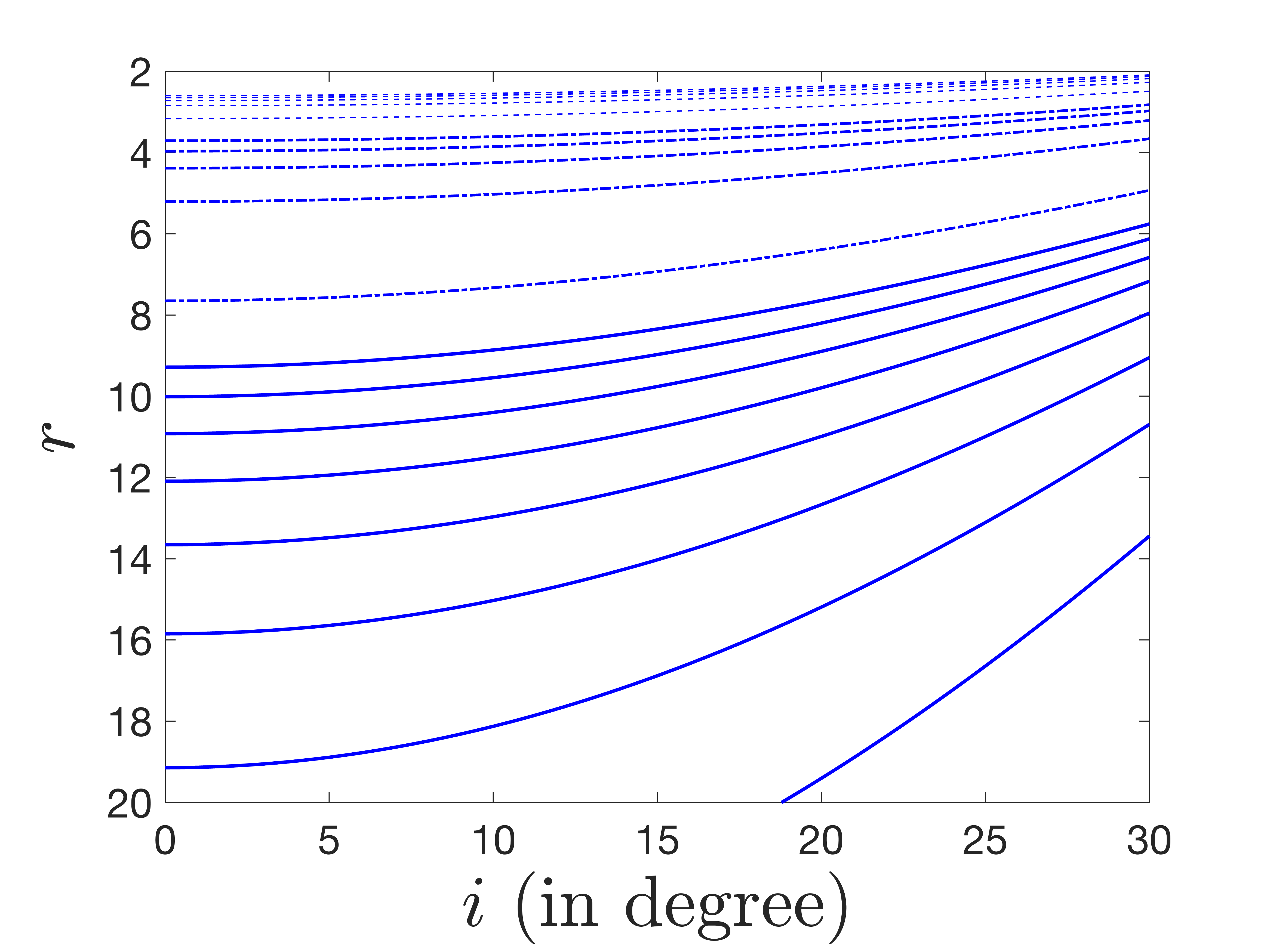}
         \caption{Contours of constant potentials $V_{\text{GMf}}$ for $a=-0.8$}
         \label{fig:GMPNP_fitting_6}
     \end{subfigure}
        \caption{Comparison between \acrshort{gm} \acrshort{pnp} and \acrshort{gmf} \acrshort{pnp}. (a),(b) Fitting the \acrshort{gm} \acrshort{pnp} $V_{\text{GM}}$ with the \acrshort{gmf} \acrshort{pnp} $V_{\text{GMf}}$; (c),(d) Variation of the specific energy of circular orbits with the radial distance $r$ for $a=\pm 0.8$ and different inclination angles for both \acrshort{gm} \acrshort{pnp} and \acrshort{gmf} \acrshort{pnp}; (e),(f) Contours of constant potentials with \acrshort{gm} \acrshort{pnp} and \acrshort{gmf} \acrshort{pnp} for $a=-0.8$, consisting of solid lines corresponding to $V_{\text{PNP}}=0.045,0.06,0.075,...,0.15$, dot-dashed lines to $V_{\text{PNP}}=0.2,0.4,0.6...,1$, and dashed lines to $V_{\text{PNP}}=2,4,6,...,10$.}
        \label{fig:GMPNP_fitting}
\end{figure*}

We can verify the quality of fitting of \acrshort{gmf} \acrshort{pnp} with \acrshort{gm} \acrshort{pnp} by doing some comparisons between them, or the derived quantities from both of them (Figure \ref{fig:GMPNP_fitting}). In Figure \ref{fig:GMPNP_fitting_1} and \ref{fig:GMPNP_fitting_2}, we show how \acrshort{gmf} \acrshort{pnp} fits with \acrshort{gm} \acrshort{pnp} for $a=\pm 0.8$ and $i=20^\circ$. For convenience, we have shown fewer points of \acrshort{gm} \acrshort{pnp} in the figures. Otherwise, we use a large number of points for efficient fitting while performing actual calculations. In Figure \ref{fig:GMPNP_fitting_3} and \ref{fig:GMPNP_fitting_4}, we look for the variation in specific energy as a function of the radial distance $r$, given by the relation
\begin{equation}
    E = \frac{r}{2} \frac{dV_{\text{PNP}}}{dr} + V_{\text{PNP}}
\end{equation}
for the circular orbits governed by \acrshort{gm} \acrshort{pnp} and \acrshort{gmf} \acrshort{pnp}. We compare the results for several values of $a$ and $i$, some of which are presented here in these figures. We also compare the contours of constant potential values for \acrshort{gm} \acrshort{pnp} and \acrshort{gmf} \acrshort{pnp} with $a=-0.8$, shown in Figure \ref{fig:GMPNP_fitting_5} and \ref{fig:GMPNP_fitting_6} respectively. The studies are similar and at par with those performed in \citep{ghosh2007}. The minute differences between the two \acrshort{pnp}s in these studies establish the quality of fitting and the effectiveness of the fitting function.

\begin{figure*} 
    
     \centering
     \begin{subfigure}[b]{0.495\textwidth}
         \centering
         \includegraphics[width=\textwidth]{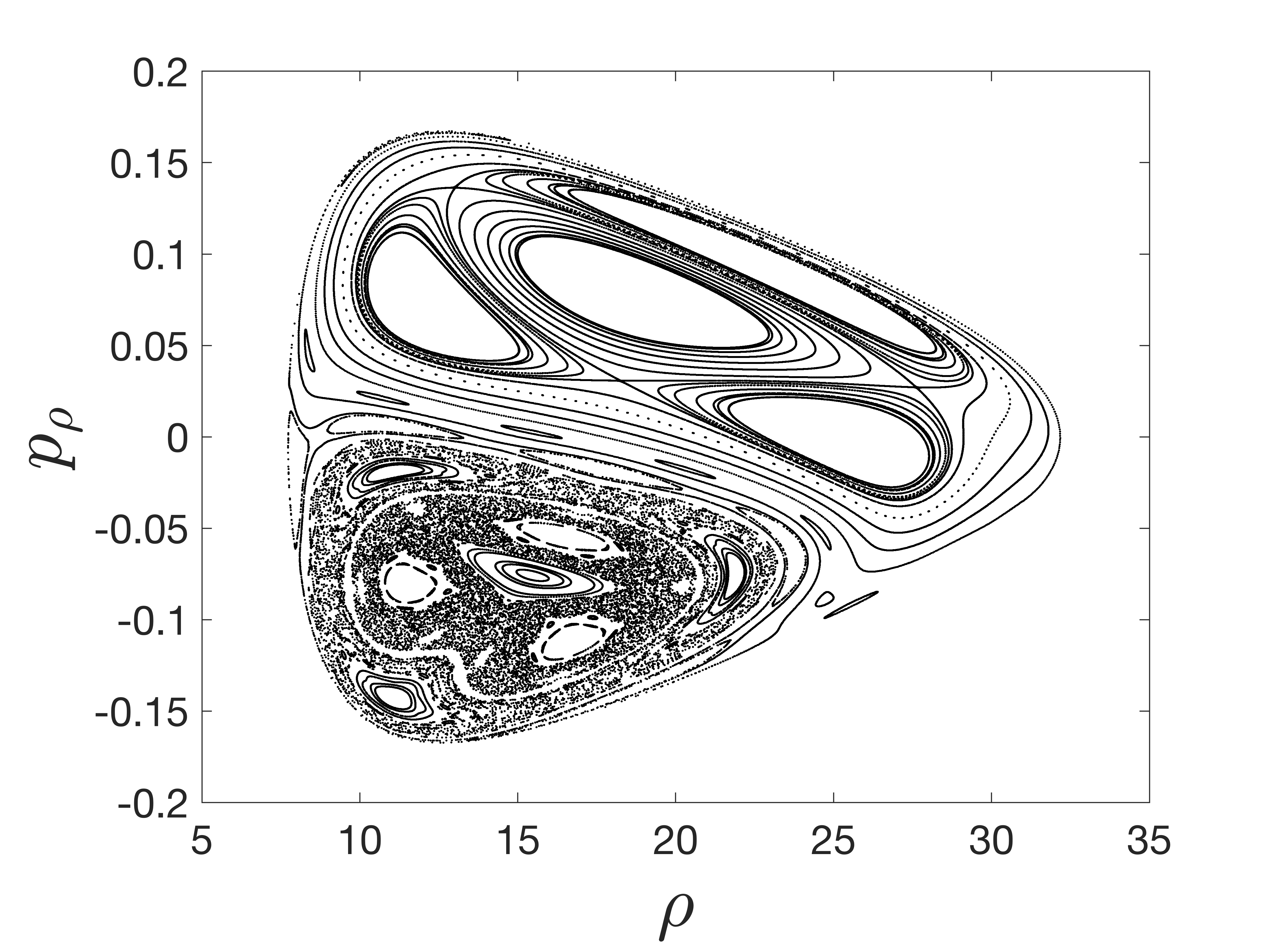}
         \caption{$a=0.2$, \acrshort{gmf} \acrshort{pnp}, Non-relativistic}
         \label{fig:PM_comparison_GMPNP_APNP_1}
     \end{subfigure}
     \hfill
     \begin{subfigure}[b]{0.495\textwidth}
         \centering
         \includegraphics[width=\textwidth]{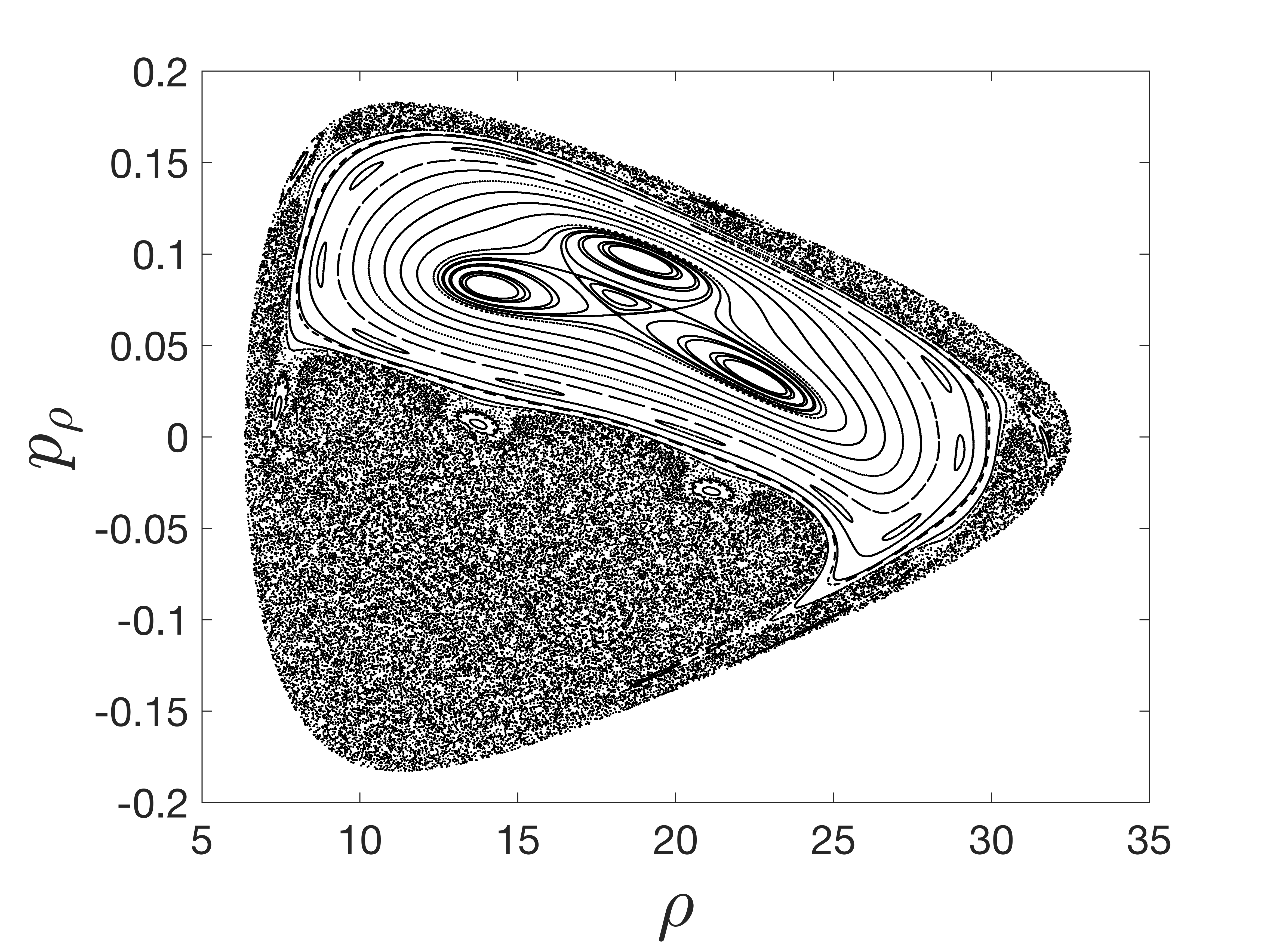}
         \caption{$a=-0.2$, \acrshort{gmf} \acrshort{pnp}, Non-relativistic}
         \label{fig:PM_comparison_GMPNP_APNP_2}
     \end{subfigure}
     \hfill
     \begin{subfigure}[b]{0.495\textwidth}
         \centering
         \includegraphics[width=\textwidth]{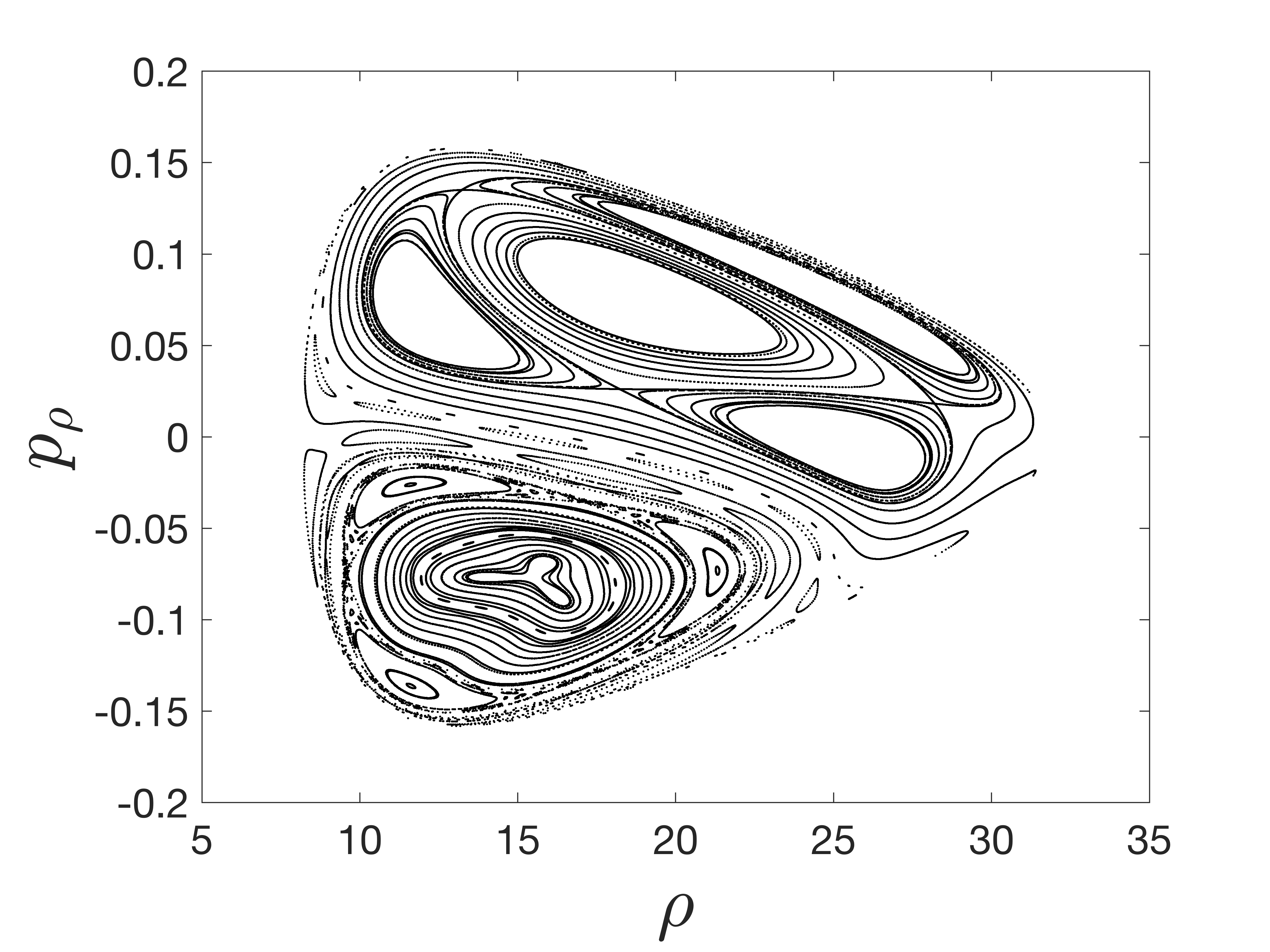}
         \caption{$a=0.2$, \acrshort{abn} \acrshort{pnp}, Non-relativistic}
         \label{fig:PM_comparison_GMPNP_APNP_3}
     \end{subfigure}
     \hfill
     \begin{subfigure}[b]{0.495\textwidth}
         \centering
         \includegraphics[width=\textwidth]{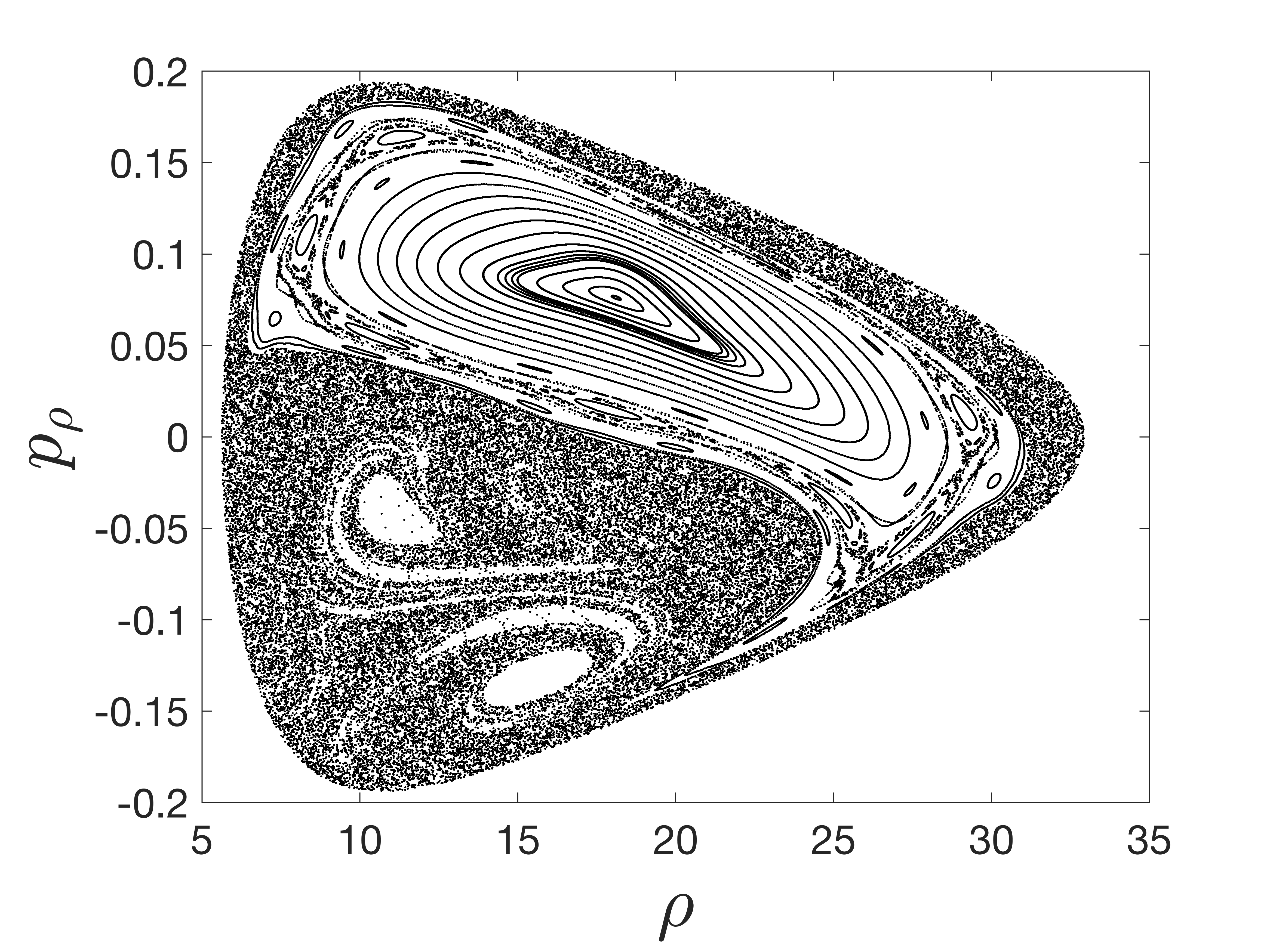}
         \caption{$a=-0.2$, \acrshort{abn} \acrshort{pnp}, Non-relativistic}
         \label{fig:PM_comparison_GMPNP_APNP_4}
     \end{subfigure}
        \caption{Poincar\'e maps on the cross-sectional plane $z=0$ for the orbits following \acrshort{gmf} \acrshort{pnp} and \acrshort{abn} \acrshort{pnp} with $a=\pm 0.2$, $i=0^\circ$, $\mathcal{D}=3\times 10^{-4}$, $E=0.975$, and $L=4.16$. The plots are evaluated without considering any special relativistic corrections.}
        \label{fig:PM_comparison_GMPNP_APNP}
\end{figure*}

We have already compared the effective potentials consisting of the \acrshort{gmf} \acrshort{pnp} with that of the \acrshort{abn} \acrshort{pnp} in section \ref{sch:comp_GMf_ABN}. Now, we must compare the Poincar\'e Maps generated using the two potentials (Figure \ref{fig:PM_comparison_GMPNP_APNP}). Apparently, we are considering the orbits on the equatorial plane in this case ($i=0^\circ$). We can observe the usual frame-dragging effect for \acrshort{gmf} \acrshort{pnp} (Figure \ref{fig:PM_comparison_GMPNP_APNP_1} and \ref{fig:PM_comparison_GMPNP_APNP_2}). The chaotic nature of the orbits for $a=-0.2$ is significantly higher than that for $a=0.2$. The frame-dragging effect is also evident in the Poincar\'e Maps corresponding to the orbits derived from \acrshort{abn} \acrshort{pnp} (Figure \ref{fig:PM_comparison_GMPNP_APNP_3} and \ref{fig:PM_comparison_GMPNP_APNP_4}). The number of regular orbits has decreased significantly in the map for $a=-0.2$ compared to that for $a=0.2$. However, it is to be noted that the degree of chaos in the map for $a=0.2$ evaluated from \acrshort{gmf} \acrshort{pnp} (Figure \ref{fig:PM_comparison_GMPNP_APNP_1}) is more than that derived from \acrshort{abn} \acrshort{pnp} (Figure \ref{fig:PM_comparison_GMPNP_APNP_3}). On the contrary, the chaotic nature of the orbits in the map for $a=-0.2$ evaluated from \acrshort{gmf} \acrshort{pnp} (Figure \ref{fig:PM_comparison_GMPNP_APNP_2}) is less compared to that derived from \acrshort{abn} \acrshort{pnp} (Figure \ref{fig:PM_comparison_GMPNP_APNP_4}). This is the consequence of the result that we presented in section \ref{sch:comp_GMf_ABN}, where we made a comparison between the two \acrshort{pnp}s (Figure \ref{fig:comparison_of_potentials_a}). The relative shift in the local minima of the effective potentials affects the chaotic behaviour of the orbits in the way we can see in Figure \ref{fig:PM_comparison_GMPNP_APNP}. 

\begin{figure*} 
    
     \centering
     \begin{subfigure}[b]{0.495\textwidth}
         \centering
         \includegraphics[width=\textwidth]{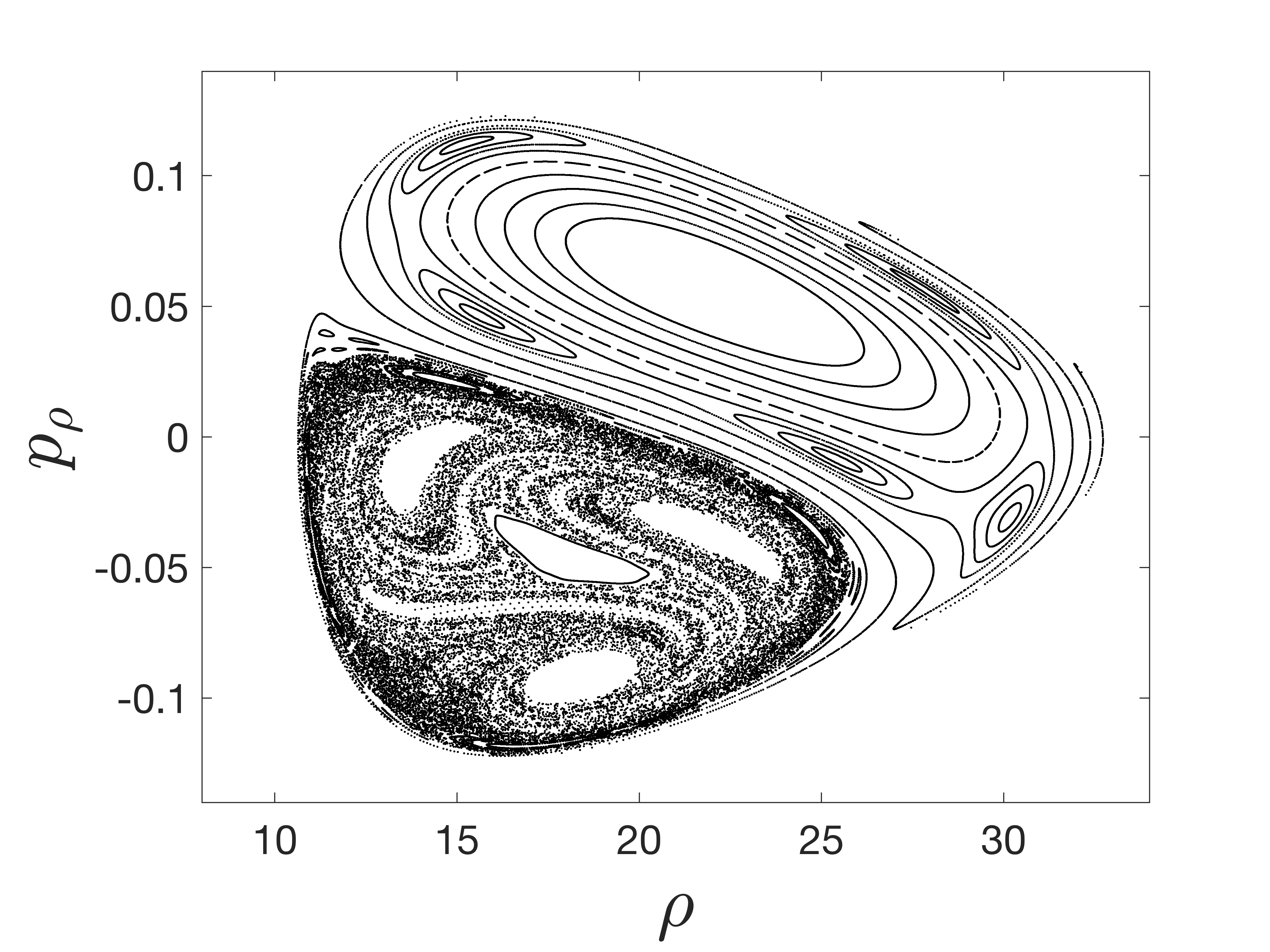}
         \caption{$a=-0.6$, $i=10^\circ$, Non-relativistic}
         \label{fig:PM_GMPNP_with_a_1}
     \end{subfigure}
     \hfill
     \begin{subfigure}[b]{0.495\textwidth}
         \centering
         \includegraphics[width=\textwidth]{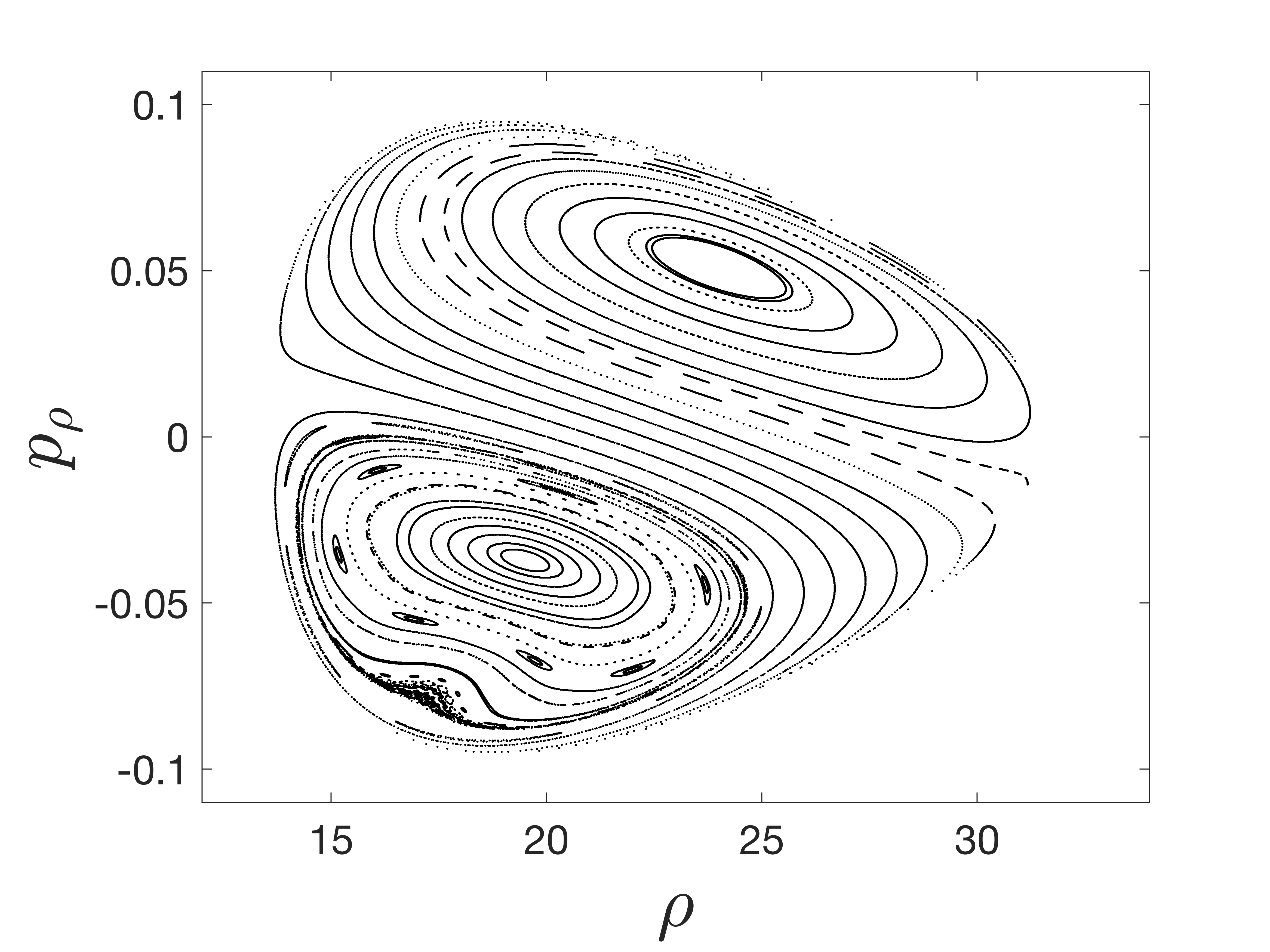}
         \caption{$a=0.6$, $i=10^\circ$, Non-relativistic}
         \label{fig:PM_GMPNP_with_a_2}
     \end{subfigure}
     \hfill
     \begin{subfigure}[b]{0.495\textwidth}
         \centering
         \includegraphics[width=\textwidth]{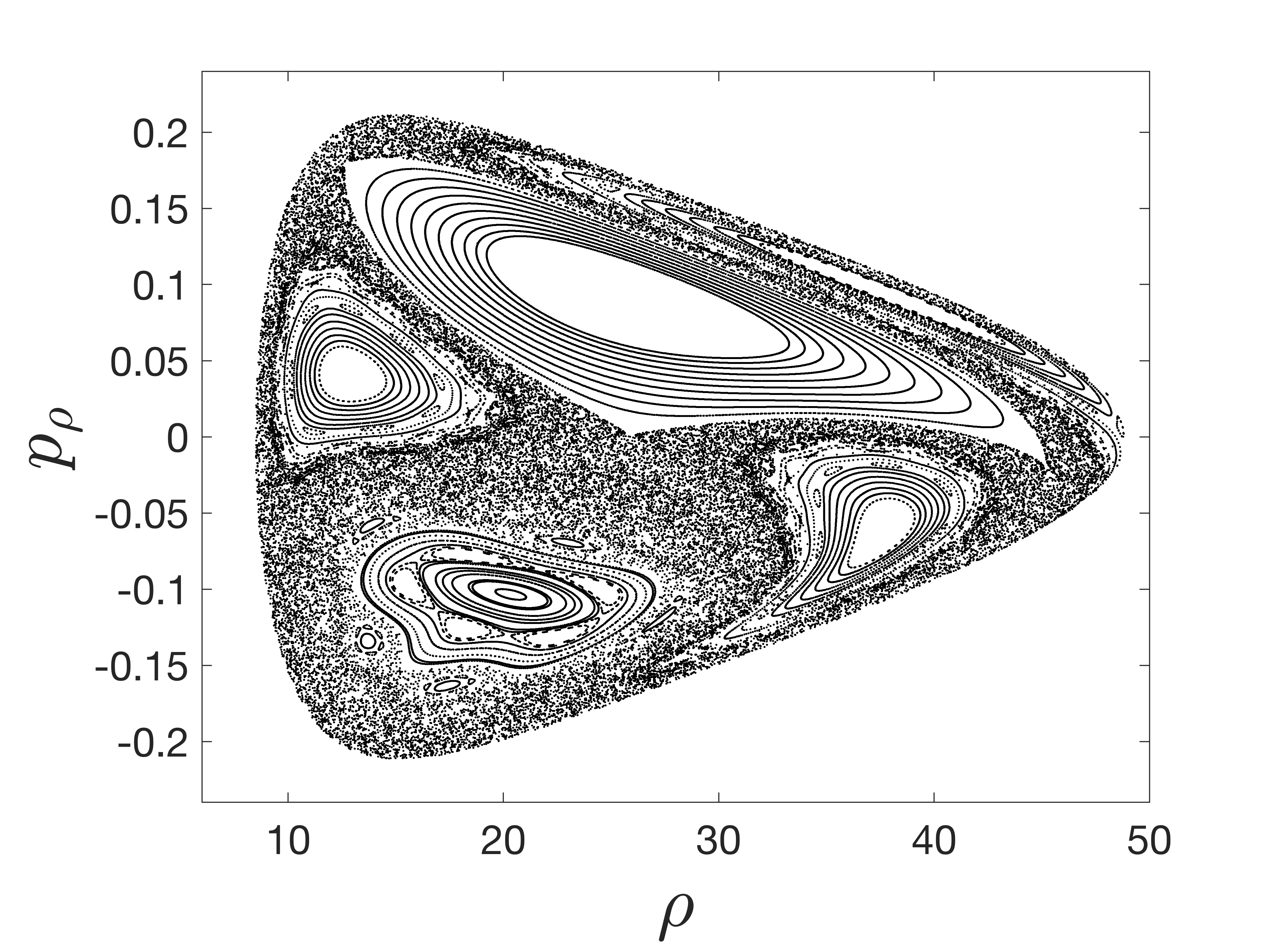}
         \caption{$a=-0.6$, $i=25^\circ$, Non-relativistic}
         \label{fig:PM_GMPNP_with_a_3}
     \end{subfigure}
     \hfill
     \begin{subfigure}[b]{0.495\textwidth}
         \centering
         \includegraphics[width=\textwidth]{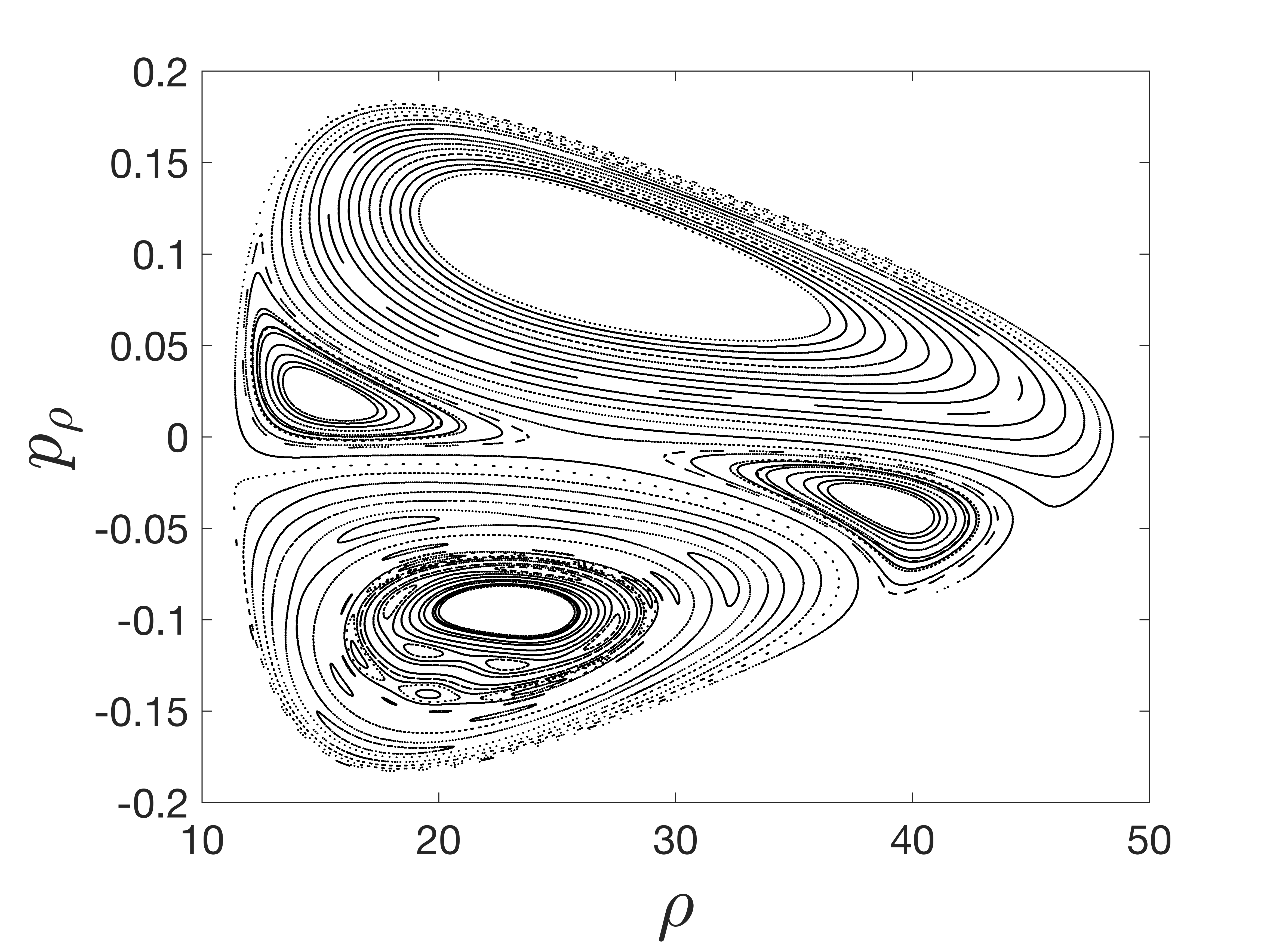}
         \caption{$a=0.6$, $i=25^\circ$, Non-relativistic}
         \label{fig:PM_GMPNP_with_a_4}
     \end{subfigure}
     \hfill
     \begin{subfigure}[b]{0.495\textwidth}
         \centering
         \includegraphics[width=\textwidth]{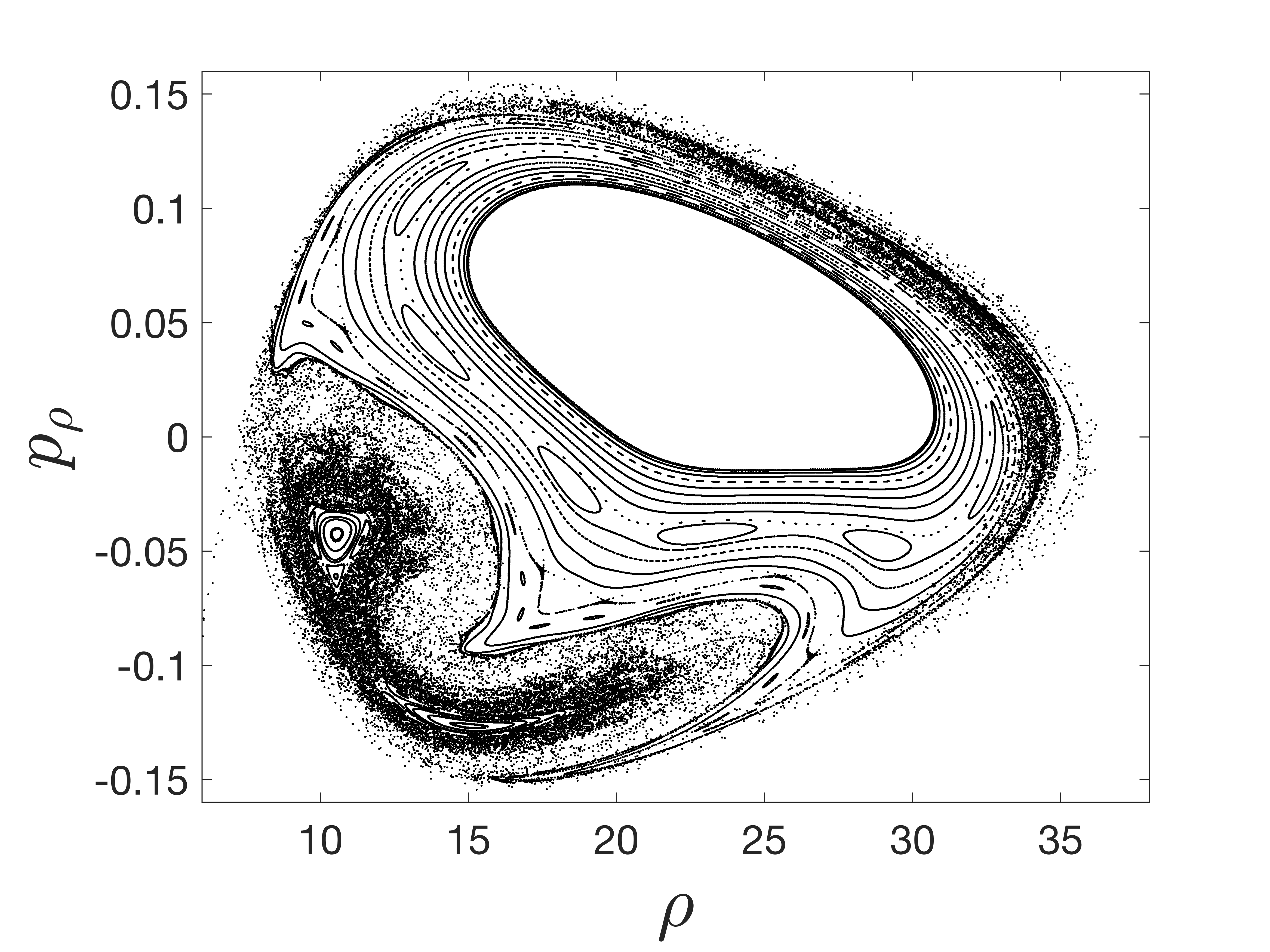}
         \caption{$a=-0.15$, $i=25^\circ$, Relativistic}
         \label{fig:PM_GMPNP_with_a_5}
     \end{subfigure}
     \hfill
     \begin{subfigure}[b]{0.495\textwidth}
         \centering
         \includegraphics[width=\textwidth]{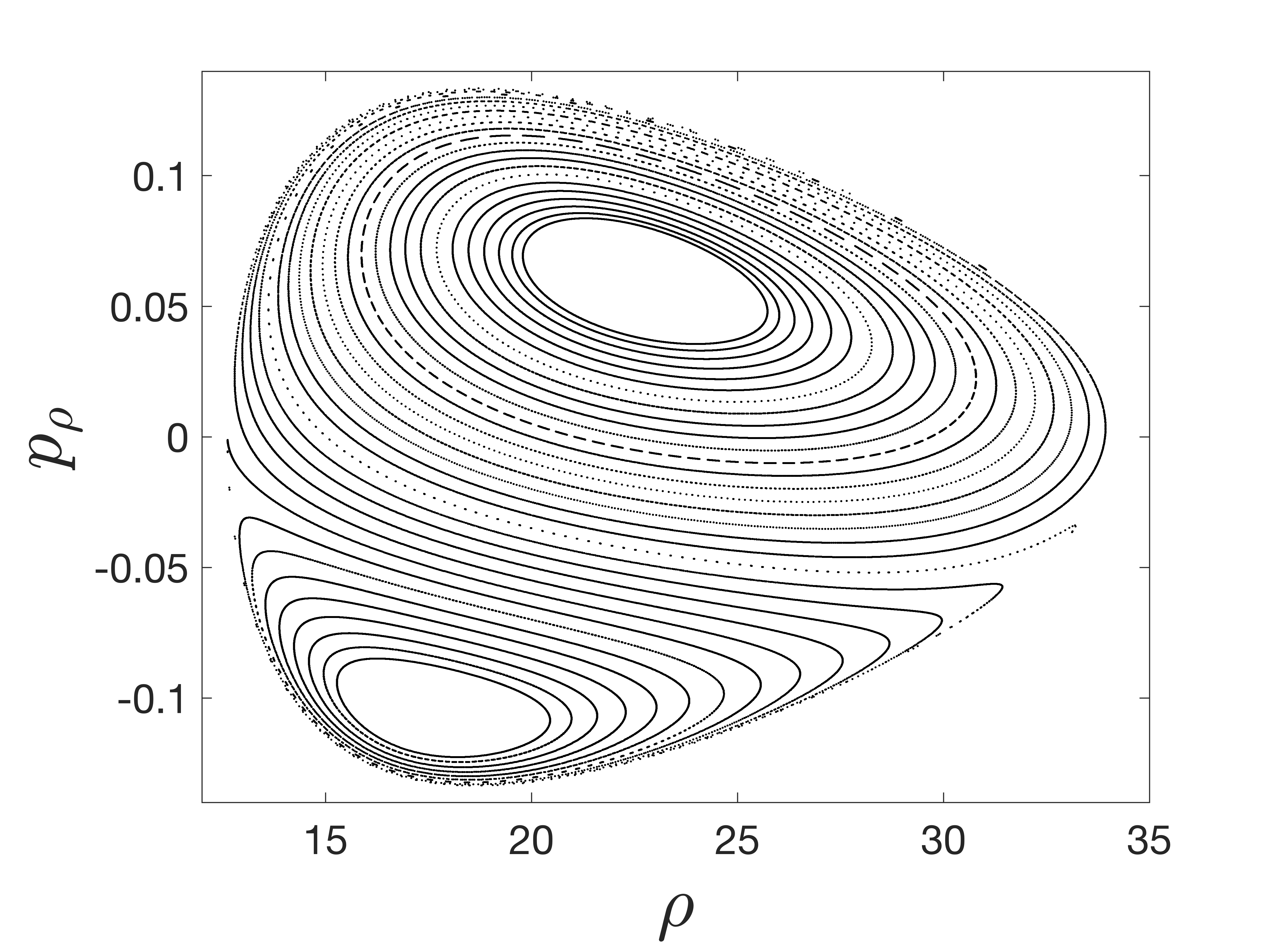}
         \caption{$a=0.15$, $i=25^\circ$, Relativistic}
         \label{fig:PM_GMPNP_with_a_6}
     \end{subfigure}
        \caption{Poincar\'e maps on the cross-sectional plane $z=0$ for the orbits governed by \acrshort{gmf} \acrshort{pnp} with different values of the Kerr parameter $a$, and fixed values of orbital parameters, such that (a)-(b) $i=10^\circ$, $E=0.976$, $L=4.90$, $\mathcal{D}=3\times 10^{-4}$, (c)-(d) $i=25^\circ$, $E=0.9734$, $L=5.82$, $\mathcal{D}=2\times 10^{-4}$, and (e)-(f) $i=25^\circ$, $E=0.965$, $L=5.835$, $\mathcal{D}=2\times 10^{-4}$.}
        \label{fig:PM_GMPNP_with_a}
\end{figure*}

To observe how the chaoticity of the inclined orbits changes with the rotation of the \acrshort{cop}, we studied the chaotic nature of the orbits for the whole range of the Kerr parameter $\left( -1 \leq a < 1 \right)$. First, we consider the orbits with a relatively smaller value of inclination ($i=10^\circ$), along with particular values of dipole coefficient $\mathcal{D}$, energy $E$, and angular momentum $L$. We observe that the orbits around the counter-rotating \acrshort{cop}s (such as Figure \ref{fig:PM_GMPNP_with_a_1}) are more chaotic than that around the co-rotating ones (such as Figure \ref{fig:PM_GMPNP_with_a_2}). For any negative value of $a$, we have examined the Poincar\'e Map plot for the corresponding positive value of $a$. The former is apparently more chaotic than the latter for all the cases. This occurs because of the usual frame-dragging effect. We have studied the gradual change in the chaoticity from $a=-1$ to $a \approx +1$. The systematic suppression of chaos with respect to the Kerr parameter is evident. We can consider a higher inclination such as $i=25^\circ$, and look for how the chaoticity changes for different values of $a$. We get to observe a similar trend in this case as well. However, the change in chaoticity is more for the higher value of the inclination angle, when the Kerr parameter changes from $a=-0.6$ (Figure \ref{fig:PM_GMPNP_with_a_3}) to $a=0.6$ (Figure \ref{fig:PM_GMPNP_with_a_4}). For the relativistic test particle, the region of chaos decreases significantly when the Kerr parameter changes from $a=-0.15$ (Figure \ref{fig:PM_GMPNP_with_a_5}) to $a=0.15$ (Figure \ref{fig:PM_GMPNP_with_a_6}). From this, we can say that the degree of chaos has a negative correlation with the Kerr parameter $a$ for any given inclination angle. However, quantitative analysis in section \ref{ch:lyapunov} will reveal more details and intricacies, where a significantly larger number of initial conditions can be considered and examined for the entire range of $a$.  

\begin{figure*} 
    
     \centering
     \begin{subfigure}[b]{0.495\textwidth}
         \centering
         \includegraphics[width=\textwidth]{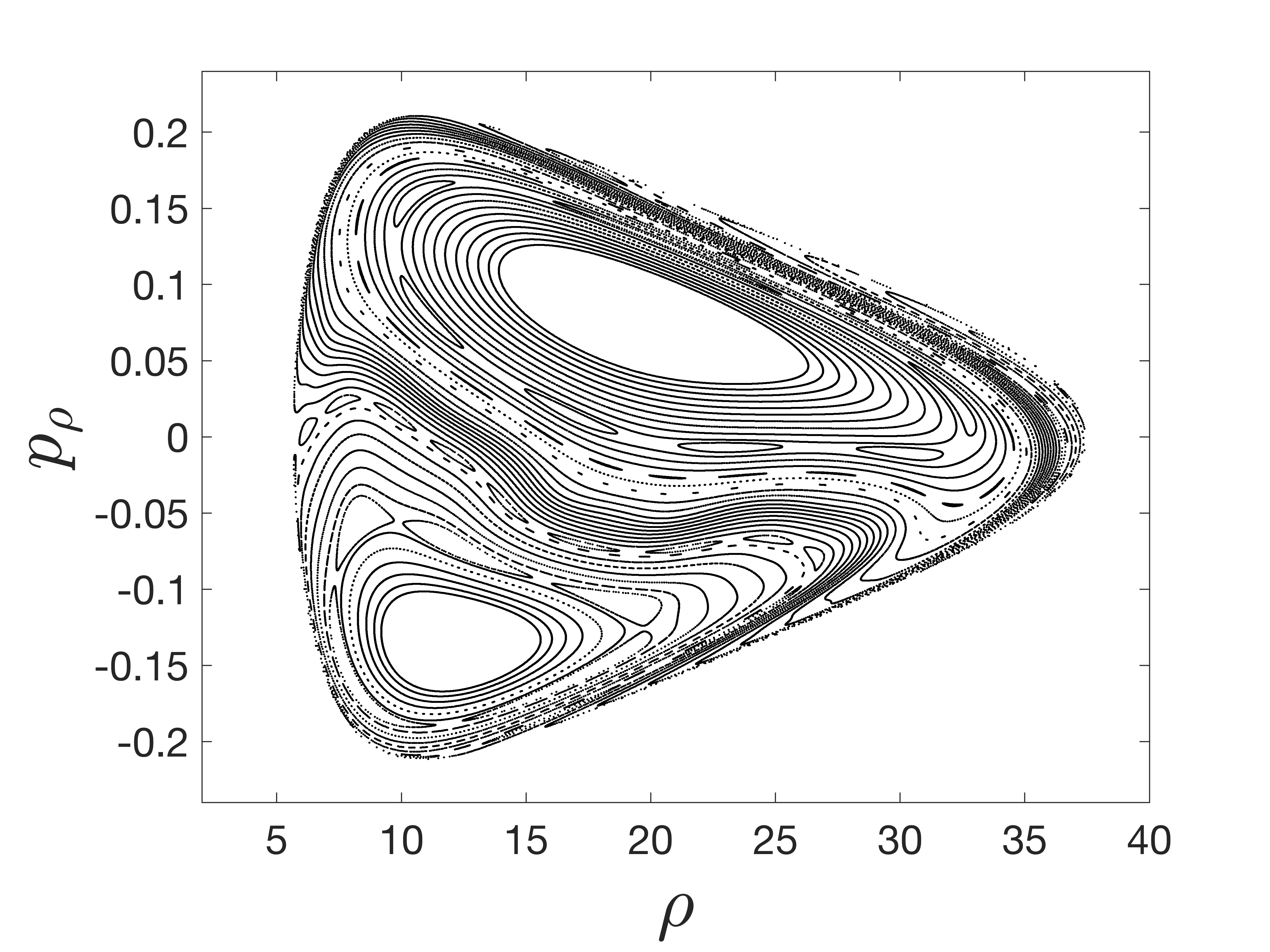}
         \caption{$i=10^\circ$, $a=-0.15$, Non-relativistic}
         \label{fig:PM_GMPNP_with_i_1}
     \end{subfigure}
     \hfill
     \begin{subfigure}[b]{0.495\textwidth}
         \centering
         \includegraphics[width=\textwidth]{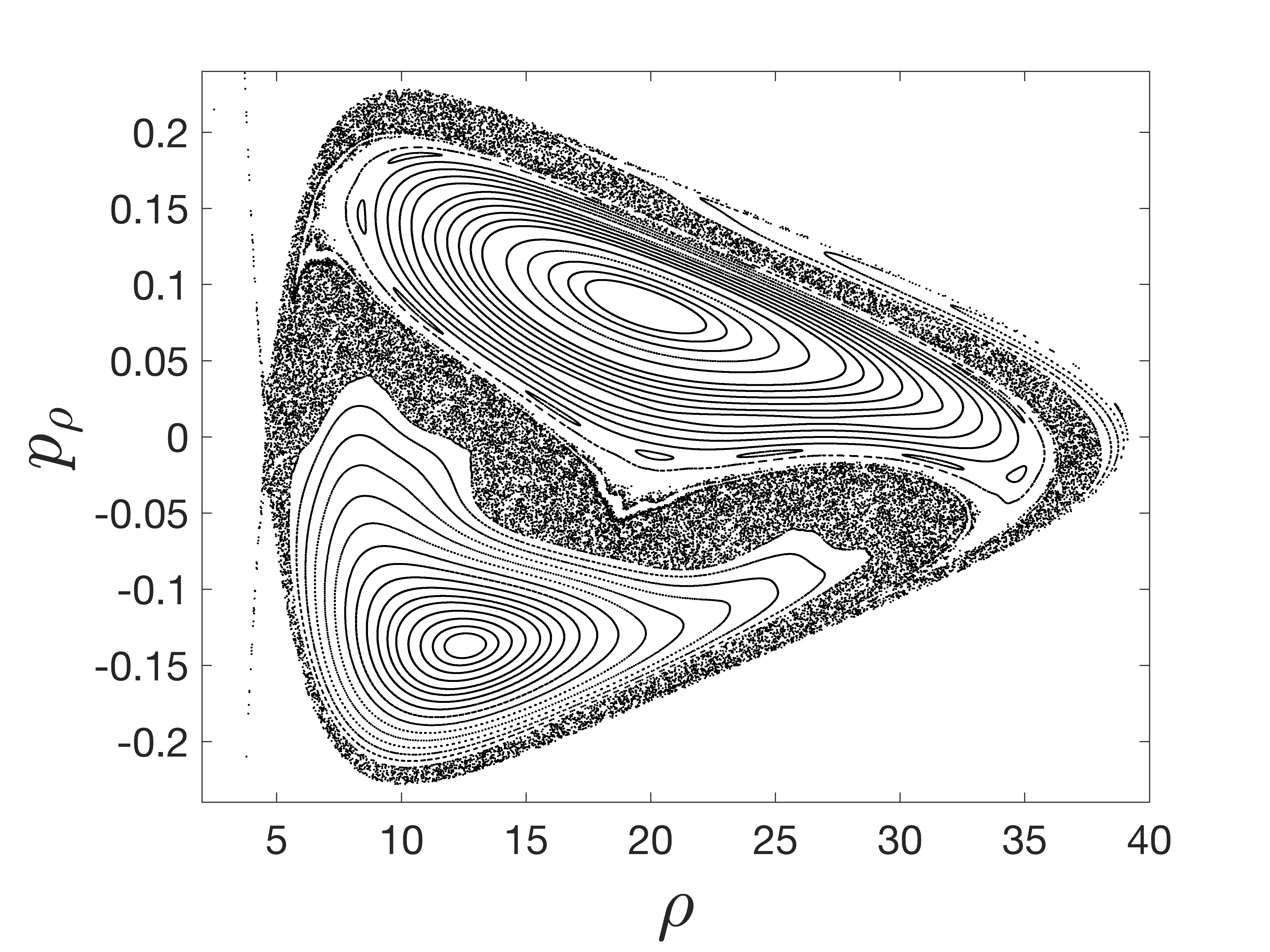}
         \caption{$i=12^\circ$, $a=-0.15$, Non-relativistic}
         \label{fig:PM_GMPNP_with_i_2}
     \end{subfigure}
     \hfill
     \begin{subfigure}[b]{0.495\textwidth}
         \centering
         \includegraphics[width=\textwidth]{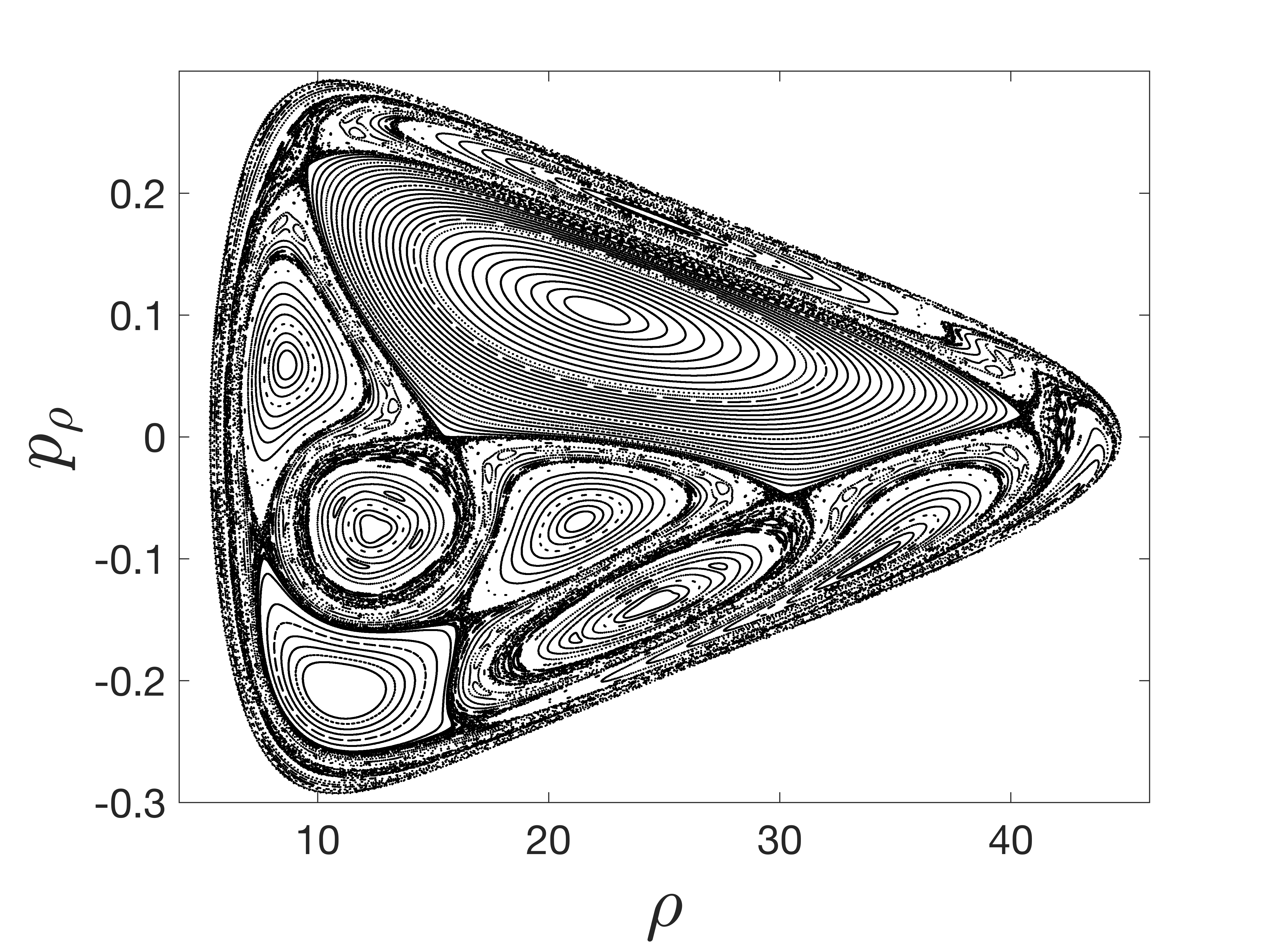}
         \caption{$i=28^\circ$, $a=-0.15$, Non-relativistic}
         \label{fig:PM_GMPNP_with_i_3}
     \end{subfigure}
     \hfill
     \begin{subfigure}[b]{0.495\textwidth}
         \centering
         \includegraphics[width=\textwidth]{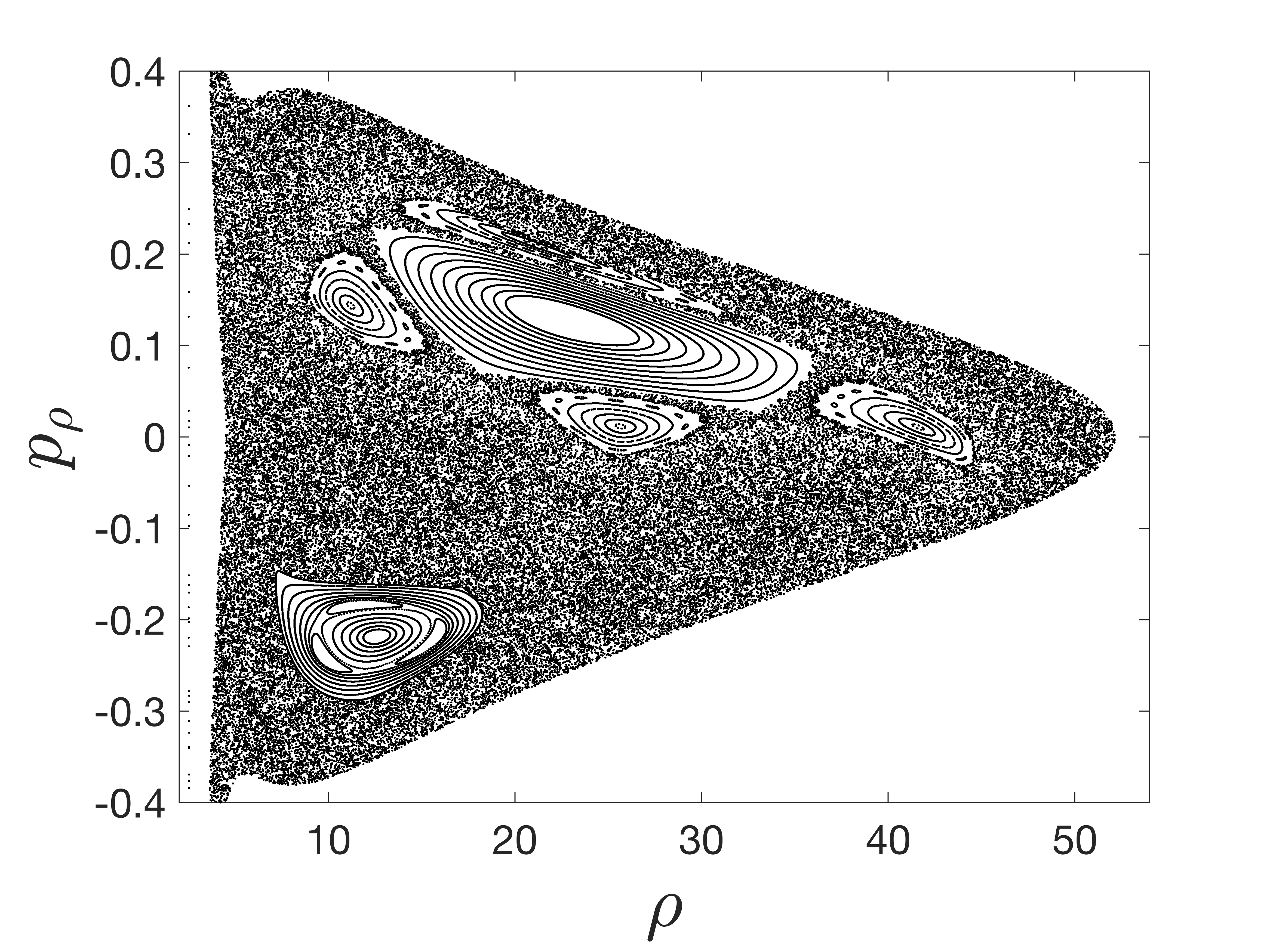}
         \caption{$i=30^\circ$, $a=-0.15$, Non-relativistic}
         \label{fig:PM_GMPNP_with_i_4}
     \end{subfigure}
     \hfill
     \begin{subfigure}[b]{0.495\textwidth}
         \centering
         \includegraphics[width=\textwidth]{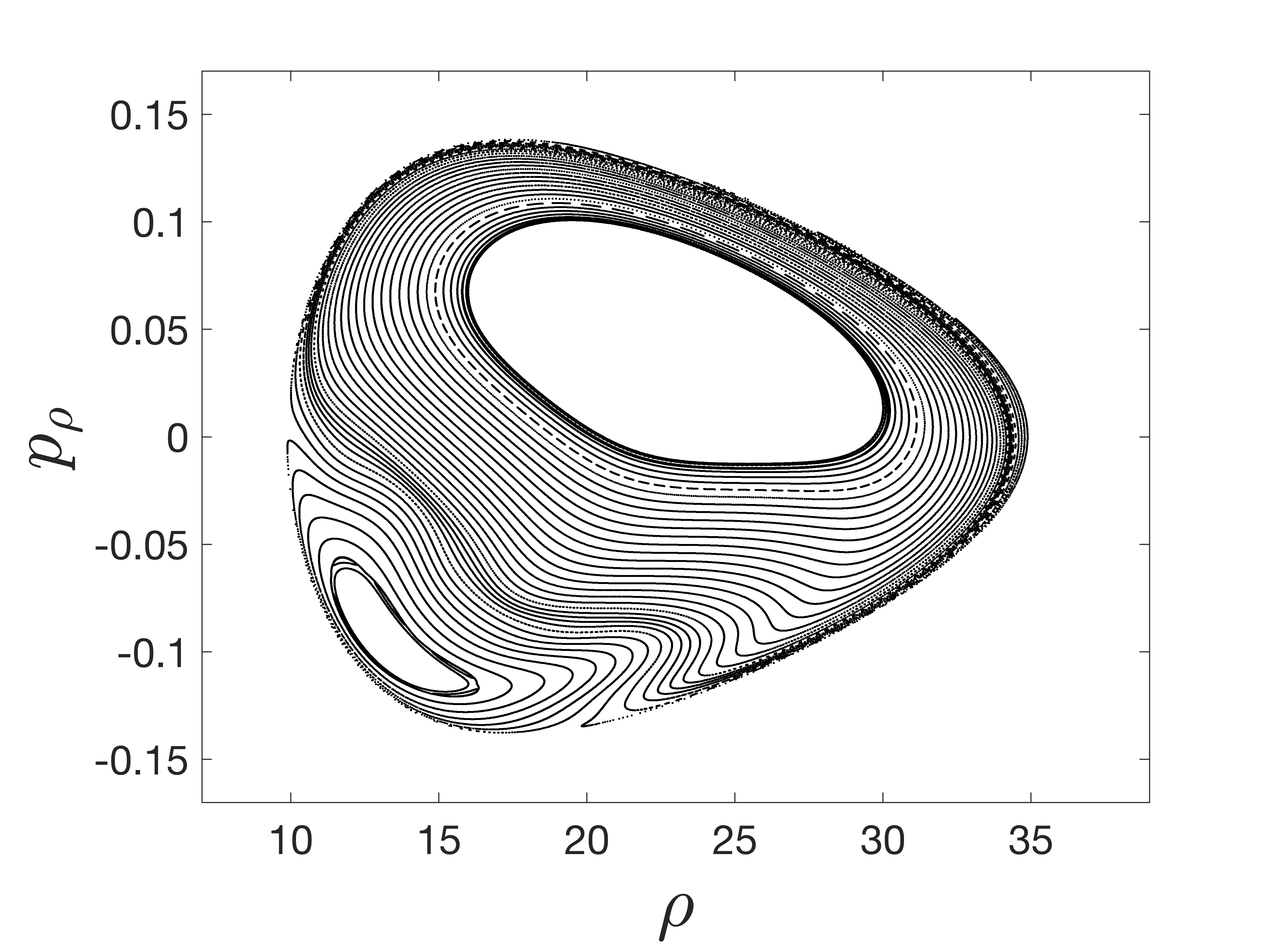}
         \caption{$i=25^\circ$, $a=-0.15$, Relativistic}
         \label{fig:PM_GMPNP_with_i_5}
     \end{subfigure}
     \hfill
     \begin{subfigure}[b]{0.495\textwidth}
         \centering
         \includegraphics[width=\textwidth]{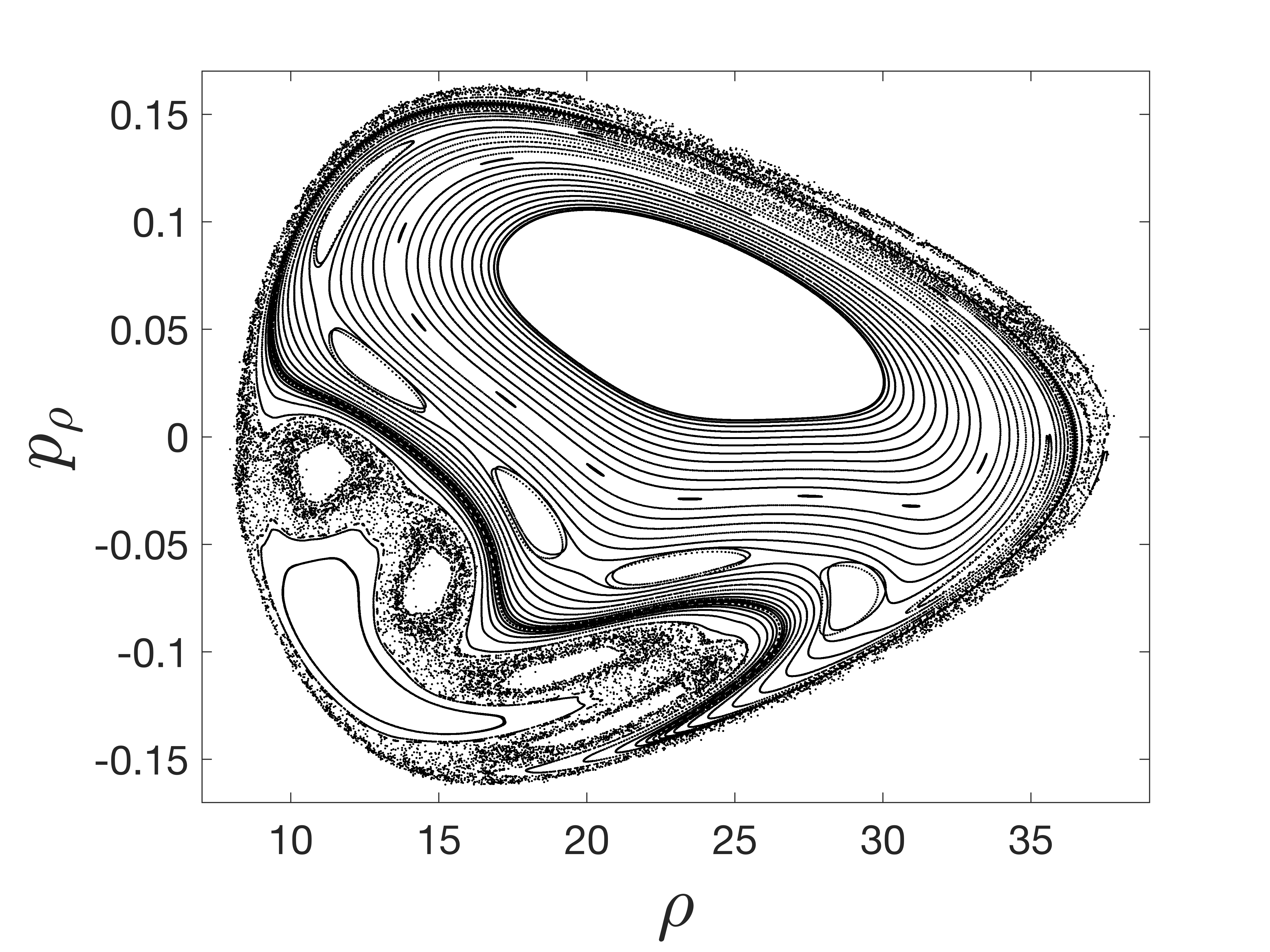}
         \caption{$i=26^\circ$, $a=-0.15$, Relativistic}
         \label{fig:PM_GMPNP_with_i_6}
     \end{subfigure}
        \caption{Poincar\'e maps on the cross-sectional plane $z=0$ for the orbits governed by \acrshort{gmf} \acrshort{pnp} with different values of the inclination angle $i$, and fixed values of orbital parameters, such that (a)-(b) $a=-0.15$, $E=0.976$, $L=4.22$, $\mathcal{D}=2\times 10^{-4}$, (c)-(d) $a=-0.15$, $E=0.965$, $L=5.48$, $\mathcal{D}=2\times 10^{-4}$, and (e)-(f) $a=-0.15$, $E=0.965$, $L=5.95$, $\mathcal{D}=2\times 10^{-4}$.}
        \label{fig:PM_GMPNP_with_i}
\end{figure*}

After finding out the dependence of chaos on the Kerr parameter $a$, we studied the change in the chaotic nature of the orbits with respect to the inclination angle $i$ for both non-relativistic and relativistic particles (Figure \ref{fig:PM_GMPNP_with_i}). In all cases, the chaotic behaviour of the orbits increases with the angle of inclination $i$. For the non-relativistic orbits with higher inclinations, the Poincar\'e Maps change drastically when the inclination angle increases from $i=28^\circ$ (Figure \ref{fig:PM_GMPNP_with_i_3}) to $i=30^\circ$ (Figure \ref{fig:PM_GMPNP_with_i_4}). In comparison to this, the change in the nature of orbits for smaller inclination angles is more gradual. A smaller region of order converts into the region of chaos when the inclination angle increases from $i=10^\circ$ (Figure \ref{fig:PM_GMPNP_with_i_1}) to $i=12^\circ$ (Figure \ref{fig:PM_GMPNP_with_i_2}). We have already observed the change to be gradual at smaller inclination angles in Figure \ref{fig:PM_with_i_NonRot} when we studied the Poincar\'e Maps corresponding to the off-equatorial orbits around the Schwarzschild-like \acrshort{cop}s. This phenomenon will also be corroborated and studied in detail using the quantitative analysis of chaos in the next section. Similar to the non-relativistic case, the chaotic nature of the relativistic orbits has a positive correlation with the inclination angle $i$. The Poincar\'e Map plot for $i=25^\circ$ (Figure \ref{fig:PM_GMPNP_with_i_5}) consists of regular orbits, many of which extinguish and turn into chaotic orbits when the inclination angle increases to $i=26^\circ$ (Figure \ref{fig:PM_GMPNP_with_i_6}). 

\section{Quantification of chaos using Maximum Lyapunov Exponents} \label{ch:lyapunov}
As we have studied the chaotic dynamics of the orbits qualitatively, we proceed to look for a quantitative way of studying chaos. There are many chaotic indicators available in the literature. We have used the Lyapunov Exponent in this regard as it is easy to implement, can efficiently distinguish between order and chaos in our system, and is very effective in studying the chaotic correlations over a long range of dynamical parameters. One can measure the exponential divergence of two neighbouring orbits while their starting point is very close to each other \citep{strogatz2007nonlinear}. The number of exponents will equal the number of dimensions of the phase space. However, in the long run (for $t \rightarrow \infty$), the maximum of the exponents dominates. This is known as the \acrshort{mle} \citep{nag2017influence, de2021beyond}. This can be calculated as
\begin{equation} \label{eq:lpnv_two_particle}
    \Lambda_{\text{max}} = \lim\limits_{\begin{subarray}{c}
      t \to \infty\\
      ||\delta x(0)|| \to 0
      \end{subarray}} \left( \frac{1}{t} \ln \frac{||\delta x(t)||}{||\delta x(0)||} \right)
\end{equation}
where $||\delta x(t)||$ is the norm of the deviation between the two neighbouring orbits in the phase space at time $t$. It can be evaluated using the Difference-Hamiltonian and the variational equations \citep{skokos2010numerical}. This method is known as the variational method. Although the method is very accurate, the calculation is intensively rigorous with the mathematically complex form of \acrshort{gmf} \acrshort{pnp}. As an alternative, we have used the two-particle method, a more convenient and less rigorous way of calculating \acrshort{mle}s \citep{de2021beyond}. 

To achieve accurate results using the two-particle method, the initial deviation has to be very small, and the precision of the computational instrument has to be high. Furthermore, we need to re-normalise the deviated orbit after every time step $\tau$ using the Gram-Schmidt renormalisation scheme \citep{tancredi2001comparison} so that the \acrshort{mle}s do not diverge very quickly. For a particular set of parameters, we can consider a large number of initial conditions evenly distributed over the allowable phase space and calculate \acrshort{mle}s for each of them. Thereafter, we can take the average of all the exponents $\left( \Lambda_{\text{av}} \right)$ and consider this as a rough quantitative measure of the overall chaos in the system. In the present work, we have used double precision for the computation of the \acrshort{mle}s. For each initial condition, the initial deviation has been taken to be $\delta x(0)=10^{-8}$, and the calculation has been carried out for $n=10^5$ iterations with renormalisation time-step $\tau=0.1$. The threshold has been found to be $\approx 1.5 \times 10^{-4}$, which means the value of $\Lambda_{\text{max}}$ is smaller than $1.5 \times 10^{-4}$ for regular orbits and it is more than $1.5 \times 10^{-4}$ for chaotic orbits.

\begin{figure*} 
    
     \centering
     \begin{subfigure}[b]{0.495\textwidth}
         \centering
         \includegraphics[width=\textwidth]{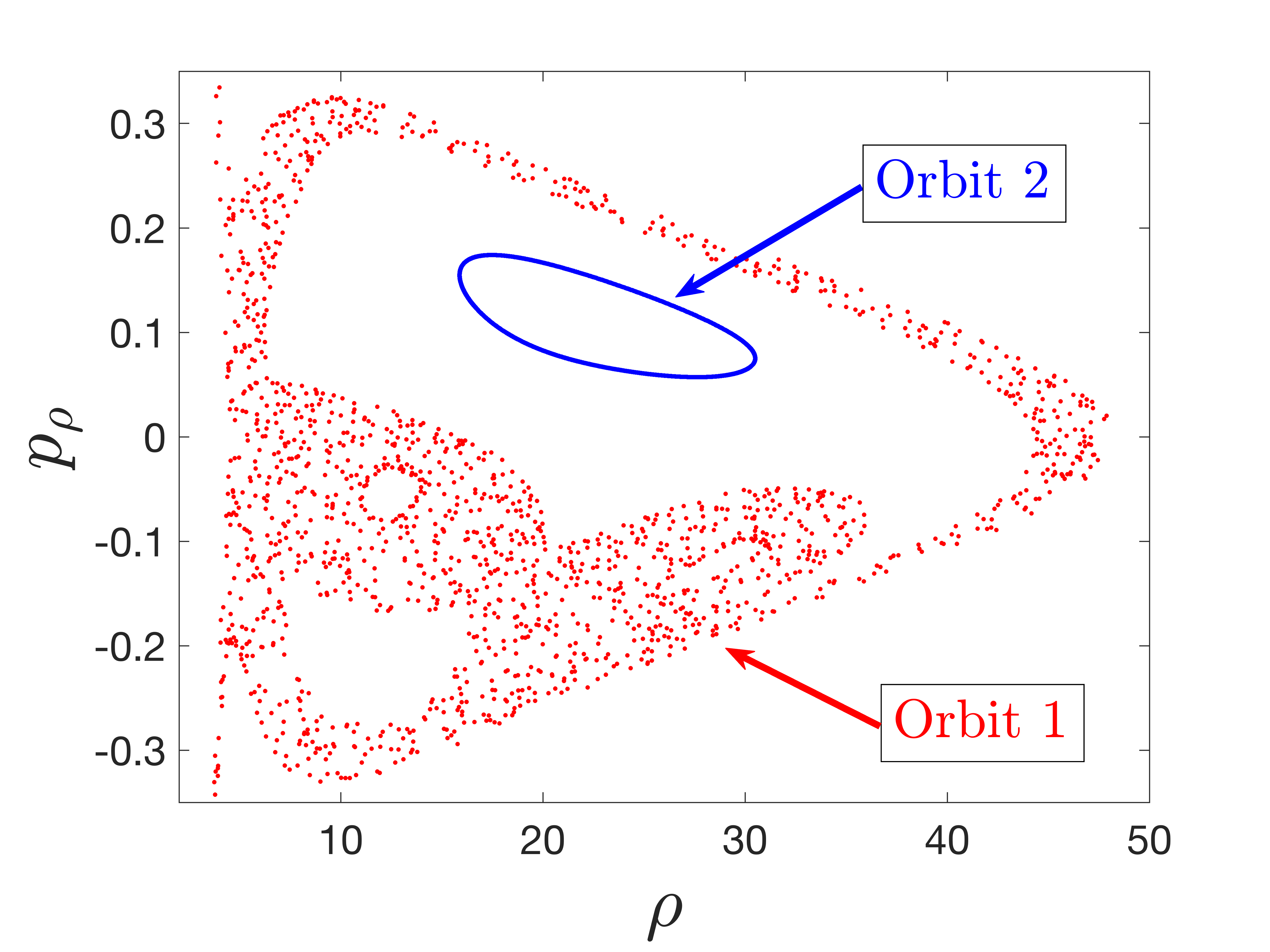}
         \caption{Poincar\'e sections for orbit 1 (chaotic) and orbit 2 (regular)}
         \label{fig:lpnv_check_1}
     \end{subfigure}
     \hfill
     \begin{subfigure}[b]{0.495\textwidth}
         \centering
         \includegraphics[width=\textwidth]{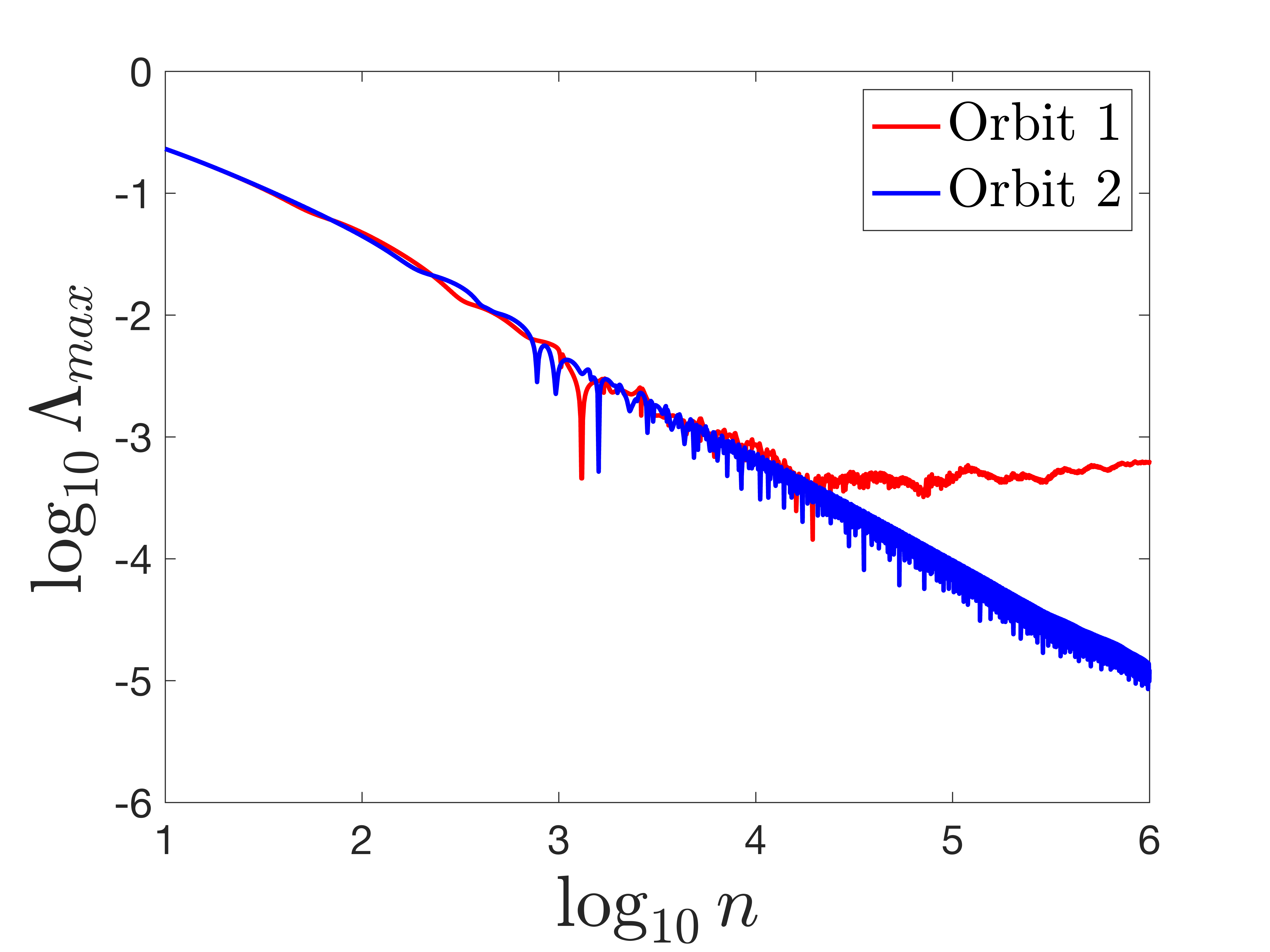}
         \caption{\acrshort{mle}s corresponding to orbit 1 (chaotic) and orbit 2 (regular)}
         \label{fig:lpnv_check_2}
     \end{subfigure}
        \caption{Poincar\'e sections and corresponding \acrshort{mle}s ($\Lambda_{\text{max}}$) for the orbits with parameters $a=-0.15$, $i=29^\circ$, $E=0.965$, $L=5.48$, and $\mathcal{D}=2\times 10^{-4}$.}
        \label{fig:lpnv_check}
\end{figure*}

Before proceeding further, we have checked if the aforementioned scheme of \acrshort{mle} is suitable to distinguish between chaos and order in our system. We have considered two orbits with the same orbital parameters $a=-0.15$, $i=29^\circ$, $E=0.965$, $L=5.48$, and $\mathcal{D}=2\times 10^{-4}$. The initial conditions for orbit 1 are $\rho = 16$, $p_\rho = -0.1$, $z = 0$, $p_z>0$, and that for orbit 2 are $\rho = 23.5$, $p_\rho = 0.065$, $z = 0$, $p_z>0$. From the Poincar\'e plot of sections (Figure \ref{fig:lpnv_check_1}), we can see that orbit 1 is chaotic in nature, whereas orbit 2 is regular. We evaluated the corresponding \acrshort{mle}s for these two orbits as a function of the number of iterations (Figure \ref{fig:lpnv_check_2}). We can observe that the scheme can efficiently distinguish between the two orbits for $\left( \log_{10} n \right) \gtrsim 5$. At $n=10^5$, the value of \acrshort{mle} for orbit 1 is above the threshold $\Lambda_{\text{max}} = 1.5 \times 10^{-4}$, and that for orbit 2 is below the threshold, which implies that orbit 1 is chaotic, whereas orbit 2 is regular. Thus, this scheme of \acrshort{mle} works efficiently to distinguish between order and chaos in our system. 

\begin{figure*} 
    
     \centering
     \begin{subfigure}[b]{0.495\textwidth}
         \centering
         \includegraphics[width=\textwidth]{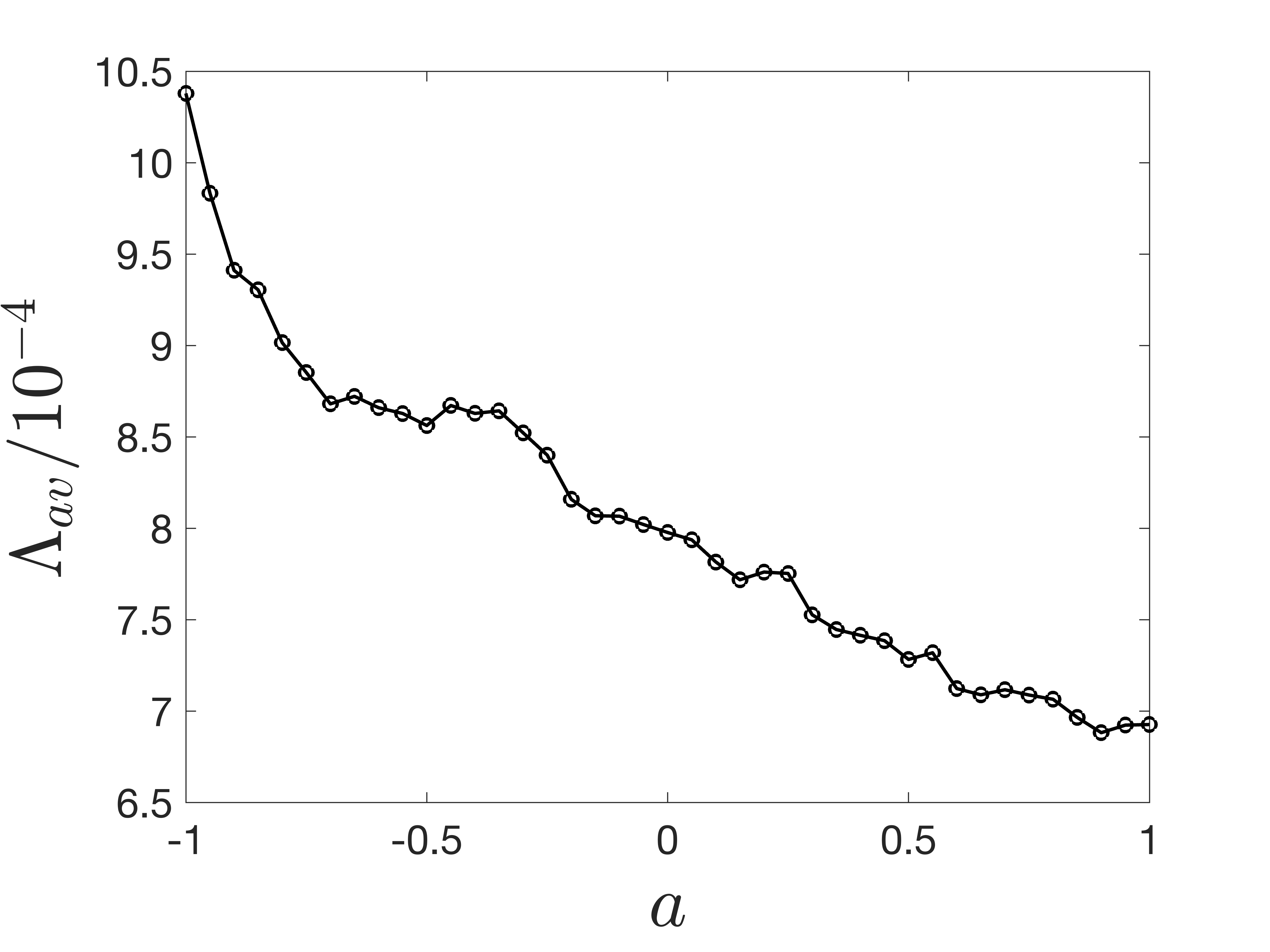}
         \caption{$i=10^\circ$, $E=0.976$, $L=4.85$, and $\mathcal{D}=3\times 10^{-4}$}
         \label{fig:lcn_with_a_1}
     \end{subfigure}
     \hfill
     \begin{subfigure}[b]{0.495\textwidth}
         \centering
         \includegraphics[width=\textwidth]{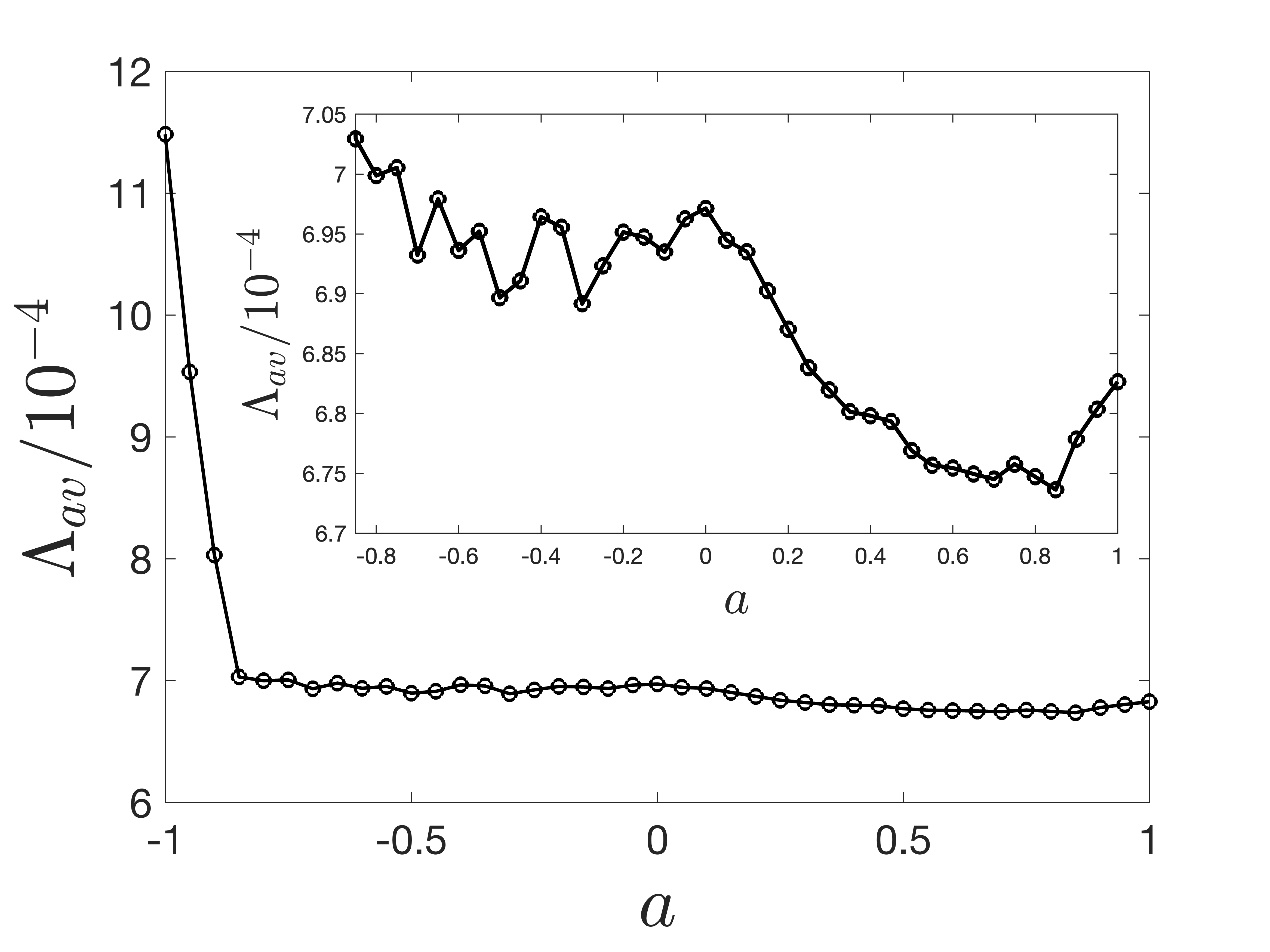}
         \caption{$i=25^\circ$, $E=0.9734$, $L=5.82$, and $\mathcal{D}=2\times 10^{-4}$}
         \label{fig:lcn_with_a_2}
     \end{subfigure}
     \hfill
     \begin{subfigure}[b]{0.495\textwidth}
         \centering
         \includegraphics[width=\textwidth]{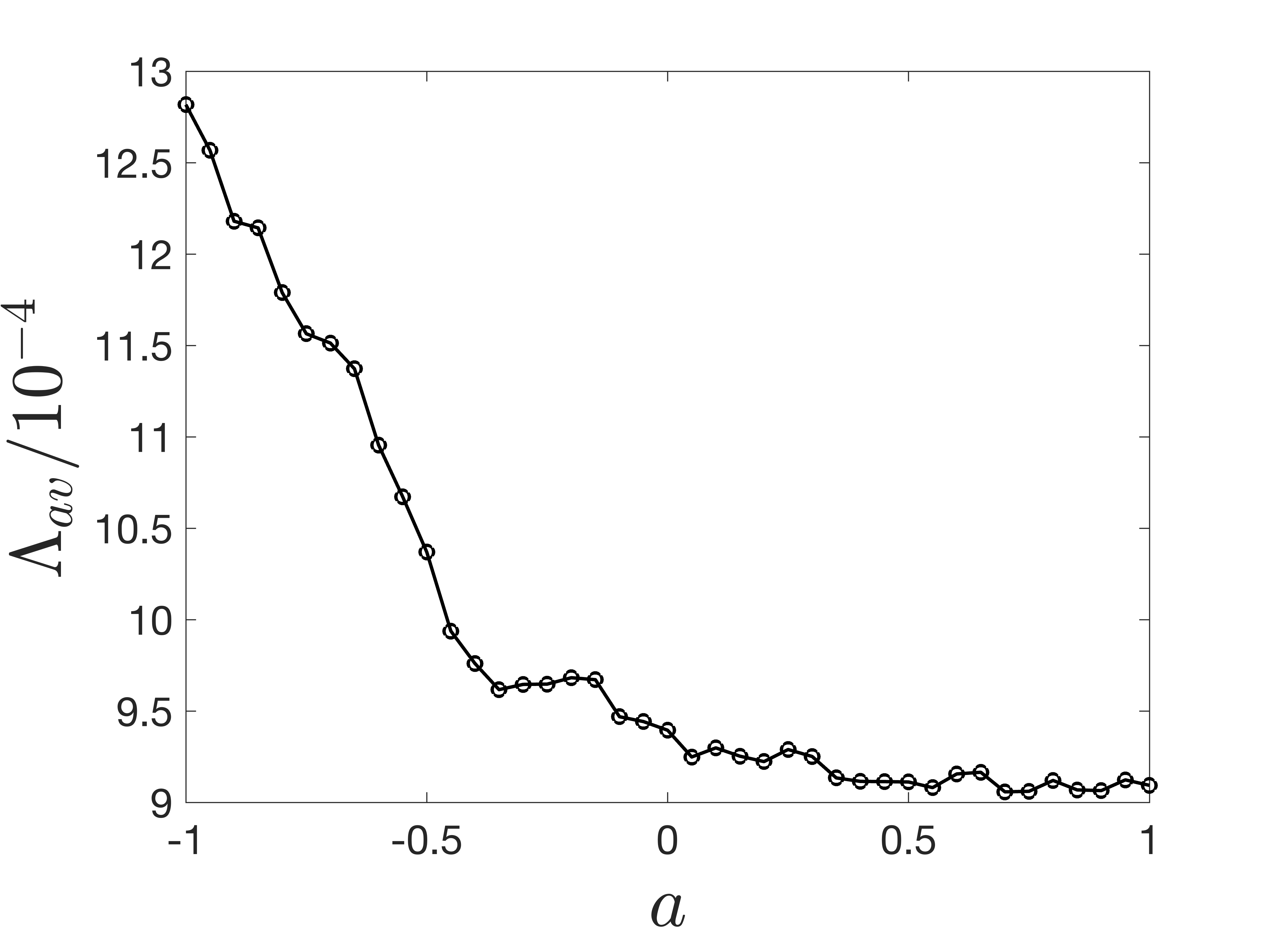}
         \caption{$i=28^\circ$, $E=0.9734$, $L=5.82$, and $\mathcal{D}=2\times 10^{-4}$}
         \label{fig:lcn_with_a_3}
     \end{subfigure}
     \hfill
     \begin{subfigure}[b]{0.495\textwidth}
         \centering
         \includegraphics[width=\textwidth]{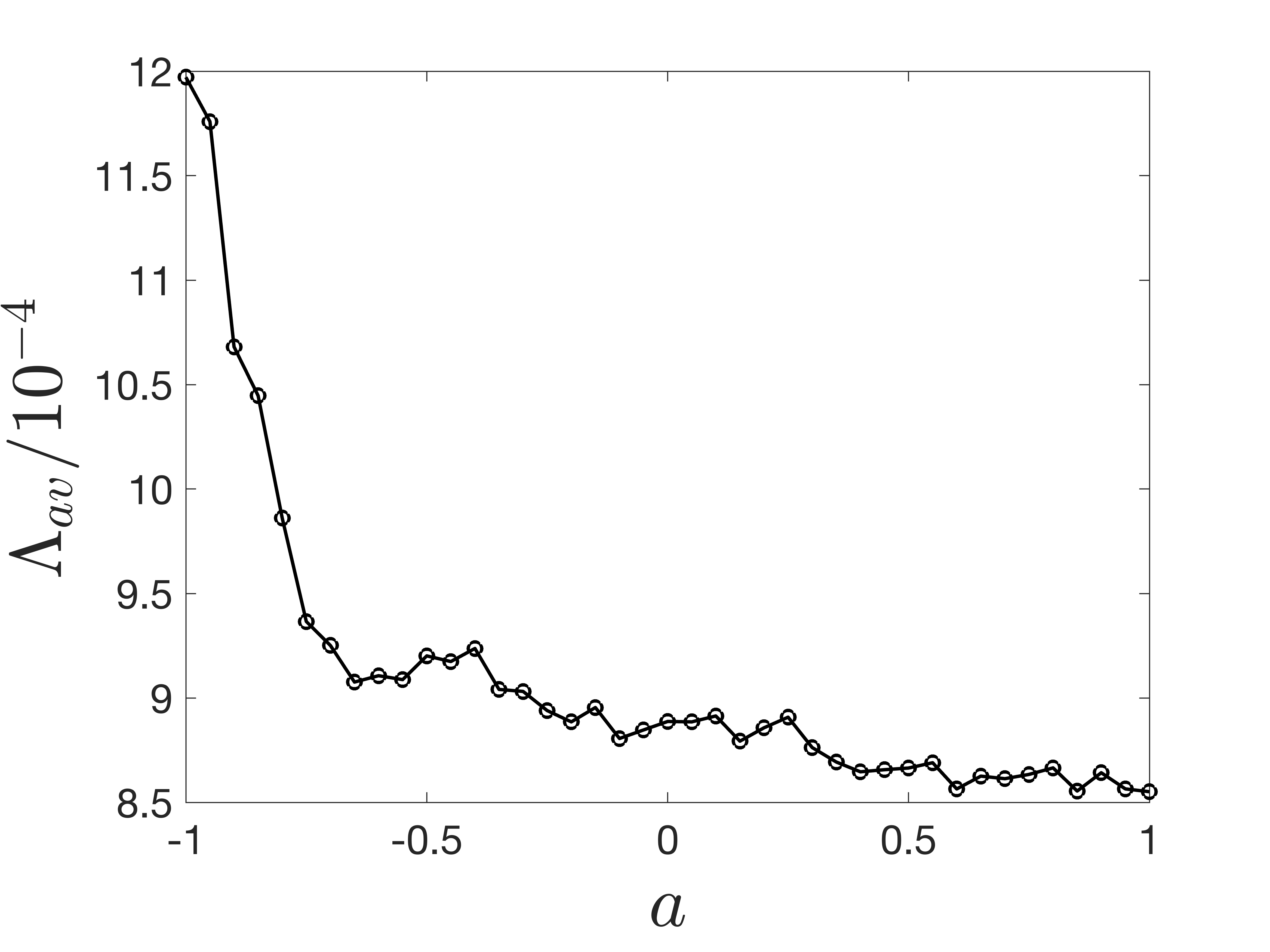}
         \caption{$i=25^\circ$, $E=0.9750$, $L=5.72$, and $\mathcal{D}=2\times 10^{-4}$}
         \label{fig:lcn_with_a_4}
     \end{subfigure}
        \caption{Maximum Lyapunov Exponent ($\Lambda_{\text{av}}$) as a function of Kerr Parameter $a$.}
        \label{fig:lcn_with_a}
\end{figure*}

By implementing \acrshort{mle} as a quantitative indicator of chaos, we have studied the correlation of chaos with the Kerr parameter $a$ of the \acrshort{cop} and the inclination angle $i$ of the orbit. First, let us consider the former. Looking at the overall trends of \acrshort{mle}s, it is evident that the chaoticity of the system has a negative correlation with the rotation parameter $a$ (Figure \ref{fig:lcn_with_a}). The inclined orbits are always more chaotic for the maximally counter-rotating \acrshort{cop}s, \textit{i.e.}, when the value of $a$ is closer to -1. For lower values of the inclination angle, such as $i=10^\circ$ (Figure \ref{fig:lcn_with_a_1}), the degree of chaos decreases gradually with the increasing value of $a$. For higher angles of inclination, the correlation becomes nuanced. It turns out that the degree of chaos weakly depends on the Kerr parameter $a$ for most of its range. However, the chaoticity rapidly increases below a threshold value of the rotation parameter $a=a_{\text{c}}$. For $i=25^\circ$ (Figure \ref{fig:lcn_with_a_2}), the threshold value is $a_{\text{c}} \approx -0.85$. When the value of $a$ is more than $a_{\text{c}}=-0.85$, the chaoticity decreases very slowly as $a$ increases. But below $a=a_{\text{c}}$, the \acrshort{mle} increases drastically as $a$ decreases. We observed this qualitatively with the Poincar\'e Maps in the previous section. 

The threshold value of the rotation parameter $a=a_{\text{c}}$ gets affected by the overall chaoticity of the system. Any change in the orbital parameters, which enhances chaoticity in the system, affects it so that the value of $a_{\text{c}}$ increases. For example, for $i=28^\circ$ (Figure \ref{fig:lcn_with_a_3}), the threshold value comes out to be $a_{\text{c}} \approx -0.45$. It is more than $a_{\text{c}} \approx -0.85$, which occurs when the inclination was $i=25^\circ$ (Figure \ref{fig:lcn_with_a_2}). It is happening because the degree of chaos is enhanced when the inclination angle increases from $i=25^\circ$ to $i=28^\circ$. It seems like the system tries to sustain its stability as long as possible. However, the enhancement of chaoticity makes the system more nonlinear, causing the system to be stable for a shorter span of the Kerr parameter. That is why the threshold value $a=a_{\text{c}}$ increases and the system retains its stable nature for a shorter range of $a$. When $a<a_{\text{c}}$, the nonlinearity in the system gets triggered, and the \acrshort{mle} shows a sharp upturn as $a$ decreases. We observe a similar phenomenon when we increase energy $E$ and decrease angular momentum $L$ by keeping the inclination angle fixed at $i=25^\circ$ (Figure \ref{fig:lcn_with_a_4}). It causes the overall chaoticity to increase, and the threshold value increases from $a_{\text{c}} \approx -0.85$ to $a_{\text{c}} \approx -0.65$.

\begin{figure*} 
    
     \centering
     \begin{subfigure}[b]{0.495\textwidth}
         \centering
         \includegraphics[width=\textwidth]{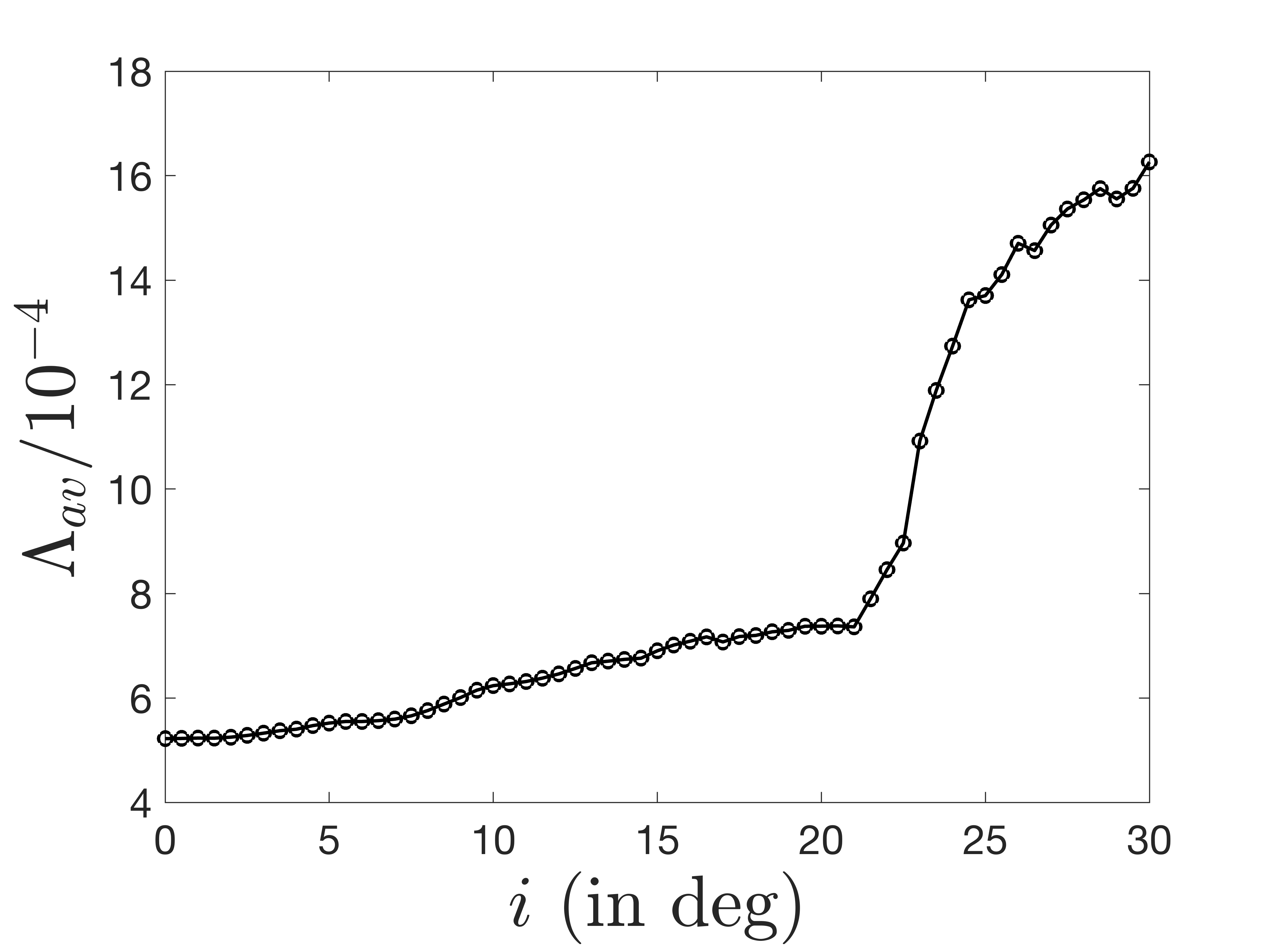}
         \caption{$a=-0.15$, $E=0.976$, $L=4.85$, and $\mathcal{D}=2\times 10^{-4}$}
         \label{fig:lcn_with_i_1}
     \end{subfigure}
     \hfill
     \begin{subfigure}[b]{0.495\textwidth}
         \centering
         \includegraphics[width=\textwidth]{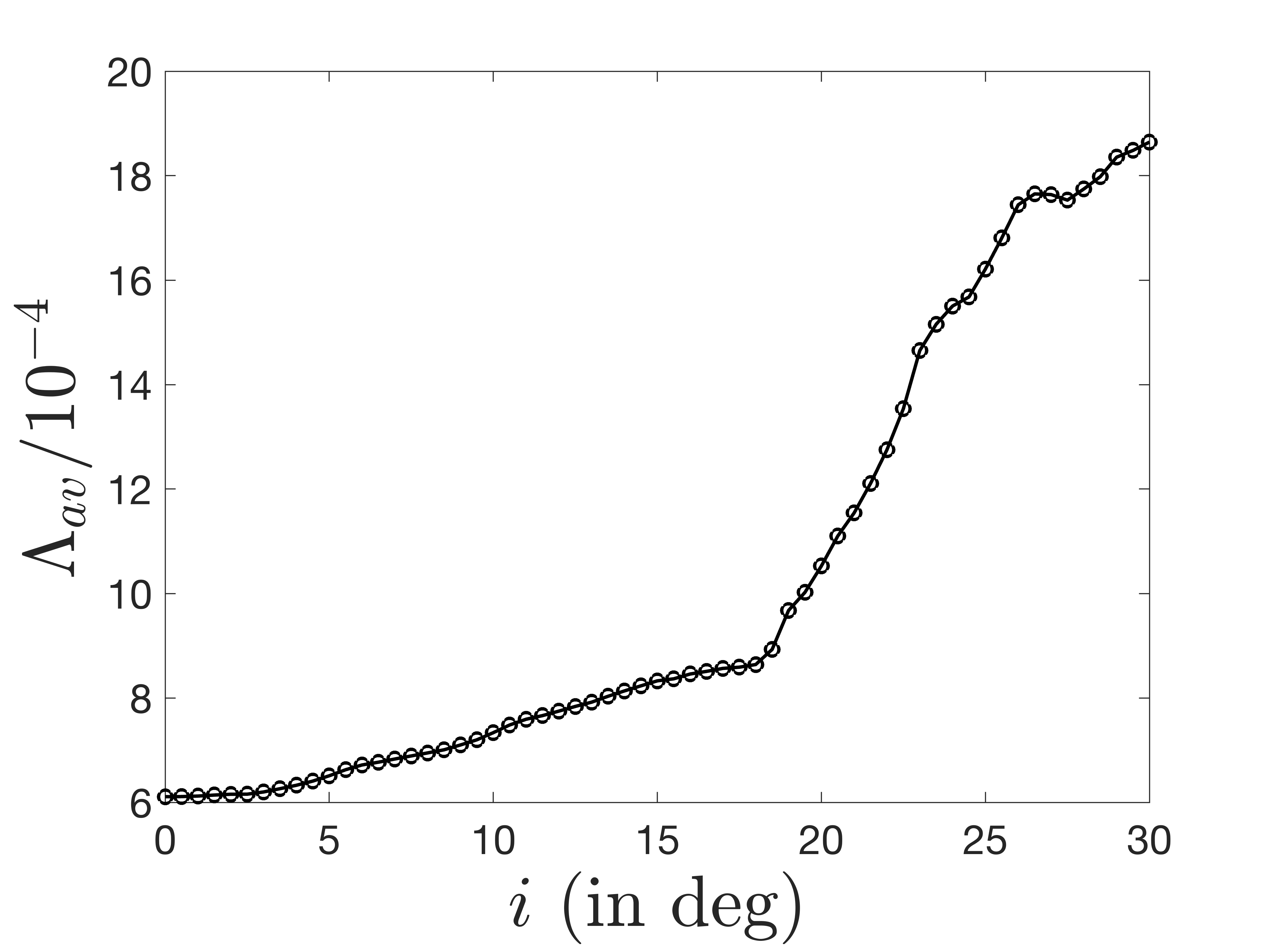}
         \caption{$a=-0.15$, $E=0.976$, $L=4.85$, and $\mathcal{D}=2.2\times 10^{-4}$}
         \label{fig:lcn_with_i_2}
     \end{subfigure}
        \caption{Maximum Lyapunov Exponent ($\Lambda_{\text{av}}$) as a function of inclination angle $i$.}
        \label{fig:lcn_with_i}
\end{figure*}

On the other hand, the dependence of chaoticity on the inclination angle $i$ is opposite to that for the rotation parameter $a$ (Figure \ref{fig:lcn_with_i}). The degree of chaos increases in the system as the orbits become more inclined with the rotation axis of the \acrshort{cop} while keeping the rest of the orbital parameters fixed. It implies that the thicker accretion disks have more chaotic orbits than the thinner ones. However, the change in the \acrshort{mle} is more rapid at the higher values of the inclination angle. This was also observed earlier while studying the chaotic behaviour of the orbits qualitatively using Poincar\'e Map plots (Figure \ref{fig:PM_GMPNP_with_i}). There is a sudden increase in the values of \acrshort{mle}s at some threshold value of the inclination angle $i=i_{\text{c}}$ (Figure \ref{fig:lcn_with_i}). For $i<i_{\text{c}}$, the chaoticity increases gradually. As the value of $i$ crosses $i_{\text{c}}$, the \acrshort{mle}s start increasing rapidly until it reaches an angle, after which the growth slows down a little, though the positive correlation still holds. This overall trend of \acrshort{mle}s with respect to the inclination angle $i$ is very consistent, and it is true for any given set of orbital parameters. 

Now, if we compare Figure \ref{fig:lcn_with_i_1} and Figure \ref{fig:lcn_with_i_2}, the threshold value has come down from $i_{\text{c}} \approx 21^\circ$ to $i_{\text{c}} \approx 18^\circ$ as the dipole coefficient increases from $\mathcal{D} = 2 \times 10^{-4}$ to $\mathcal{D} = 2.2 \times 10^{-4}$. These results imply that the value of $i_{\text{c}}$ has an anti-correlation with the degree of chaos in the system. As the chaoticity gets enhanced, the nonlinearity in the system increases, and the value of $i_{\text{c}}$ gets lowered. The explanation of stability, in this case, is similar to what we saw while discussing the dependence of chaoticity on the rotation parameter for the higher values of inclination angles. As before, it seems like the system is trying to sustain its stability as long as possible. But the enhancement in the nonlinearity due to any change in the orbital parameters triggers the system early at a lower value of the inclination angle, and the chaoticity starts increasing rapidly. Thus, the value of $i_{\text{c}}$ gets lowered. 

\section{Conclusions} \label{ch:conclusion}
In the present work, we have studied the chaotic behaviour of the off-equatorial orbits around a pseudo-Newtonian \acrshort{cop} using the generalised force presented in \citet{ghosh2007}. Because of the complex mathematical form of the pseudo-Keplerian force, we have prescribed a numerical method in which a fitting function is used to generate the \acrshort{pnp} and implement it in the analysis of orbital dynamics. To incorporate a more realistic scenario, we have introduced an artificial dipolar perturbative term corresponding to an asymmetrically placed hollow halo of matter around the \acrshort{cop}. To study the chaotic dynamics of the off-equatorial orbits around the \acrshort{cop}, we have implemented the Poincar\'e Maps and the \acrshort{mle}s as the indicators of chaos. Where Poincar\'e maps help to visualise chaos qualitatively, \acrshort{mle}s quantify the degree of chaos. 

At first, we studied stable circular orbits and observed how their stability gets affected by the Kerr parameter $a$ and the angle of inclination $i$. We have found that the \acrshort{isco}s are significantly affected by the Kerr parameter compared to the inclination angle. Thereafter, we studied the chaotic dynamics of the generic equatorial orbits governed by the \acrshort{gmf} \acrshort{pnp} and compared them with those governed by the \acrshort{abn} \acrshort{pnp}. We saw that for the orbits around the co-rotating \acrshort{cop}s, the \acrshort{gmf} \acrshort{pnp} induces more chaos into the system. For the orbits around the counter-rotating \acrshort{cop}s, the effect is the opposite, \textit{i.e.}, the \acrshort{abn} \acrshort{pnp} makes the equatorial orbits more chaotic compared to the \acrshort{gmf} \acrshort{pnp}.

While studying the correlation of chaos with respect to the rotation parameter $a$, we observed that the chaoticity decreases as the value of $a$ increases. It is maximum for the orbits around the maximally counter-rotating \acrshort{cop}s. The degree of chaos for lower values of the inclination angles shows a consistent negative correlation with $a$. Hence, the chaoticity gradually decreases as $a$ increases. However, for higher orbital inclinations, the degree of chaos shows a weak dependence on $a$ at its higher values. Only when $a$ decreases below a threshold value $a_{\text{c}}$, it begins to increase rapidly as $a$ gets lowered. The value of $a_{\text{c}}$ has a positive correlation with the degree of chaos present in the system. If the chaoticity gets enhanced because of the change of any orbital parameters, the value of $a_{\text{c}}$ increases.

We have also established that the chaoticity increases as the orbits become more inclined with the equatorial plane. This has been shown for both Schwarzschild and Kerr-like \acrshort{cop}s. The change in the chaotic nature of the orbits is more rapid at higher inclination angles than the lower ones, in which cases the changes are more gradual. The \acrshort{mle}s show a sudden sharp upturn at a particular threshold value of the inclination angle $i=i_{\text{c}}$. For orbits with inclination $i>i_{\text{c}}$, the change in the degree of chaos is more rapid, and for $i<i_{\text{c}}$, the chaoticity changes gradually. Therefore, we can state that for $i>i_{\text{c}}$, even a tiny change in the inclination angle can affect the chaotic behaviour of the orbits significantly. Furthermore, the value of $i_{\text{c}}$ anti-correlates with the nonlinearity in the system. It increases when the overall degree of chaos decreases due to any corresponding change in the orbital parameters. 

In future, we would like to perform an in-depth analysis of the dynamics of chaos in the current system using other indicators of chaos, namely the Fast Lyapunov Index or FLI \citep{froeschle1997fast, wang2012dynamics}, Small Alignment Index or SALI \citep{skokos2001alignment, skokos2004detecting}, and General Alignment Index or GALI \citep{skokos2007geometrical}, which are more sensitive to chaos than \acrshort{mle}, making them more efficient to distinguish between order and chaos. We would also like to implement the \acrshort{pnp} to study the off-axis motion of a test particle in a restricted three-body system, where the other two bodies are pseudo-Newtonian compact object binaries.

\section*{Data Availability}

No data was used for the research described in the article.


\section*{Acknowledgements}

The authors would like to thank Dr. Sankhasubhra Nag for his helpful suggestions and discussions. The authors would also like to thank Ms. Roopkatha Banerjee, Mr. Samantak Kundu, and Mr. Jayanta Jana for their help in the computational work. In addition, the author SD would also like to acknowledge Dr. Arindam Dey and Dr. Mousumi Mukherjee for taking time out for useful discussions. The author SR would like to acknowledge the Intramural Research Fund provided by St. Xavier's College, Kolkata (sanction No. IMSXC2022-23/005) for the support in this work. Last but not least, the authors would like to acknowledge the anonymous referees for their valuable comments and suggestions. We want to dedicate this paper to all the researchers who continue to research despite facing innumerable professional and personal hardships just for the love of science.




\end{document}